\documentclass[12pt]{iopart}
\usepackage{iopams}
\usepackage{graphicx}
\usepackage{mathrsfs}

\eqnobysec

\begin{document}

\title[Bosenova and Axiverse]{Bosenova and Axiverse}

\author{Hirotaka Yoshino$^1$, Hideo Kodama$^{1,2}$}

\address{$^1$ Institute of Particle and Nuclear Studies,
KEK, Tsukuba, Ibaraki, 305-0801, Japan}
\address{$^2$ Department of Particle and Nuclear Physics,
Graduate University for Advanced Studies, Tsukuba 305-0801, Japan}
\ead{hyoshino@post.kek.jp}
\vspace{10pt}
\begin{indented}
\item[]April 28, 2015
\end{indented}

\begin{abstract}
We report some new interesting features of the dynamics of a string axion field (i.e., a (pseudo-)scalar field with tiny mass with sine-Gordon-type self-interaction) around a rotating black hole in three respects. First, we revisit the calculation of the growth rate of superradiant instability, and show that in some cases, overtone modes have larger growth rates than the fundamental mode with the same angular quantum numbers when the black hole is rapidly rotating. 
Next, we study the dynamical evolution of the scalar field caused by the nonlinear self-interaction,
taking attention to the dependence of the dynamical phenomena
on the axion mass and the modes.  
The cases in which two superradiantly unstable modes are excited simultaneously
are also studied. 
Finally, we report on our preliminary simulations for gravitational wave emission from the dynamical axion cloud in the Schwarzschild background approximation. Our result suggests that fairly strong gravitational wave burst is emitted during the bosenova, which could be detected by the ground-based detectors if it happens in Our Galaxy or nearby galaxies. 
\end{abstract}

\pacs{04.30.Db, 04.70.Bw,14.80.Va,11.25.Wx}

\vspace{2pc}
\noindent{\it Keywords}: black hole, string axion, gravitational wave

\submitto{\CQG}

%
%

%
%
\section{Introduction}

The {\it axiverse} scenario provides us with an interesting possibility
to find evidence for string theories through observation in the cosmological
and astrophysical contexts \cite{Arvanitaki:2009, Arvanitaki:2010}.
String theories (M theory) are formulated in 10 (11) spacetime
dimensions, and the extra dimensions should be compactified to realize our four spacetime
dimensions. Many moduli appear as a result and some of them are expected
to behave as (pseudo-)scalar fields with tiny masses from four-dimensional point of view.
Although such exotic fields are in hidden sector, we may detect signals
from such fundamental fields through weak coupling to standard model
matter/fields or through coupling to gravity (see \cite{Kodama:2011} for an overview).

If such a scalar field with tiny mass exists in nature, what happens
around a black hole? If the black hole is rotating, there is an
ergoregion around the event horizon where the time-translation Killing vector $\partial_t$ becomes spacelike.
Because of this property, the spatial density of the Killing energy
with respect to $\partial_t$  of the scalar field
can become negative in the ergoregion,
although the local proper energy density with respect to a time-like vector
is always positive if the dominant energy condition is satisfied.
The Killing energy density becomes negative when the superradiant condition,
\eref{Eq:superradiant-condition} in \sref{Sec:III}, is satisfied.
Hence, for a quasibound state mode satisfying the superradiant condition,
the Killing energy flux towards the horizon becomes negative.
Because the total Killing energy is conserved and represents the total energy
of the system measured by an observer at spatial infinity,
this implies that the field outside the ergoregion grows exponentially.
This is called the {\it ``superradiant instability''}, and
it is one of the methods to extract energy
from a rotating black hole analogous to the famous Penrose process.
The superradiant instability effectively happens if the Compton wavelength of
the axion mass $\mu$ has the order of the gravitational radius of a black hole, and its typical time scale is about one minute for a solar-mass black hole, and about one year for an intermediate scale massive black hole such as the Sgr A$^*$ at the centre of Our Galaxy.
The superradiant instability was studied also before the appearance of the
axiverse scenario by theoretical motivations
\cite{Detweiler:1980,Zouros:1979,Furuhashi:2004,Strafuss:2004,Cardoso:2005,Dolan:2007}
(see also \cite{Rosa:2009,Rosa:2012,Dolan:2012} for recent works and
\cite{Brito:2015} for a review).
The axiverse scenario has provoked renewed interests in this topic because it suggests
that the superradiant instability may happen in our real universe,
and furthermore, signals from an axion field may be observed by gravitational wave detectors.

If we take account of the possibility that astrophysical black holes
wear clouds of a string axion field, there are many issues to be studied.
Up to now, some work has been done including simulations of scalar field as a test field \cite{Witek:2012,Yoshino:2012},
properties of gravitational wave emissions \cite{Yoshino:2013,Yoshino:2014,Arvanitaki:2014},
simulations including gravitational backreactions \cite{Okawa:2014},
the evolution of black hole parameters \cite{Arvanitaki:2014,Brito:2014}, and so on.
Our primary interest in the present paper is the effect of the
nonlinear self-interaction of the axionic scalar field and its impact
on the gravitational wave emission. Here, the nonlinear self-interaction
means that the potential of the axion field becomes periodic, typically
described as
%
\begin{equation}
U=\mu^2f_{\rm a}^2\left[1-\cos(\Phi/f_{\rm a})\right].
\label{Eq:Potential}
\end{equation}
%
This type of the potential arises by the nonperturbative effect for
the QCD axion, and a similar mechanism is expected to work
also for a string axion. In our previous paper \cite{Yoshino:2012}, we
performed numerical
simulations of a scalar field with the potential \eref{Eq:Potential}, i.e.,
the sine-Gordon field, as a test field in the Kerr background spacetime
for a specific choice of system parameters.
There, we numerically demonstrated that
the self-interaction causes gradual concentration of the axion field configuration
with its superradiant growth and eventually leads to a dynamical collapse of the configuration
associated with an infall of positive energy towards the black hole and
an outflow towards the far region.  We call this violent phenomenon  the {\it ``bosenova''}
in analogy with the bosenova explosion observed in experiments
of condensed matter physics \cite{Donley:2001}.

%
\begin{figure}[tbh]
\centering
\includegraphics[width=0.6\textwidth,bb=0 0 1024 768]{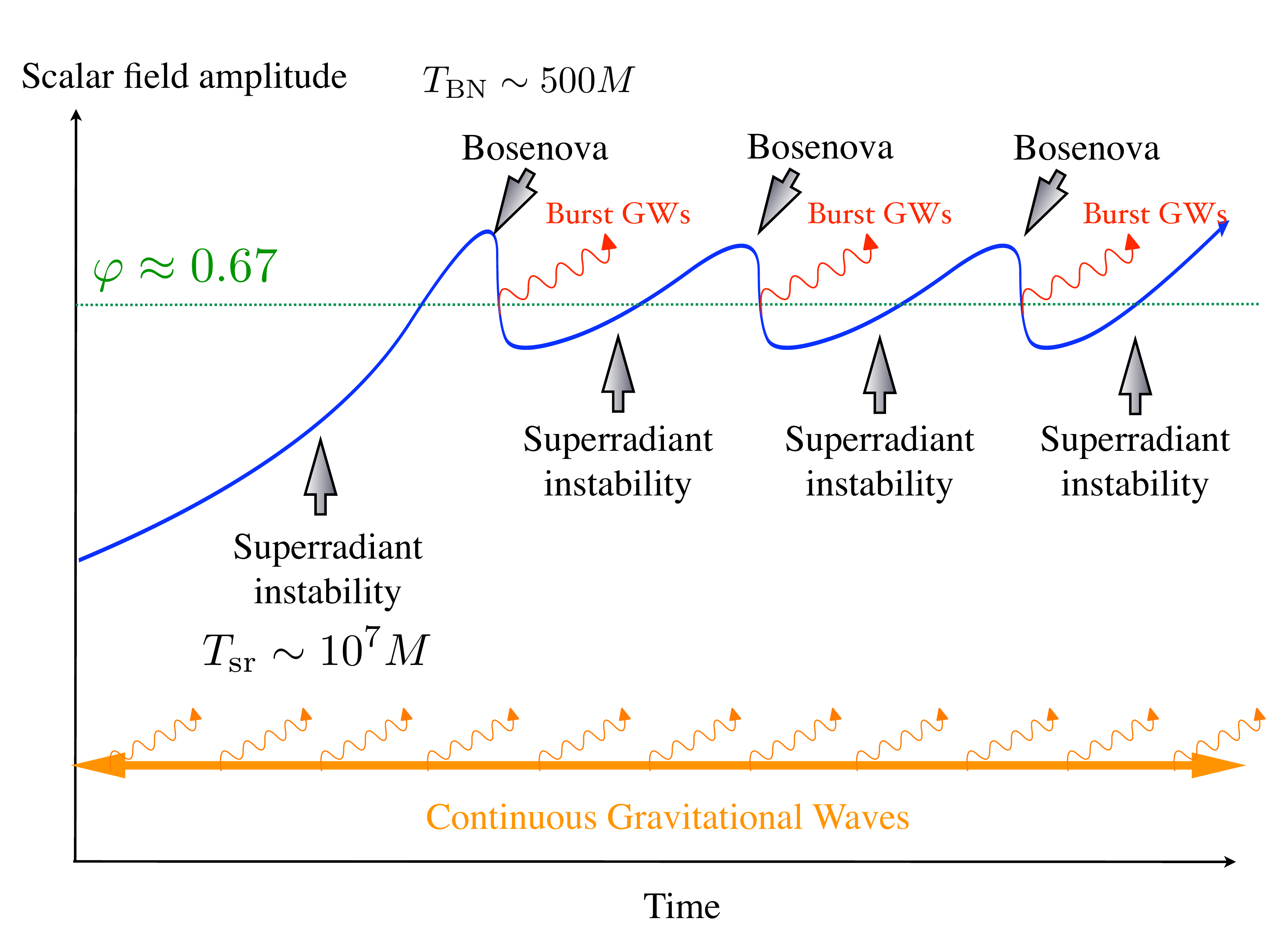}
\caption{
Schematic picture for time evolution of the
scalar cloud energy and emission of gravitational waves. }
\label{schematic-figure}
\end{figure}
%

\Fref{schematic-figure} summarizes the overall picture of phenomena
in the time evolution of an axion cloud with the self-interaction
suggested from our previous results~\cite{Yoshino:2012,Yoshino:2013}.
In the first stage, the energy exponentially grows by the superradiant instability.
At some energy for which the field amplitude becomes $\varphi:=\Phi/f_{\rm a}\sim 1$,
the nonlinear self-interaction becomes important and causes a sudden collapse of the cloud
followed by the infall of the positive energy to the black hole and an outflow
towards the far region. This bosenova event continues for a period $\sim 1000M$,
and then, the system settles to the superradiant phase again.
Bosenova typically happens when the axion cloud extracts a tiny portion
of the black hole rotation energy. After that, bosenova and
superradiant growth happen alternatively.
During each bosenova event, the energy extraction from the black hole
effectively stops, and instead, a part of the axion cloud energy extracted
from the black hole in the precedent superradiant phase is put back to the black hole.
Thus, some part of the rotation energy extracted from the black hole
is recycled.
In the meantine, bosenova is expected to generate burst-type gravitational waves. Further,
continuous gravitational waves are generated by oscillation of the axion field.
Thus, the long-time evolution of the black hole  is determined by the energy loss
by these gravitational wave emissions, if the interaction with surrounding plasma
and magnetic fields are neglected.

In the present paper, we explore these processes more deeply.
In particular, we focus our attention to the following three problems.
First, we revisit the calculation of the superradiant growth rates of the
quasibound states of a massive Klein-Gordon field (without nonlinear
self-interaction).
Although several calculations have been performed for
the fundamental modes (i.e., the modes with zero radial quantum number),
to our knowledge no report has been made for overtone modes (i.e., modes
with nonzero radial quantum number). Using the Leaver's method \cite{Leaver:1985},
we show that there are some cases
where an overtone mode has the largest growth rate, and thus,
consists of the primary component of the axion cloud, depending
on the system parameters and the angular quantum numbers $\ell$ and $m$.
In order to prove that this result is not produced by a mode confusion or an error,
we check our results by highly accurate time-domain simulations.

Next, we extend our simulations on bosenova in our previous paper~\cite{Yoshino:2012}, 
which dealt with the case with a single unstable mode of a specific type initially,  
to a wider class of initial conditions and to more complicated situations. 
To be specific, we perform simulations for several axion mass values and mode parameters
different from that in \cite{Yoshino:2012}. For an axion cloud
in the $\ell = m = 2$ mode, 
our results indicate the possibility that the bosenova collapse does not happen
in contrast to the $\ell = m = 1$ case. Instead, 
a steady outflow of the axion field will be formed at the end point of the superradiant instability. 
We also study the effects of nonlinearity in the case that two unstable modes are excited.

Finally, we address the problem of gravitational wave emissions from the
axion cloud. In our previous papers~\cite{Yoshino:2013}, we calculated
the amplitudes of continuous gravitational waves emitted from this system ignoring the
nonlinear self-interaction (i.e., from quasi-bound states of massive Klein-Gordon field)
and discussed their observational consequences.
The next step is to calculate the emission of gravitational waves from
 a dynamical axion cloud in the presence of nonlinear self-interaction.
Because we already have the time-domain data of the dynamical evolution of the scalar field,
the main task is to solve the perturbation equations for the gravitational field
with the prescribed source term in the Kerr background.
We attack this problem by solving the Teukolsky master equation in the time domain.
At this moment of time, we have finished developing a code to simulate
gravitational waves generated by a scalar field in a Schwarzschild spacetime,
and the extension to the Kerr case is still under progress.
In the present paper, we report the results of test simulations for gravitational wave emissions
from quasibound states of a massive scalar field with self-interaction
around a Schwarzschild black hole. Although this is an artificial setup
in the sense that this system does not cause the superradiant instability
and the initial configuration is prepared by hand, we can obtain a lot of
implications for gravitational waves emitted during the realistic bosenova.

The present paper is organized as follows. In the next section, 
we briefly overview the basics of axions including both QCD and string axions 
from the effective field theory point of view, and thereby explain the meaning 
of the basic axion parameters and the origin of the axion potential.  Then, in \sref{Sec:III},
after briefly overviewing the basics of superradiant instability and
the method to evaluate the instability growth rate, 
we show our numerical results on the growth rates of the axion cloud
by superradiant instability taking attention to overtone modes.
In \sref{Sec:IV}, we study the effects of the nonlinear self-interaction
on the basis of the results of our simulations.
In \sref{Sec:V}, we explain our approach for calculating gravitational
waves emitted from the axion cloud around a Kerr black hole, and
show preliminary results of our simulations in the case of the Schwarzschild black hole.
\Sref{Sec:VI} is devoted to a summary and discussions. In \ref{Appendix:A},
a supplementary explanation on the pseudospectral method for calculating
scalar field behaviour is given. In \ref{Appendix:B},
we give the supplementary informations on the gravitational wave calculation.

%
%
\section{Axions and Axiverse}
\label{Sec:II}

In this section, we briefly overview basic features of axion on the basis of model-independent arguments in order to explain the meaning of the axion parameters and the origin of self-interactions.

\subsection{General effective action for axions}

In this paper, we define an axion as a Goldstone field/boson associated with a spontaneous breaking  of a chiral shift or ${\rm U}(1)$ symmetry. Let us denote such a chiral transformation with the angle parameter $\lambda$ by $\rme^{\rmi\lambda Q_5}$ symbolically. Then, when this symmetry is spontaneously broken, the associated Goldstone field $\phi$ transforms as \cite{Weinberg.S2010B}
%
\begin{equation}
\phi \rightarrow \phi + \lambda f_{\rm a},
\label{ChiralTrf:axion}
\end{equation}
%
where $f_{\rm a}$ is a constant called the {\em axion decay constant} in analogy with the pion decay constant $F_\pi$ for the pion fields as the pseudo-Goldstone fields for the spontaneous breaking of the approximate chiral ${\rm SU}(2)$ symmetry. The axion decay constant is normalized by the condition that $\phi$ has the canonical kinetic term $-(e/2) (\nabla\phi)^2$ ($e=\sqrt{-g}$) and that $\lambda$ has the period $2\pi$.

A symmetry is chiral when it transforms fields with distinction of left and right components asymmetrically. In four spacetime dimensions, spinor fields are the only fields with chirality. Such a chiral transformation can be generally written
%
\begin{equation}
\Psi \mapsto \rme^{\rmi\lambda t \gamma_5} \Psi,
\label{ChiralTrf:spinor}
\end{equation}
%
where $t$ is a hermitian matrix characterizing the transformation. In higher dimensions, bosonic fields can have a chirality. An important example is the 5-form flux in the type IIB supergravity in 10 spacetime dimensions \cite{Green&Schwarz&Witten1987,Polchinski.J1998}.

In four spacetime dimensions, this invariance determines the coupling of the axion field to the spinor fields uniquely in the leading order as follows. First, because the action of the system is invariant under the chiral transformation, the transformation \eref{ChiralTrf:spinor} of the spinor fields with fixed $\phi$ produces the same change in the action as the transformation $\phi \rightarrow \phi -\lambda f_{\rm a}$ with fixed $\Psi$ when $\lambda$ is a constant parameter. Hence, if we redefine $\Psi$ by 
\eref{ChiralTrf:spinor} with $\lambda$ replaced by the field $\lambda=\phi/f_{\rm a}$, then $\phi$ disappears from the terms that do not contain a derivative of $\Psi$. However, the kinetic term of the spinor fields produces a new term as
%
\begin{equation}
-\rmi\bar{\Psi} \gamma^\mu D_\mu \Psi \mapsto
-\rmi\bar{\Psi} \gamma^\mu D_\mu \Psi -\rmi \bar{\Psi}\gamma^\mu
\left(\frac{\rmi}{f_{\rm a}}\partial_\mu\phi t \gamma_5\right)\Psi.
\end{equation}
%
From this, it follows that after the redefinition \eref{ChiralTrf:spinor} of the spinor field with $\lambda=\phi/f_{\rm a}$, the $\phi$-dependent part of the Lagrangian is given by
%
\begin{equation}
e^{-1}{\mathscr L}_{\phi,0}= -\frac12 (\nabla\phi)^2 + \frac1{f_{\rm a}} \partial_\mu\phi \bar{\Psi} \gamma^\mu\gamma_5 t \Psi,
\label{axion:lagrangian:tree}
\end{equation}
%
if the Lagrangian does not contain a derivative term of $\Psi$ of higher dimensions. In this new representation, the chiral transformation does not act on the spinor field $\Psi$ and just shift the axion field.

The action \eref{axion:lagrangian:tree} is not correct in quantum theory. It is because due to the chiral anomaly, the effective Lagrangian transforms as \cite{Bell&Jackiw1969,Adler.S1969,Weinberg.S2010B}%
\footnote{In $\Tr(tt_at_b)$ in this expression, the left and right chirality components of the spinor fields are treated as independent field components.}
%
\begin{equation}
\fl
e^{-1}\delta {\mathscr L}_{\rm eff}= \lambda {\mathscr P};\quad
{\mathscr P}=\sum_{a,b} \Tr(tt_a t_b) \frac1{16\pi^2} F^a\cdot \tilde{F}^b
\equiv\sum_{a,b}\Tr(tt_at_b) \frac1{64\pi^2} \epsilon^{\mu\nu\lambda\sigma}F^a_{\mu\nu}F^b_{\lambda\sigma},
\end{equation}
%
where we have used the notation in which the coupling of the gauge field $A^a=A^a_\mu dx^\mu$ to the fermion reads
%
\begin{equation}
D_\mu \Psi = \left(\partial_\mu - \rmi A_\mu^a t_a\right) \Psi,
\end{equation}
%
and the gauge kinetic term can be written
%
\begin{equation}
e^{-1}{\mathscr L}_A=-\sum_a \frac1{2g_a^2} F^a\cdot F^a;\quad
F^a= \rmd A^a -\frac{\rmi}{2}f^a_{bc} A^b\wedge A^c.
\end{equation}
%
Here, it is understood that the index $a$ runs over all the components of all gauge fields. So, the gauge coupling constants $g_a$ whose index belongs to the same gauge group should be identical.

When we generalize the transformation to a local transformation with $\lambda=\lambda(x)$, because of the shift symmetry of the original Lagrangian, the effective Lagrangian changes as
%
\begin{equation}
e^{-1}\delta {\mathscr L}_{\rm eff} = J^\mu_5 \partial_\mu \lambda + \lambda {\mathscr P},
\end{equation}
%
where $J^\mu_5$ is the current corresponding to the chiral transformation $\rme^{\rmi\lambda t\gamma_5}$. In the path-integral expression for the partition function $Z$ of the system, this change is produced just by the variable redefinition corresponding to the local chiral transformation\cite{Fujikawa.K1984,Fujikawa.K&Suzuki2004B}:
%
\begin{equation}
Z = \int [\rmd\Psi \rmd\bar{\Psi}\cdots] \rme^{\rmi S}
=\int [\rmd\Psi' \rmd\bar{\Psi}'\cdots] \rme^{\rmi S'}
=\int [\rmd\Psi \rmd\bar{\Psi} \cdots] \rme^{\rmi(S +\delta S)}.
\end{equation}
%
Hence, its expectation value has to vanish:
%
\begin{equation}
\left\langle{\nabla_\mu J^\mu_5}\right\rangle={\mathscr P}
\end{equation}
%

This modification of the conservation law for the chiral current can be made consistent with the field equations by adding the term $(\phi/f_{\rm a}){\mathscr P}$ to the tree level Lagrangian \eref{axion:lagrangian:tree}. By this modification, the additional change of the effective action due to the chiral anomaly is reproduced in the tree level. Thus, generalizing to the multiple axion case and including the gravitational Chern-Simons (CS) term from the chiral anomaly, we obtain the following general effective action for axions:
%
\begin{eqnarray}
\fl
e^{-1}{\mathscr L}_{\rm a} = -\frac12 \sum_{\alpha\beta}K^{\alpha\beta} \nabla\phi_\alpha\cdot\nabla \phi_\beta
  + \sum_\alpha \frac1{f_\alpha}\partial_\mu \phi_\alpha (\bar{\Psi} \gamma^\mu \gamma_5 t_\alpha \Psi)
  \nonumber\\
   + \sum_\alpha \frac{\phi_\alpha}{f_\alpha}
  \left(\sum_{a,b}\frac{\xi^\alpha_{ab}}{16\pi^2} F^a\cdot \tilde{F}^b
    + \frac{\Tr(t_\alpha)}{568\pi^2} R_{\mu\nu\lambda\sigma}\tilde{R}^{\mu\nu\lambda\sigma} \right),
\end{eqnarray}
%
where $t_\alpha$ is the matrix defining the chiral transformation $\rme^{\rmi\lambda^\alpha t_\alpha \gamma_5}$, $\phi_\alpha$ is the corresponding axion field and $\xi^\alpha_{ab}=\Tr(t_\alpha t_at_b)$. In the multi axion case, the kinetic term of the axion fields may not be diagonalized if $K^{\alpha\beta}$ depends on other fields. In that case, the definition of the axion decay constant $f_\alpha$ become ambiguous.

Here, note that if the chiral transformations $\rme^{\rmi\lambda^\alpha t_\alpha \gamma_5}$ commute with the gauge transformations, $\xi^\alpha_{ab}$ vanishes unless both $a$ and $b$ belong to Abelian factors or to the same non-Abelian factor of the full gauge group because $\Tr(U^{-1}t_\alpha t_a t_b U)=\Tr(t_\alpha t_a t_b)$ for any gauge transformation $U$. Hence, for axions, $\xi^\alpha_{ab}=\xi^\alpha_A \delta_{ab}$ except for the indices $a$ and $b$ belonging to ${\rm U}(1)$ factors of the full gauge group. If we consider a pseudo-Goldstone boson associated with an approximate symmetry, this rule does not apply and a mixing between a non-Abelian gauge field and an Abelian or non-Abelian gauge field can appear in the CS terms. The most famous example of such mixing is the CS term proportional to $\pi^0 F\cdot \tilde{F}$ giving the main decay channel $\pi^0\rightarrow 2\gamma$, where $F$ is the EM field. This term arises because $\pi^0$ is a pseudo-Goldstone boson associated with the approximate chiral ${\rm SU}(2)$ symmetry in the $(u,d)$ quark sector. In this chiral dynamics approach, $\pi^0$ corresponds to the same transformation matrix $\tau_3$(=the Pauli matrix $\sigma_3$) as that for the third component of the weak ${\rm SU}(2)$ gauge field, and  $\xi=\Tr(\tau_3 Y \tau_3)=1$. This produces the CS coupling of $\pi^0$ to the EM field and the $Z$ boson field.

\subsection{Potential and self-interactions of axion}

As we saw above, the Lagrangian of axions taking account of perturbative quantum corrections can be determined uniquely by the structure of the associated chiral transformations, or equivalently by the chiral currents $J_{5(\alpha)}$,  except for the kinetic metric $K^{\alpha\beta}$ and axion decay constants $f_\alpha$.  This kinetic metric and $f_\alpha$ are independent of the axion fields but may depend on other fields and sensitive to the kinetic structure of the boson sector of the original theory.

In this perturbative level, the theory preserves the shift symmetries exactly. In particular, there appears no mass term or potential for axions.  When non-perturbative effects of non-Abelian gauge fields in strong coupling regions  or stringy instantons are taken into account, however, the system acquires a non-trivial potential and as a consequence, the shift symmetries are broken to discrete symmetries. Although the exact calculation of these non-perturbative effects are out of our reach, they are usually estimated by the two methods \cite{Weinberg.S2010B}: the instanton approach and the effective  Lagrangian approach for chiral symmetry breaking.

We first discuss non-perturbative effects of gauge field instantons. When the axion fields are uniform, the CS action for a $G={\rm SU}(n)$ gauge field $A=A^a t_a$ with $\Tr(t_a t_b)=\delta_{ab}/2$,
%
\begin{equation}
S_{\rm CS}=\theta p_1;\quad
   p_1 =\int \frac{1}{8\pi^2} \Tr(F\wedge F),
\end{equation}
%
is a topological quantity proportional to the 1st Pontrjagin number $p_1$ of the gauge bundle, where $\theta$ is expressed in terms of some constant $\theta_0$ and the axion fields as
%
\begin{equation}
\theta= \theta_0 + \sum_\alpha \frac{\xi^\alpha}{f_\alpha} \phi_\alpha ;\quad
\Tr(t_\alpha t_a t_b)=\xi^\alpha \delta_{ab}.
\label{thetabyphi}
\end{equation}
%
In the Lorentzian spacetime, there exists no classical solution for which $p_1$ is bounded, while in the Euclidean space obtained by the Wick rotation, there may exist a classical solution with finite $p_1$. Such a solution is called an instanton and give a non-trial contribution in the Euclidian path integral formulation.

Topologically, instanton solutions in four dimensions are classified by $\pi_3(G)$. Because $\pi_3({\rm U}(1))=0$, there exists no instanton for an Abelian field, hence the CS term induces no potential. In contrast, because $\pi_3({\rm SU}(n))\cong {\mathbb Z}$($n\ge2$), all non-Abelian ${\rm SU}(n)$ gauge field has instanton solutions topologically classified by the instanton number $n\in {\mathbb Z}$. Among these solutions, those with self-dual or anti-self-dual field strength have the least action for a given instanton number, hence give the dominant contribution in the path integral.  Explicit solutions are known for some cases. The most famous one is the self-dual BPST solution with $p_1=1$ for the ${\rm SU}(2)$ gauge field \cite{Belavin.A&&1975}, which is also an instanton solution for the  ${\rm SU}(n)$ gauge field with $n>2$ by the standard inclusion ${\rm SU}(2)\subset {\rm SU}(n)$. Because the BPST solution has the moduli freedom corresponding to the spacetime translation and the dilation, its contribution to the partition function $Z$  can be expressed as
%
\begin{equation}
Z_1= \tilde{\Lambda}^5 \int \rmd^4x \int \rmd R~\rme^{-S_{\rm E}},
\end{equation}
%
where $\tilde{\Lambda}$ is some mass scale and $R$ is the size of the instanton. For the self-dual solution, the Euclidian instanton action $S_{\rm E}$ is determined by the coupling constant and the instanton number as
%
\begin{equation}
S_{\rm E} = \int \rmd^4x \frac1{g^2}\Tr ({}*\!  F\wedge F)= \pm \frac1{g^2}\int \rmd^4x \Tr(F\wedge F)= \frac{8\pi^2}{g^2}|p_1|.
\end{equation}
%

If we assume that a general instanton solution with $p_1=n$  can be approximated by the linear superpositions of $p$ BPST solutions and of $q$ anti-BPST solutions with $n=p-q$ ({\em the dilute gas approximation}), the total contribution of the instantons to $Z$ can be written
%
\begin{equation}
Z_{\rm inst} = \sum_{p,q\ge0} \frac{Z_1^p} {p!} \frac{Z_{-1}^q}{q!} \rme^{\rmi (p-q)\theta}
 = \exp\left[ \tilde{\Lambda}^5\int \rmd^4x \int \rmd R~\rme^{-8\pi^2/g^2} (\rme^{\rmi\theta}+\rme^{-\rmi\theta})\right].
\end{equation}
%
Hence, the instantons produces the non-perturbative potential
%
\begin{equation}
V= - \Lambda^4 \cos\theta + \mathrm{const.}
=-\Lambda^4 \cos\left(\theta_0+\sum_\alpha \xi^\alpha \phi_\alpha/f_\alpha\right) + \mathrm{const.},
\end{equation}
%
where
%
\begin{equation}
\Lambda^4 = 2 \tilde{\Lambda}^5 \int \rmd R~\rme^{-8\pi^2/g^2}.
\end{equation}
%

From this formal expression, we see that instantons with large size $R$ produces large contribution if the gauge field becomes strongly coupled in the IR region, but we cannot calculate this strong coupling effect properly. In the QCD axion case, $\Lambda$ is of the order of the pion mass as we see below. In the case in which there exists more than one gauge fields with strong coupling regions, the axion potential is given by the sum of the contributions from all such gauge fields. Thus, generically, the potential is given by
%
\begin{equation}
V= -\sum_A \Lambda_A^4 \cos\left(\theta_{A,0}+ \sum_\alpha \xi^\alpha_A\phi_\alpha/f_\alpha\right)+\mathrm{const.}
\end{equation}
%
Here, $\theta_{A,0}$'s are phases determined by the CP phase of the fermion mass matrix and  the $\theta$-angle of the vacuum of each non-Abelian gauge field.

Note that if the number $N_{\rm a}$ of independent axions is smaller than the number $N_{\rm s}$ of strongly coupled gauge fields, some of $\theta_{A,0}$'s may not be set to zero by the constant shift of the axion fields. In this case, CP invariance is not recovered in some gauge sector by the Peccei-Quinn type mechanism. In particular, a strongly coupled gauge field in the hidden sector may interfere the QCD sector to cause a CP violation larger than the experimental upper limit. In contrast, in the case $N_{\rm a}\ge N_{\rm s}$, $\theta_{A,0}$ can be always set to zero by shifts of the axion fields as far as the effects of instantons of four-dimensional gauge fields are taken into account. However, these residual shift symmetries may be broken by stringy effects as discussed below.

The masses of the axions can be calculated by expanding $V$ with $\theta_{A,0}=0$ with respect to $\phi_\alpha$ and diagonalizing the quadratic terms
%
\begin{equation}
V_2= \frac{1}{2}\sum_{\alpha\beta} \left(\sum_A \Lambda_A^4\xi^\alpha_A\xi^\beta_A\right)
\frac{\phi_\alpha\phi_\beta}{f_\alpha f_\beta}.
\end{equation}
%
From this expression, we see that the axion mass is roughly given by
%
\begin{equation}
 m_{\rm a} \sim \frac{\Lambda^2}{f_{\rm a}}
\end{equation}
%
where $\Lambda$ is the largest strong coupling energy scale of the gauge fields coupled to the axion and $f_{\rm a}$ is the axion decay constant.

Another important consequence of this formula is that if $N_{\rm a}>N_{\rm s}$, $N_{\rm a}-N_{\rm s}$ axions should be massless. This problem is serious for the axiverse scenario because $N_{\rm s}$ cannot be become so large. We can easily show that the kinematic mixing of the axions does not change the situation. This problem can be avoided if we take into account string instanton effects \cite{Conlon.J2006,Svrcek.P&Witten2006}

The second method to estimate the non-perturbative effects is to use the effective field theory for composite mesons corresponding to the pseudo-Goldstone bosons associated with a spontaneous breaking of non-anomalous approximate chiral symmetry. For example, let us consider the chiral ${\rm SU}(2)$ symmetry of the QCD system with two quarks $u$ and $d$. First, let us move the $\theta$ phase \eref{thetabyphi} for the color ${\rm SU}(3)$ gauge field to the quark mass matrix sector by the chiral transformation $(u,d)\rightarrow (\rme^{\rmi y_u \theta \gamma_5}u, \rme^{\rmi y_d\theta \gamma_5}d)$ with $y_u+y_d=-1$. Then, the quark mass terms changes as
%
\begin{equation}
\rmi m_u \bar{u} u + \rmi m_d \bar{d} d ~\rightarrow~
\rmi m_u \bar{u} \rme^{2\rmi y_u \theta \gamma_5} u + \rmi m_d \bar{d} \rme^{2\rmi y_d \theta\gamma_5} d,
\end{equation}
%
and the kinetic term of the quarks produces the additional axion quark coupling
%
\begin{equation}
\partial_\mu \theta (y_u \bar{u} \gamma^\mu\gamma_5 u + y_d \bar{d} \gamma^\mu\gamma_5 d).
\end{equation}
%
Now, in the strong coupling regime, the products of quark fields acquire non-vanishing vacuum expectation values as
%
\numparts
\begin{eqnarray}
&& -\rmi \left\langle{\bar{u} u}\right\rangle=
   -\rmi \left\langle{\bar{d} d}\right\rangle=v_c\cos\left(2\pi^0/f_\pi\right),\\
&& -\rmi \left\langle{\bar{u} \gamma_5 u}\right\rangle=
    \rmi \left\langle{\bar{d} \gamma_5 d}\right\rangle=-\rmi v_c\sin\left(2\pi^0/f_\pi\right),\\
&& \left\langle{\bar{u} \gamma_\mu \gamma_5 u}\right\rangle =-\left\langle{\bar{d} \gamma_\mu \gamma_5 d}\right\rangle
 =\frac12 f_\pi \partial_\mu \pi^0,
 \end{eqnarray}
 \endnumparts
%
where $\pi^0$ is the field of the neutral pion and $f_\pi$ is the pion decay constant.
Then, the quark mass term produces the potential
%
\begin{equation}
V_{{\rm a}\pi}=-v_c m_u \cos\left(y_u\theta-2\pi^0/f_\pi\right)
-v_c m_d \cos\left(y_d\theta+2\pi^0/f_\pi\right),
\end{equation}
%
and the derivative coupling between the axions and quarks gives rise to the axion-$\pi^0$ kinetic mixing
%
\begin{equation}
\frac12 \left\{ \sum_\alpha \frac{\nabla_\mu\phi_\alpha}{f_\alpha}(z_u^\alpha-z_d^\alpha)
 + \nabla_\mu \theta (y_u-y_d)\right\}f_\pi \nabla^\mu \pi^0,
\end{equation}
%
where we have set $t_\alpha u=z^\alpha_u u$ and $t_\alpha d=z^\alpha_d d$.

In the case of a single axion, which is supposed to be the QCD axion, the value of $(y_u,y_d)$ is uniquely determined by the requirements that $y_1+y_2=-1$ and that this axion-pion mixing disappears. Then, from  $V_{{\rm a}\pi}$, we obtain the standard formula for the axion mass and the pion mass in terms of $v_c$:
%
\numparts
\begin{eqnarray}
m_\pi^2 &\simeq & 4v_c\frac{m_u + m_d}{f_\pi^2},\\
m_{\rm a}^2 &\simeq & v_c \frac{\xi^2}{f_{\rm a}^2}\frac{m_u m_d}{m_u+m_d}
\simeq \left(\frac{\xi f_\pi}{2f_{\rm a}}\right)^2 \frac{m_u m_d}{(m_u + m_d)^2}m_\pi^2,
\end{eqnarray}
\endnumparts
%
where we have assumed the invisible axion condition $f_{\rm a}\gg f_\pi$.

In contrast, if two or more axions couple to the color ${\rm SU}(3)$ gauge fields or the quarks $u$ and $d$, we cannot eliminate the $\pi^0$-axion mixing by the choice of $y_u$ and $y_d$. This mixing, however, can be eliminated by just shifting the axion fields other than the QCD axion by some constant multiple of $\pi^0$ field. Thus, the axion-pion mixing does not provide additional mass to axions other than the QCD axion, and just yield a tiny correction to the $\pi^0$ mass. Actually, we can show that the kinetic mixing between axions and mesons does not change the number of massless axions \cite{Cicoli.M&Goodsell&Ringwald2012}.

%
\begin{figure}
\centerline{
\includegraphics[width=0.8\textwidth,bb=0 0 591 402]{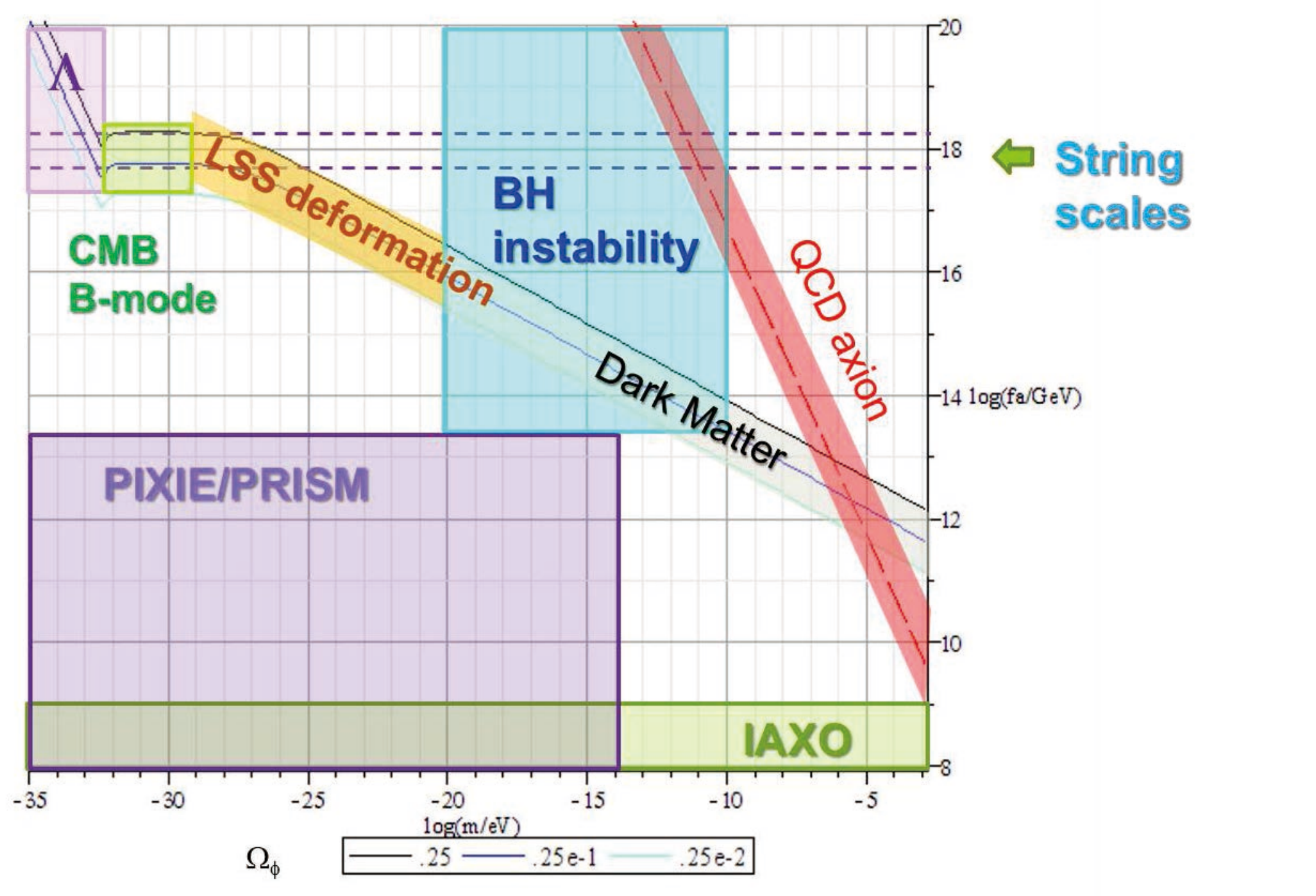}
}
\caption{Various cosmophysical phenomena that may be caused by string axions and constraints by future experiments in the plane of the axion parameters $(m_{\rm a},f_{\rm a})$. The CMB experiments PIXIE \cite{PIXIE}/PRISM\cite{PRISM} and the axion helio scope experiment IAXO \cite{IAXO:whitepaper2014} utilize the CS coupling of an axion to the EM field, $g_{{\rm a}\gamma}\boldsymbol{E}\cdot\boldsymbol{B}$. Hence, they do not directly give constraints on $f_{\rm a}$. In this figure, we have translated the expected sensitivity on $g_{{\rm a}\gamma}$ to the constraint on $f_{\rm a}$ by the relation $g_{{\rm a}\gamma}\approx \alpha/(\pi f_{\rm a})$. }
\label{fig:axion:ma-fa}
\end{figure}
%

\subsection{Axiverse}

In string theory, the compactification of extra dimensions can produce a large number of axions if the internal space has a rich topological structure. This is because the generalized gauge symmetry for form fields contained in the original ten or eleven dimensional supergravity  turn to a shift symmetry of zero modes.  As an example, let us take up the RR 2-form field $C_2 =(1/2) C_{MN}dx^M dx^N$ in the type IIB supergravity in ten-dimensional spacetime. When it is compactified to the product of a four-dimensional spacetime $X_4$ and a six-dimensional Calabi-Yau manifold $Y_6$, the zero mode of $C_2$ can be expanded in terms of harmonic 2-forms $\chi^\alpha_2$ on $Y_6$ up to the gauge transformation $C_2 \rightarrow C_2 + d\sigma_1$ as
%
\begin{equation}
C_2 \sim \omega_2(x) + \sum_\alpha \phi_\alpha(x) \chi_2^\alpha,
\end{equation}
%
where $\omega_2(x)$ and $\phi_\alpha(x)$ are a 2-form and functions on $X_4$, respectively. Because a closed form is locally exact, the theory is also invariant under a transformation $ C_2 \rightarrow C_2 + \beta_2$, where $\beta_2$ is an arbitrary closed form. This implies that the theory is invariant under the shift of the fields, $\phi_\alpha(x) \rightarrow \phi_\alpha (x) + c_\alpha$ with $c_\alpha$ being constants. Further, we can confirm that the term of the form $\sum_{\alpha,j,k} \partial_\mu \phi_\alpha y^{jk}_\alpha \bar{\psi}_j \gamma^\mu \gamma_5 \psi_k$ appears in the effective four-dimensional theory after compactification, where $\psi_j$'s are spinor fields on $X_4$ corresponding to the zero modes of the gravitino field in the internal directions and the dilatino field on $Y_6$. This implies that the current of this shift symmetry $J_{5(\alpha)}^\mu = \sum_{j,k}y^{jk}_\alpha \bar{\psi}_j \gamma^\mu \gamma_5 \psi_k$ is chiral, and $\phi_\alpha$'s become axion fields according to the general arguments in the previous subsections. Thus, we obtain $b_2(Y_6)$ axions from $C_2$, where $b_2$ is the second Betti number.

In the original type IIB theory, there exists no matter gauge fields, and no CS term appears. However, if we introduce D-branes, the CS D-brane action \cite{Harvey.J2005}
%
\begin{equation}
S_{\rm CS}= 2\pi \int_{B^{p+1}} C \wedge \Tr~\rme^{-\frac{B}{\ell_s^2}+ \frac{F}{2\pi}}\frac{\sqrt{\hat{A}(TB)}}{\sqrt{\hat{A}(NB)}},
\end{equation}
%
produces the CS interaction of the axions to gauge fields and the gravitational field on the branes after compactification. This is natural because this CS action is obtained from the anomaly free condition.

All string axions should be described by the effective theory in four dimensions explained in the previous subsections. In particular, their roles in cosmophysical phenomena and experiments are essentially determined  by its mass $m_\alpha$ and axion decay constant $f_\alpha$(see \fref{fig:axion:ma-fa}) \cite{Kodama:2011}. We can probe the compactification structure and string theory behind it mainly through the distribution of the values of these quantities.

From the perspective of searching phenomena that may be caused by axions, it is important to know  plausible predictions of theory on the parameter distribution. As we mentioned in the previous subsection, we can give non-perturbative mass  by instanton effects of low energy gauge fields only to $N_{\rm s}$ axions where $N_{\rm s}$ is the number of gauge fields with strong coupling regimes including those in the hidden sector. Hence, in the axiverse perspective, most of string axions get mass by string or brane instanton effects. In this case, in the mass formula $m_{\rm a} \sim \Lambda^2/f_{\rm a}$,  $f_{\rm a}$ is generally given by $f_{\rm a}\sim M_{\rm pl}/S_{\rm E}$, and $\Lambda^2\sim M^2 \rme^{-S_{\rm E}}$ where $S_{\rm E}$ is the instanton Euclidean action \cite{Svrcek.P&Witten2006,Conlon.J2006}. Svrcek and Witten estimated $S_{\rm E}$ by considering the interference of string axions to the QCD theta phase and obtained the constraint $S_{\rm E} \gtrsim 200$, i.e. $f_{\rm a}\lesssim 10^{16}{\rm GeV}$. On the basis of this result, the axiverse paper \cite{Arvanitaki:2009} argued that $m_{\rm a}$ is distributed uniformly with respect to $\log m_{\rm a}$ because $m_{\rm a}\propto \rme^{-S_{\rm E}}$.

As we have shown in \fref{fig:axion:ma-fa}, various experiments that may detect axions or provide constraints on the axion parameters are planned. These experiments, however, require a sufficient abundance of axions or rather strong coupling of axions and photons, hence a rather small value for the axion decay constant. In contrast, the superradiant instability of an axion field around a rotating black hole that we discuss in the present paper does not require any abundance because the seed can be zero point oscillations of the axion field. Further, the phenomena caused by the instability becomes more violent and energetic for a larger $f_{\rm a}$ as we will see.  Hence, this phenomenon enables us to probe the parameter region complimentary to other experiments.

%
\section{Superradiant instability}
\label{Sec:III}

In this section, we study the growth rate of the superradiant instability
for a massive Klein-Gordon field without nonlinear self-interaction.
We focus on overtone modes.

\subsection{Formulation}

We consider a Kerr black hole whose metric is given by
%
\begin{eqnarray}
\fl
\rmd s^2 = -\frac{\Delta - a^2\sin^2\theta}{\Sigma}\rmd t^2
-\frac{4Mar\sin^2\theta}{\Sigma}\rmd t\rmd \varphi
\nonumber\\
+\left[\frac{(r^2+a^2)^2-\Delta a^2\sin^2\theta}{\Sigma}\right]
\sin^2\theta \rmd \varphi^2
+\frac{\Sigma}{\Delta}\rmd r^2
+\Sigma \rmd \theta^2,
\label{Metric-Kerr-ST}
\end{eqnarray}
%
with
%
\begin{equation}
\Sigma = r^2+a^2\cos^2\theta, \qquad \Delta = r^2+a^2-2Mr,
\end{equation}
%
in the Boyer-Lindquist coordinates. Here, $M$ is the black hole mass,
and the angular momentum  is given by $J=Ma$. We often
use the nondimensional rotation parameter $a_*:=a/M$.
The event horizon is located at $r=r_+:=M+\sqrt{M^2-a^2}$, and it
rotates with the angular velocity $\Omega_{\rm H} = a/(r_+^2+a^2)$.
The important property of the Kerr black hole is the existence
of the ergoregion where the time coordinate basis $\xi^\mu=\left(\partial_t\right)^\mu$ becomes
spacelike, i.e. $\xi^\mu\xi_{\mu}=g_{tt}>0$. Because of this property, the Killing energy of
a particle/boson field with respect to $\partial_t$ can be negative in the ergoregion.

In this section, we study the behaviour of massive Klein-Gordon field,
whose behaviour is governed by the massive Klein-Gordon equation,
%
\begin{equation}
\nabla^2\Phi - \mu^2{U}^\prime =0
\label{massive-KG}
\end{equation}
%
with $U=\Phi^2/2$. The separation of variables is possible in this case as
%
\begin{equation}
\Phi=2\mathrm{Re}\left[\rme^{-\rmi\omega t}R(r)S(\theta)\rme^{\rmi m\phi}\right].
\end{equation}
%
The equations for $S(\theta)$ and $R(r)$ become
%
\begin{equation}
\frac{1}{\sin\theta}
\frac{\rmd}{\rmd\theta}
\left(\sin\theta\frac{\rmd S}{\rmd\theta}\right)
+\left[
-k^2a^2\cos^2\theta-\frac{m^2}{\sin^2\theta}+\Lambda_{\ell m}
\right]S=0,
\label{Eq:angular_Teukolsky}
\end{equation}
\begin{equation}
\frac{\rmd}{\rmd r}\left(\Delta
\frac{\rmd R}{\rmd r}\right)
+\left\{
\frac{\left[(r^2+a^2)\omega-am\right]^2}{\Delta}
-\tilde{\Lambda}_{\ell m}
-\mu^2r^2
\right\}R=0,
\label{Eq:radial_Teukolsky}
\end{equation}
%
where
%
\numparts
\begin{eqnarray}
 k^2=\mu^2-\omega^2,\\
 \tilde{\Lambda}_{\ell m}=\Lambda_{\ell m}+a^2\omega^2-2am\omega.
\end{eqnarray}
\endnumparts
We normalize each angular mode function $S$ as
%
\begin{equation}
\int_0^\pi \rmd\theta \sin\theta
|S|^2=\frac1{2\pi}
\end{equation}
%

In order to study the behaviour of $R$ near the horizon,
it is useful to introduce the tortoise coordinate $r_*$ by
%
\begin{equation}
\rmd r_* = \frac{(r^2+a^2)}{\Delta} \rmd r.
\end{equation}
%
In this coordinate, the horizon is located at $r_*=-\infty$,
and from the regularity of $\Phi$ around the future horizon
in the future Finkelstein coordinates, we find that the behaviour
of the radial function $R(r_*)$ becomes
%
\begin{equation}
R\sim \rme^{-\rmi(\omega-m\Omega_{\rm H})r_*}
\label{Eq:ingoing-condition}
\end{equation}
%
for ingoing waves.

Our problem is as follows. We seek solutions that fall off at infinity
and satisfy the ingoing boundary condition \eref{Eq:ingoing-condition} near the horizon.
Since the two boundary conditions are imposed, the problem becomes an
eigenvalue problem. There is a close analogy between our case and the hydrogen
atom in the quantum mechanics. In the hydrogen atom, the regularity at the origin
and the falling-off condition at infinity are imposed, and as a result,
the energy levels are discretized. Similarly,
in our problem, the angular frequency $\omega$ is discretized. Since there is
a flux towards the horizon, the eigenfrequency becomes a complex number
in general,
%
\begin{equation}
\omega = \omega_{\rm R}+\rmi \omega_{\rm I}.
\end{equation}
%
If the imaginary part $\omega_{\rm I}$ is negative, the field decays in time, and the system
is stable. In contrast, if $\omega_{\rm I}$ is positive, the field grows exponentially,
and the system is unstable. This instability is called ``superradiant instability''.

In order to understand the mechanism of the superradiant instability,
it is helpful to introduce the Killing energy $E(t)$.
The energy-momentum tensor of a scalar field is
%
\begin{equation}
T_{\mu\nu}=\nabla_\mu\Phi\nabla_\nu\Phi
-\frac12 g_{\mu\nu}
\left(\nabla_\rho\Phi\nabla^\rho\Phi+2U(\Phi)\right),
\label{Eq:Energy-momentum-tensor-scalar}
\end{equation}
%
and the conserved energy current $P^\mu$ is introduced
by
%
\begin{equation}
P^\mu = -T^{\mu\nu}\xi_\nu.
\end{equation}
%
After some algebra, the energy contained in the part $r_*\ge r_*^{\rm (in)}$
of the $t=\mathrm{const.}$ hypersurface is found to be
%
\begin{equation}
\eqalign{
{E}(t)
=\int_{t={\rm const}, r_*^{\rm (in)}\le r_*}  P^\mu \rmd\Sigma_\mu
\cr
\fl\quad
= \frac12
\int_{r_*^{\rm (in)}}^{\infty}\left[
(r^2+a^2)\Phi_{,r_*}^2
+\frac{\Sigma\Delta}{r^2+a^2}
\left(
-g^{tt}\dot{\Phi}^2
+g^{\phi\phi}{\Phi}_{,\phi}^2
+g^{\theta\theta}{\Phi}_{,\theta}^2
+2 U(\Phi)
\right)
\right]\rmd r_* \rmd\Omega,}
\end{equation}
%
with $\rmd\Omega=\sin\theta \rmd\theta \rmd\phi$ and $\dot{\Phi}=\rmd \Phi/\rmd t$.
On the other hand, the energy flux towards the horizon
with respect to the time coordinate observed at $r_*=r_*^{\rm (in)}$ is
%
\begin{eqnarray}
F_E(t) = \int_{r_*=r_*^{\rm (in)}} \Phi_{,r_*}\dot{\Phi}(r^2+a^2)\rmd\Omega.
\end{eqnarray}
For these formulas, the energy conservation holds:
\begin{equation}
\dot{E}+F_E=0.
\label{Eq:energy-conservation}
\end{equation}

Now, we direct our attention to the near-horizon region
and assume that the field behaves as \eref{Eq:ingoing-condition}.
The Killing energy density with respect to the tortoise coordinate $\rmd E/\rmd r_*$
and the flux towards the horizon, $F_E$, are
%
\begin{equation}
\frac{\rmd E}{\rmd r_*} \approx F_E \sim 2\omega (\omega-m\Omega_{\rm H})(r_+^2+a^2).
\end{equation}
%
Therefore, if the superradiant condition
%
\begin{equation}
\omega<m\Omega_{\rm H}
\label{Eq:superradiant-condition}
\end{equation}
%
is satisfied, negative energy distributes in the neighborhood of the horizon and
it falls into the black hole. Then, because of the energy conservation \eref{Eq:energy-conservation},
the energy outside of the horizon increases. This explains the
origin of the superradiant instability, i.e. an unstable mode with a positive $\omega_{\rm I}$.

For a fixed $a_*$, the system can be specified by a nondimensional parameter $M\mu$
if we measure lengths in the unit of $M$.
For  $M\mu\ll 1$ and $M\mu\gg 1$, the eigenvalue problem
can be approximately solved by the matched asymptotic expansion method \cite{Detweiler:1980}
and by the WKB method \cite{Zouros:1979}, respectively.
In the matched asymptotic expansion method for $M\mu\ll 1$,
the equation is solved in the near-horizon region and in the far region,
and they are matched to each other in the matching region.
The solution in the far region is the same as that of the hydrogen atom
in quantum mechanics, except that $e^2$ is replaced by $M\mu$.
Therefore, a quasibound state of a massive scalar field
can be regarded as a huge gravitational atom, and
the solutions are labeled by the angular quantum numbers
$\ell$ and $m$, and the radial quantum number $n_{\rm r}=0,1,2,...$ that
characterizes the oscillation in the radial direction (or
equivalently, the principal quantum number $n_{\rm p}=\ell + 1 +n_{\rm r}$).
In this paper, we call the solution with $n_{\rm r}=0$ the fundamental mode,
and the solutions with $n_{\rm r}\ge 1$ the overtone modes.

In the parameter region $M\mu\sim 1$, this eigenvalue problem has to be solved numerically,
and some work has been done on this problem \cite{Furuhashi:2004,Cardoso:2005,Dolan:2007}.
Here, we follow the method by Dolan \cite{Dolan:2007}, where
the radial function $R(r)$ is assumed to have the form of
the infinite series,
%
\begin{equation}
R(r) = (r-r_+)^{-\rmi\sigma}(r-r_-)^{\rmi\sigma + \chi-1}
\rme^{-kr}\sum_{n=0}^{\infty}a_n\left(\frac{r-r_+}{r-r_-}\right)^n,
\label{Eq:radial-infinite-series}
\end{equation}
%
with
%
\begin{eqnarray}
\sigma = -\frac{2Mr_+}{r_+-r_-}\left(\omega-m\Omega_{\rm H}\right)
 \quad
\textrm{and} \quad \chi=\frac{M(2\omega^2-\mu^2)}{k}.
\end{eqnarray}
%
Here, $\mathrm{Re}(k)>0$ is required in order to make the radial function
fall off at the distant region, $r\gg M$. This expression correctly reproduces
the field behaviour near the horizon and at infinity.
Substituting the expression \eref{Eq:radial-infinite-series}
into the radial equation \eref{Eq:radial_Teukolsky}, we have a three-term
recurrence relation for $a_n$: 
%
\numparts
\begin{equation}
a_1=-\frac{\beta_0}{\alpha_0}a_0,
\label{ThreeTerm1}
\end{equation}
\begin{equation}
\alpha_na_{n+1}+\beta_na_n+\gamma_{n}a_{n-1}=0,
\label{ThreeTerm2}
\end{equation}
\endnumparts
%
where
%
\numparts
\begin{eqnarray}
{\alpha}_n &=& (n+1)[n+1-2\rmi\sigma],
\\
{\beta}_n &=& -2n^2 + 2n
\left[
-1+2\rmi\sigma - 2bk -\frac{M}{k}(\mu^2-2\omega^2)
\right]+{\beta}_0,
\\
{\gamma}_n &=& (n-1)\left[ n+1 -2\rmi\sigma - \frac{2M}{k}(2\omega^2-\mu^2)
\right]
+{\gamma}_1,
\end{eqnarray}
\endnumparts
%
with $b=\sqrt{M^2-a^2}$ and 
%
\numparts
\begin{eqnarray}
\fl
{\beta}_0 = a^2k^2 - 2M(M+b)(\mu^2-2\omega^2)
+4M^2\omega^2 - (1-2\rmi M\omega)\left[1+\frac{\rmi}{b}(am-2M^2\omega)\right]
\nonumber\\
+\left[
-2k-\frac{M}{bk}(\mu^2-2\omega^2)
\right]
\left[
b(1-2\rmi M\omega)+\rmi(am-2M^2\omega)
\right]-\Lambda_{\ell m},
\end{eqnarray}
\begin{eqnarray}
\fl
{\gamma}_1 = \frac{M^2(\mu^2-2\omega^2)^2}{k^2}
-\frac{\rmi M(\mu^2-2\omega^2)}{bk}\left[-am+2M^2\omega+2\rmi b(1-\rmi M\omega)\right]
\nonumber\\
+(1-2\rmi M\omega)\left[1+\frac{\rmi}{b}(am-2M^2\omega)\right].
\end{eqnarray}
\endnumparts
%
Although the appearance of these formulas differs from that of \cite{Dolan:2007}, 
it can be checked that they exactly agree.
From the three-term recurrence relations~\eref{ThreeTerm1}
and \eref{ThreeTerm2}, we obtain the following equation
for $\omega$ in the form of the continued fraction:
%
\begin{equation}
0=\beta_0-\frac{\alpha_0\gamma_1}{\beta_1-}
\frac{\alpha_1\gamma_2}{\beta_2-}
\frac{\alpha_2\gamma_3}{\beta_3-}\cdots,
\end{equation}
%
which can be solved numerically
using the Newton-Raphson method. In the numerical computation, we took account of 
the first $10^4$ terms. 
One subtlety in this method is that
we have to calculate the angular eigenvalue $\Lambda_{\ell m}$ simultaneously.
We avoided this problem by using the approximate formula for small $ak$ derived by Seidel \cite{Seidel:1988}
(see also \cite{Berti:2005}),
which is the series expansion up to sixth order. Seidel's approximate formula
gives fairly accurate value in the problem below, and
the error from the truncation of the series expansion
is of $O(10^{-7}\omega_{\rm I})$.

In our previous paper \cite{Kodama:2011}, we developed a numerical code to solve this
problem, and some results are shown in \cite{Kodama:2011,Yoshino:2013}.
In the present paper, we report numerical results for overtone modes obtained by the same numerical code.
As far as we know, all works up to now took attention just to the
fundamental mode $n_{\rm r}=0$, and no results for the overtone modes $n_{\rm r}\ge 1$
have been reported.

\subsection{Numerical results}

The overtone modes are obtained by
just changing the initial guess in the Newton-Raphson method.
To be more specific, we first choose a value of $M\mu$ for which the imaginary
part of the quasibound state solution is small.
Then, we solve the equation for $\omega$ with various initial guess for $\omega/\mu$,
which is taken to be a real number, by gradually increasing from a small value to unity.
Typically, a quasibound state solution with a larger number of $n_{\rm r}$ has a larger real part
of the eigenfrequency $\omega_{\rm R}$.
After several solutions are obtained, we determine the
radial quantum number $n_{\rm r}$ by drawing the wave function explicitly.
The sequence with a fixed $n_{\rm r}$ can be obtained by repeating the
process adopting the solution for some $M\mu$ as the initial guess for a slightly larger/smaller
value of $M\mu$.

Although the Leaver's method has the ability to determine the solution
with arbitrary accuracy in principle, in some cases
the solution becomes inaccurate by hitting the machine accuracy.
This tends to happen as $n_{\rm r}$ or $a_*$ is increased and $M\mu$ is decreased.
Such data are omitted from the figures below.

%
\begin{figure}[tbh]
\centering
\includegraphics[width=0.49\textwidth,bb= 0 0 305 215]{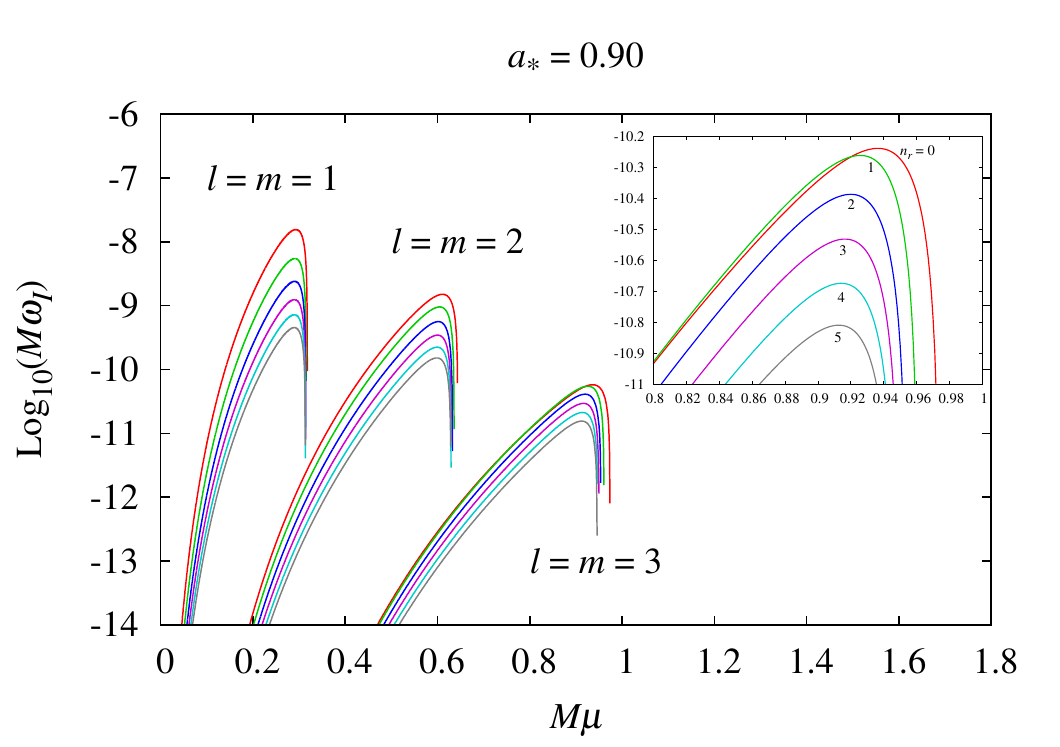}
\includegraphics[width=0.49\textwidth,bb= 0 0 305 215]{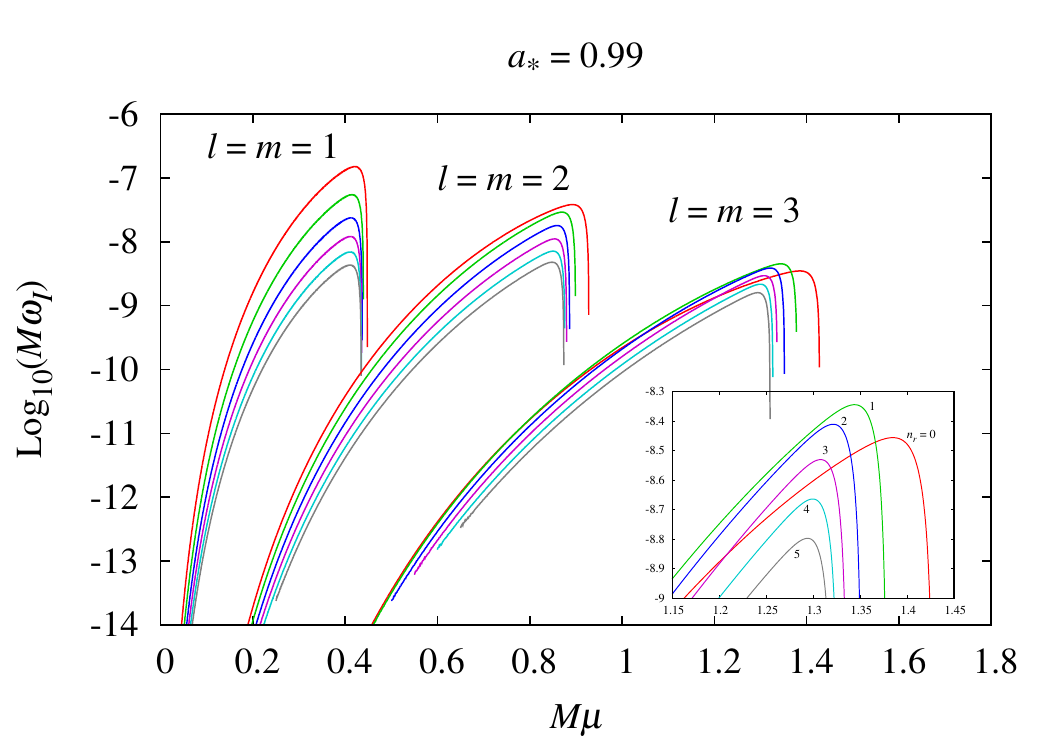}
\includegraphics[width=0.49\textwidth,bb= 0 0 305 215]{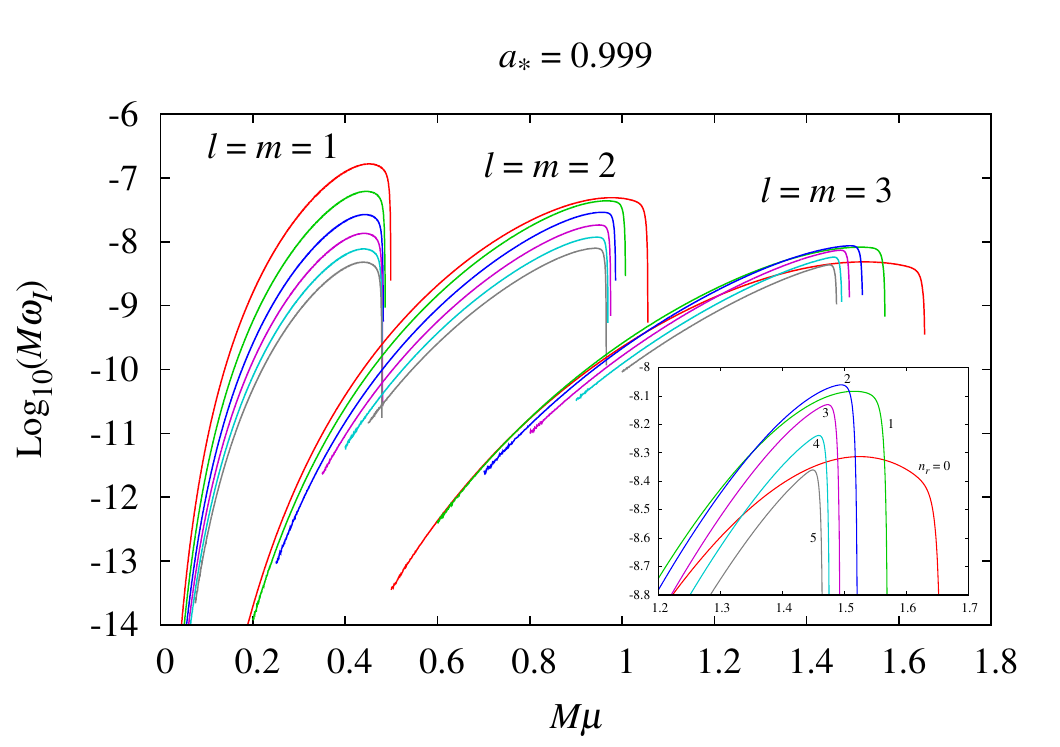}
\caption{The imaginary part of the boundstate frequency, $M\omega_{\rm I}$ (i.e.,
  growth rate of the superradiant instability), 
  for $n_{\rm r}=0$, $1$, $2$, $3$, $4$, and $5$ 
  (red, green, blue, purple, sky blue, and grey, respectively, online) 
  and $\ell = m = 1,2,$ and $3$ for the case
  $a_* = 0.90$ (upper left), $0.99$ (upper right), and $0.999$ (bottom).
  Each inset highlights the peaks of six curves for $\ell = m = 3$.}
\label{Fig:growth-rate}
\end{figure}
%

\Fref{Fig:growth-rate} shows the
values of the imaginary part of the boundstate frequency, $M\omega_{\rm I}$,
for $a_*=0.90$ (upper left), $0.99$ (upper right), and $0.999$ (bottom).
In each panel, the results for the modes with $\ell=m=1,2,$ and $3$ and
$n_{\rm r}=0,...,5$ are shown.
In the case of $\ell= m = 1$, the growth rate of $n_{\rm r}=0$ is greater
than that of $n_{\rm r}=1$, and they
are different by a factor of a few. For $a_*=0.99$
and $M\mu=0.40$, for example, the two growth rates are
different by a factor of $2.6$.  The difference between the two growth rates is
smaller for $\ell=m=2$: For $a_*=0.99$
and $M\mu=0.80$, for example, the difference is $22$\%.

An especially interesting thing happens for $\ell = m = 3$,
as shown in the inset of each figure. In this case, there is a parameter
region of $M\mu$ for which the growth rate of the overtone mode
with $n_{\rm r}=1$ becomes larger than that of the fundamental mode $n_{\rm r}=0$.
This happens in the domain $0.771 \lesssim M\mu\lesssim 0.920$ for $a_*=0.90$
and in the domain $0.772\lesssim M\mu\lesssim 1.366$ for $a_*=0.99$.
It is also worth mentioning that for $a_*=0.99$, there are regions where
the growth rates of the $n_{\rm r}=2$ and $3$ modes are
larger than that of the fundamental mode $n_{\rm r}=0$, although in these regions
the most unstable mode is the $n_{\rm r}=1$ mode.
When the black hole is close to an extremal one, there appears a domain where
the higher overtone mode with $n_{\rm r}=2$ has the greatest growth rate: For $a_*=0.999$, this
happens for $1.353\lesssim M\mu\lesssim 1.508$. On the other hand, the growth rate of the
$n_{\rm r}=1$ mode becomes largest in the domains
$0.773\lesssim M\mu \lesssim 1.352$ and $1.509 \lesssim M\mu\lesssim 1.564$.
In the other domains, $M\mu\lesssim 0.772$ and $1.565\lesssim M\mu\lesssim 1.656$,
the $n_{\rm r}=0$ mode has the largest growth rate.

Up to now, most discussions
on the evolution of axion cloud have been done under the assumption
that the $n_{\rm r}=0$ mode is most unstable.
However, our result indicates that when the black hole is rapidly rotating,
an overtone mode may grow first in the $\ell = m = 3$ mode.
Since this result is very important, we will check its validity
using a different method below.

%
\begin{figure}[tbh]
\centering
\includegraphics[width=0.49\textwidth,bb= 0 0 360 252]{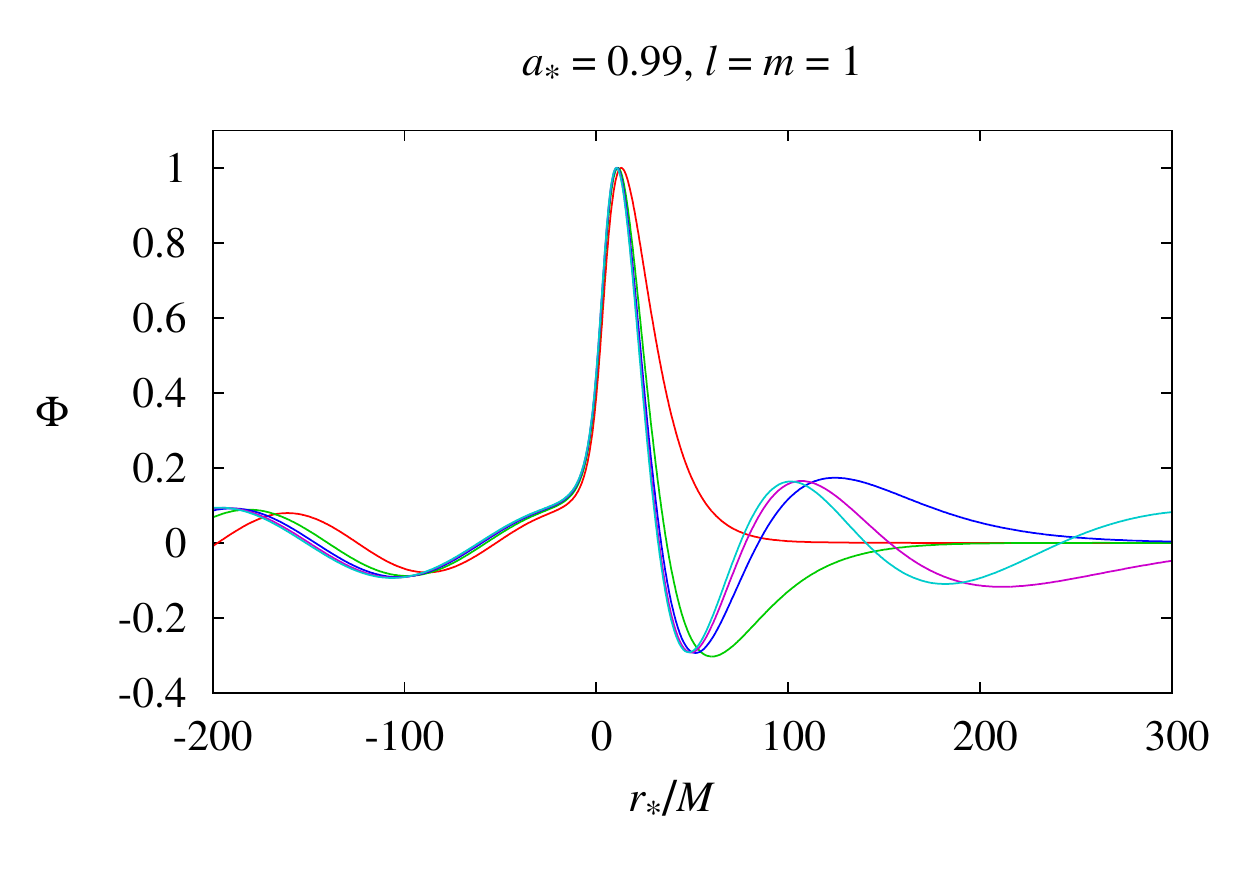}
\includegraphics[width=0.49\textwidth,bb= 0 0 360 252]{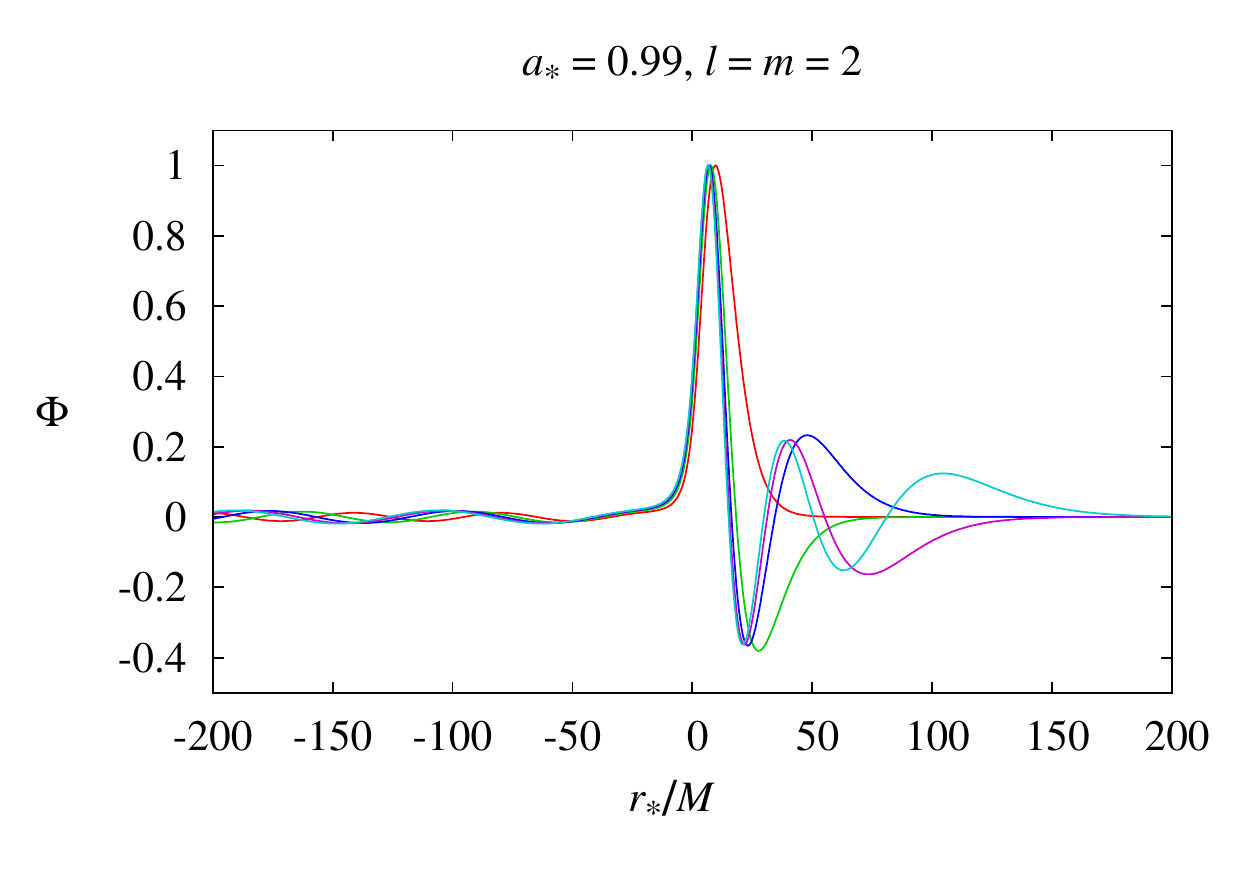}
\includegraphics[width=0.49\textwidth,bb= 0 0 360 252]{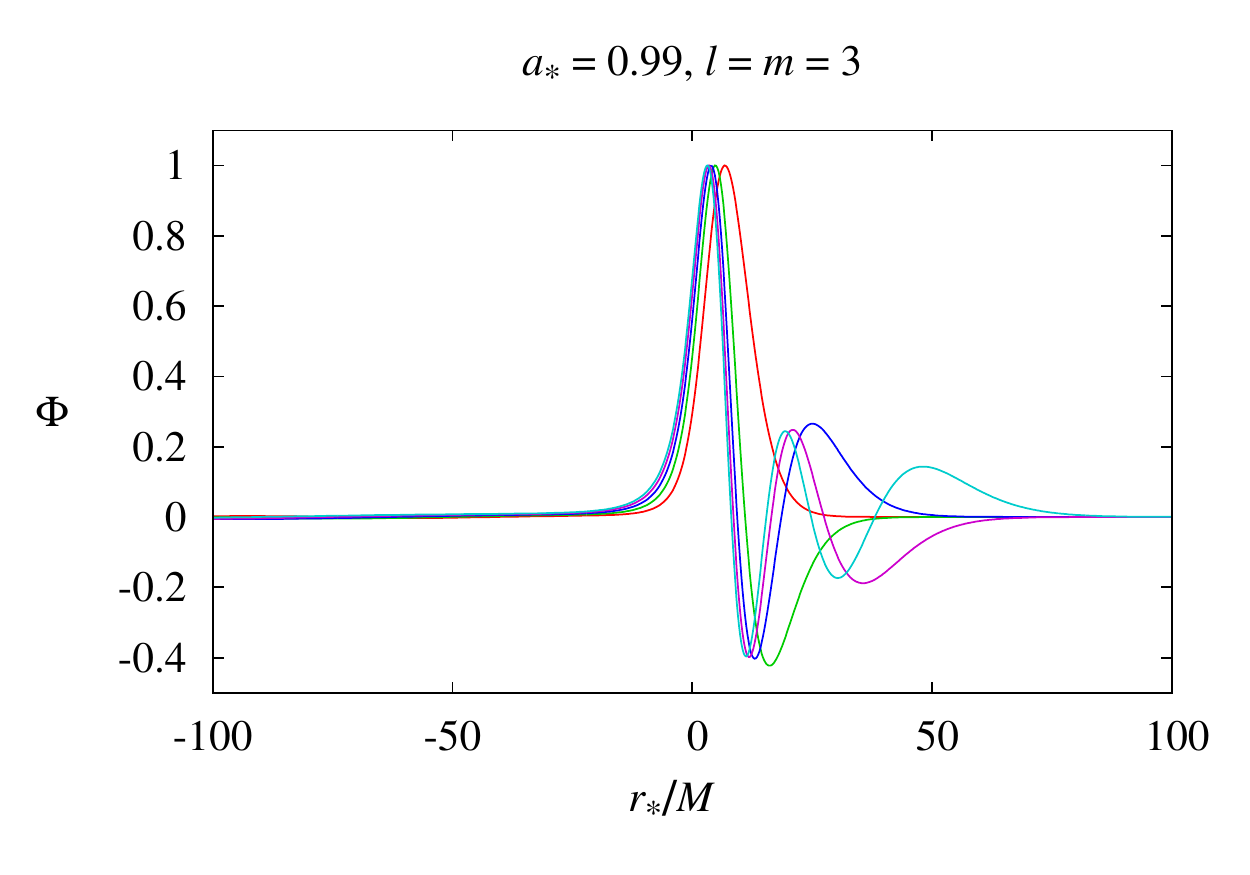}
\caption{Snapshots of the quasibound states of the modes $\ell = m = 1$ (upper left),
  $2$ (upper right), and $3$ (bottom) for $M\mu=0.40$, $0.80$, and $1.30$, respectively.
The black hole rotation parameter is $a_*=0.99$. Five curves
in each figure correspond to $n_{\rm r} = 0$, $1$, $2$, $3$, and $4$ 
(red, green, blue, purple, and sky blue, respectively, online).}
\label{Fig:wavefunctions}
\end{figure}
%

\Fref{Fig:wavefunctions} shows snapshots of the wavefunctions
for $\ell = m = 1$ (upper left), $2$ (upper right), and $3$ (bottom)
in the case $a_*=0.99$.
The values of $M\mu$ are $0.40$, $0.80$, and $1.30$, respectively,
and the modes with radial quantum numbers $n_{\rm r}=0,...,4$ are shown.
The field configuration for $r_*/M\gtrsim 0$ is very analogous to the
wavefunction of the hydrogen atom: The $n_{\rm r}=0$ mode has one local maximum,
the $n_{\rm r}=1$ mode has one local maximum and one local minimum, the $n_{\rm r}=2$ mode
has two local maxima and one minimum, and so on.
The position of the first peak slightly becomes closer to the horizon
as $n_{\rm r}$ is increased, and the mode with larger $n_{\rm r}$
tends to distribute also at the distant region.

\subsection{Time evolution for $\ell = m = 3$}

Because the result for the case $\ell=m=3$ may be counter-intuitive, we have carefully checked the
correctness of the numerical method. First, we compared our numerical data
with those obtained by Dolan \cite{Dolan:2007} for the case $n_{\rm r}=0$.  The two results agree
very well. Next, we calculated the time evolution of the quasibound states
for the situation where the growth rate of $n_{\rm r}=1$ is larger than that of $n_{\rm r}=0$,
starting with the initial data generated by the Leaver's method.
As an example, we chose the system parameters as $a_*=0.99$ and $M\mu=1.35$.
This provides us with a good test because if the result of the Leaver's method is
generated by some mistake or error, the field configuration would include
some spurious data, and then, the field
would relax to a different state and the growth rate would change in the time evolution.

For this purpose, we used our code based on
the pseudospectral method described in the next section.
Our code is sixth-order in the radial direction and fourth-order in the time direction, and
the grid differencing with $\Delta r_* = 0.2M$ and $\Delta t/\Delta r_*=0.05$
turned out to be sufficient to achieve required accuracy.
In order to evaluate the growth rate in the time-domain simulation,
we used the two formulas,
%
\begin{equation}
\omega_{\rm I}^{\rm (1)} = -\frac{F_E}{2E(t)},
~~\textrm{and}~~
\omega_{\rm I}^{\rm (2)} = \frac{\dot{E}}{2E(t)}.
\label{Eq:omega-I}
\end{equation}
%
Here, $E(t)$ denotes the total energy from $r_*=-100M$ to $r_*=300M$,
$\dot{E}:=\rmd E/\rmd t$, and
$F_E$ denotes the flux towards the horizon at $r_*=-100M$.
The latter
formula for $\omega_{\rm I}^{\rm (2)}$ holds because the energy is
quadratic in the field $\Phi$ for the Klein-Gordon field, and $\dot{E}$
is evaluated from the numerical result of $E(t)$.
The former formula for $\omega_{\rm I}^{\rm (1)}$ holds because of the
energy conservation \eref{Eq:energy-conservation}.
Although the energy conservation must hold exactly,
they slightly differ in the numerical simulation because of the numerical error.

%
\begin{figure}[tbh]
\centering
\includegraphics[width=0.45\textwidth,bb= 0 0 360 252]{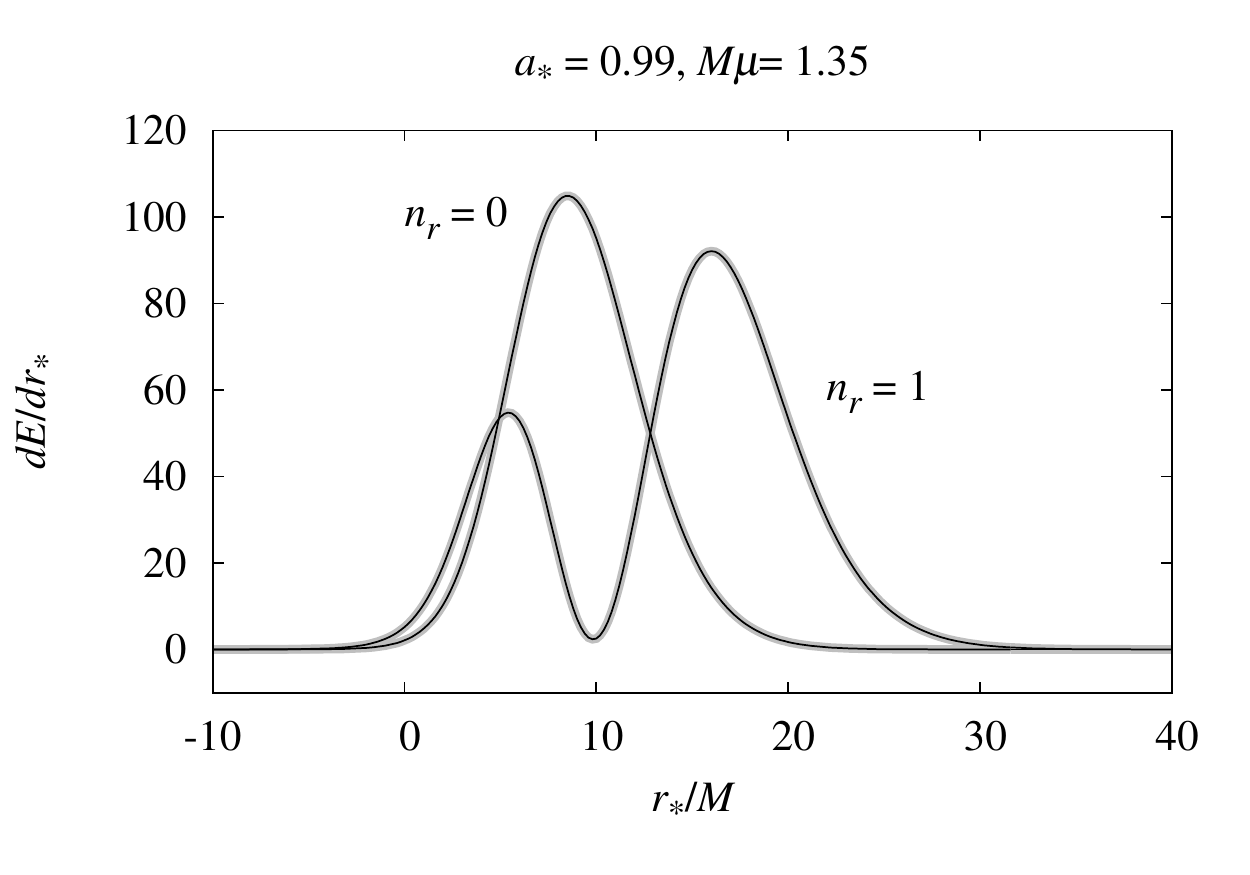}
\includegraphics[width=0.45\textwidth,bb= 0 0 360 252]{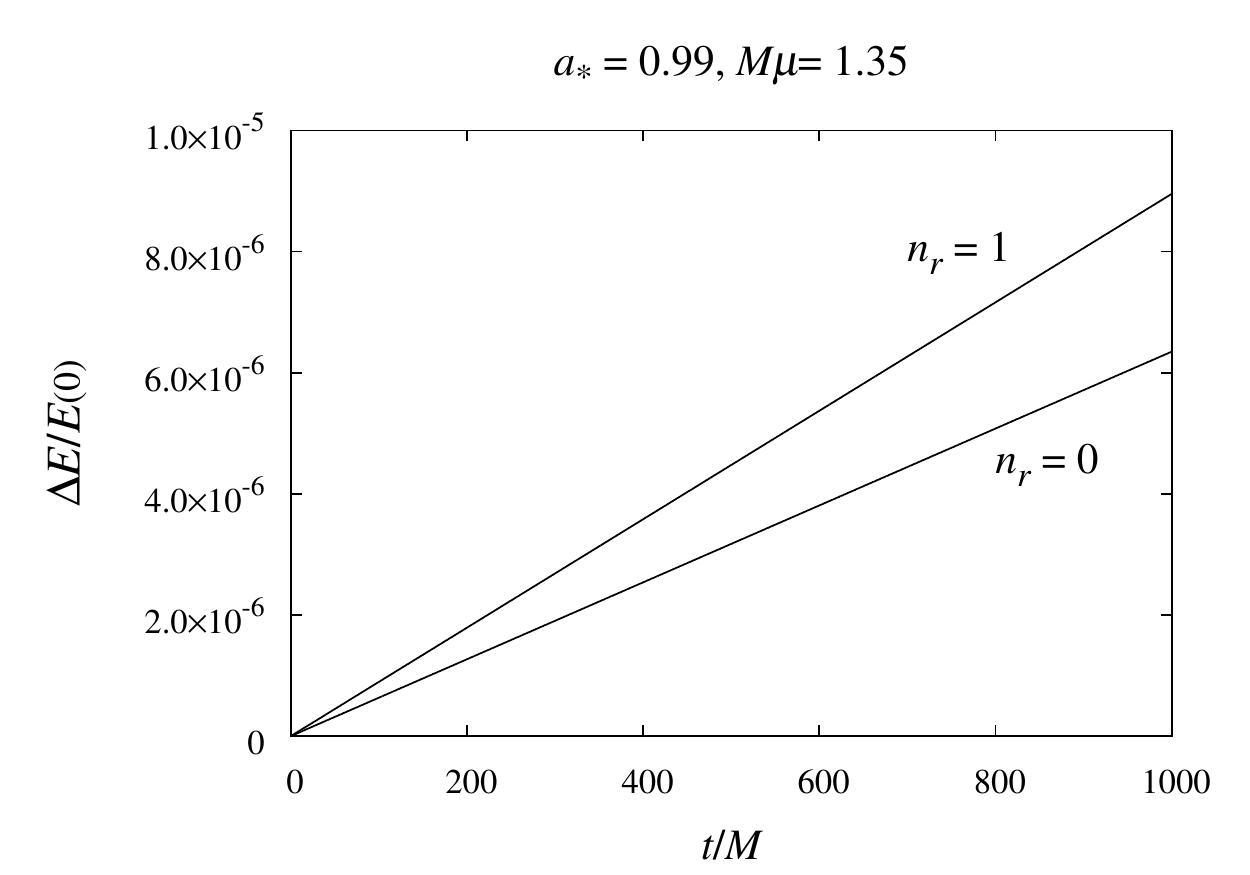}
\caption{Results for the time-domain simulation
of the quasibound states in the modes $\ell=m=3$ and $n_{\rm r}=0$ and $1$.
The system parameters are $a_*=0.99$
and  $M\mu = 1.35$.
Left panel: Comparison of energy density with respect to the tortoise
coordinate $r_*/M$ at  $t=0$ (thick grey lines) and $t=1000M$ (thin black lines)
for $n_{\rm r}=0$ and $1$. The two lines approximately coincide for each $n_{\rm r}$.
Right panel: Time evolution of increase in the energy $\Delta E$ normalized by
the initial total energy $E(0)$ for $n_{\rm r}=0$ and $1$.
The energy increase rate for $n_{\rm r}=1$ is larger compared to that for $n_{\rm r}=0$.}
\label{Fig:lm3-confirmation}
\end{figure}
%

\Fref{Fig:lm3-confirmation} shows the result for the time evolution
up to $t=1000M$.
The left panel shows the energy density $\rmd E/\rmd r_*$ with respect to the tortoise coordinate
at time $t=0$ (thick grey lines) and $1000M$ (thin black lines) for the cases $n_{\rm r}=0$ and $1$.
For each of $n_{\rm r}$, the energy distribution scarcely changes and
we interpret this result as the evidence for the absence of a spurious contribution from numerical errors. Next, let us look at the
energy amount more closely.
The right panel shows the increase in the energy $\Delta E:=E(t)-E(0)$ normalized
by the initial amount of the energy $E(0)$.
The numerical data of the energy $E(t)$ show a clear signal of growth in time,
and the growth rate for $n_{\rm r}=1$ is larger than that for $n_{\rm r}=0$.

%
\begin{table}[tbh]
\caption{\label{Table:growth-rate}Comparison between the growth rate $M\omega_{\rm I}$
of the mode $\ell=m=3$ and $n_{\rm r}=0$ and $1$ for the case $M\mu=1.35$ and $a_*=0.99$,
evaluated by the three different methods.
They are indicated by
$\omega_{\rm I}^{\rm (L)}$, $\omega_{\rm I}^{\rm (1)}$, and $\omega_{\rm I}^{\rm (2)}$
(see text for definitions). All the three methods give consistent values
for each of $n_{\rm r}=0$ and $1$.}
\begin{indented}
\item[]\begin{tabular}{@{}cccc}
\br
~~~~$n_{\rm r}$~~~~ & $M\omega_{\rm I}^{\rm (L)}$ & $M\omega_{\rm I}^{\rm (1)}$ & $M\omega_{\rm I}^{\rm (2)}$\\
\mr
$0$ & $3.1763\times 10^{-9}$ & $3.1754\times 10^{-9}$ & $3.1736\times 10^{-9}$\\
$1$ & $4.4789\times 10^{-9}$ & $4.4784\times 10^{-9}$ & $4.4758\times 10^{-9}$ \\
\br
\end{tabular}
\end{indented}
\end{table}
%

\Tref{Table:growth-rate} summarizes the growth rate $\omega_{\rm I}^{\rm (L)}$ obtained by the
Leaver's method, and $\omega_{\rm I}^{(1)}$
and $\omega_{\rm I}^{(2)}$ by the time evolution. During the time evolution, $\omega_{\rm I}^{(1)}$ and $\omega_{\rm I}^{(2)}$
approximately stay constant, with typical fluctuation of $10^{-13}$ and $10^{-11}$,
respectively. All of these values are consistent for each of $n_{\rm r}=0$ and $1$. This
result is the evidence for the correctness of the results by the Leaver's method:
For $\ell=m=3$, an overtone mode can have a greater growth rate
compared to the fundamental mode when the black hole is rapidly rotating.

This result may seem strange at first. However,
when the radial quantum number $n_{\rm r}$ is increased,
the position of the first peak shifts towards the horizon as shown in \fref{Fig:wavefunctions},
and therefore, the tunneling effect becomes stronger.
In terms of $\omega_{\rm I}^{(1)}$ of \eref{Eq:omega-I},
the numerator $-F_E$ becomes larger if we compare with a fixed value of the first peak.
On the other hand, since a larger amount of the energy distributes at the distant region
for larger $n_{\rm r}$,
the denominator $2E(t)$ also becomes larger.
Therefore, these two effects compete with each other, and it is nontrivial
whether $\omega_{\rm I}$ decreases or not when $n_{\rm r}$ is increased, especially
when $M\mu$ is large.

\subsection{Summary of this section}

In this section, we revisited the calculation of the superradiant instability growth rate
for quasi-bound states of the massive Klein-Gordon field
taking account of the overtone modes.
For $\ell = m = 1$ and $2$,
the growth rate decreases by a factor when the radial quantum number
$n_{\rm r}$ is increased by one. For $\ell = m = 3$, there appears a parameter
region of $M\mu$ where the overtone mode with $n_{\rm r}=1$ gives the
largest growth rate rather than the fundamental mode with $n_{\rm r}=0$.
If the black hole is spinning very rapidly, $a_*=0.999$, there appears a region
where the mode with $n_{\rm r}=2$ gives the largest growth rate.

Our results suggest the importance of overtone modes in studying the evolution
of the axion cloud around a rotating black hole, especially when the angular quantum
numbers $\ell$ and $m$ are large.
Most of the studies on the axion cloud up to now are based on the assumption
that the most unstable mode is the fundamental mode $n_{\rm r}=0$.
For example, in our previous work \cite{Yoshino:2013}, we calculated the gravitational wave
emission rate from the axion clouds in the modes $\ell = m = 1, 2,$ and $3$,
and we used the quasibound state solutions of the fundamental mode $n_{\rm r}=0$
also for $\ell = m = 3$. In order to make the calculation more realistic,
a reconsideration using the most unstable mode is required,
although we postpone it as a future problem.

Also, when $\ell$ and $m$ are one or two,
there is a possibility that the overtone modes contribute to the axion cloud
since the typical time scale is not very different from that of the fundamental mode.
In the next section where the effect of nonlinear self interaction
of axion cloud is studied, we also take attention to the axion cloud
that consists of superposition of the modes $n_{\rm r}=0$ and $1$.

%
%
\section{Axion Bosenova}
\label{Sec:IV}

In this section, we study the non-linear evolution of a massive scalar filed due to self-interaction
in the Kerr-background spacetime.
We adopt the equation \eref{massive-KG} but with the
potential $U$ replaced by \eref{Eq:Potential}. It is quite useful to
normalize $\Phi$ with the decay constant $f_{\rm a}$,
%
\begin{equation}
\varphi=\Phi/f_{\rm a}.
\label{Eq:normalized-scalar}
\end{equation}
%
The equation becomes the sine-Gordon equation,
%
\begin{equation}
\nabla^2\varphi  - \mu^2\sin\varphi = 0.
\label{Eq:Sine-Gordon}
\end{equation}
%
When $\varphi\ll 1$, the field behaviour is well described by the massive Klein-Gordon equation,
and the simple superradiant instability discussed in the previous section takes place.
However, when $\varphi$ grows to order unity, the effect of nonlinear self-interaction becomes
important.

In our previous paper \cite{Yoshino:2012}, we reported only the result
for the specific quasibound state with $\ell=m=1$ and $n_{\rm r}=0$
for the system parameters $a_*=0.99$ and $M\mu=0.40$.
In this section, we show the results of our simulations
for a wider class of initial conditions and to more complicated situations
including  mutimode excitations.

\subsection{Numerical method}

The three-dimensional (3D) code to simulate the dynamics of a scalar field
in the Kerr background was developed in our previous paper \cite{Yoshino:2012}.
Since then we also have developed another code to simulate the same system
based on the pseudospectral method that is sometimes
very powerful as we have seen
in the previous section (but depending on the situation).
We explain these codes briefly one by one.

\subsubsection{3D code}

The 3D code solve \eref{Eq:Sine-Gordon} by the finite differencing method
both in the radial and angular directions. The subtle point is that
the numerical instability rapidly develops
if we use the Boyer-Lindquist coordinates $(t, r_*, \theta, \phi)$. Our prescription
is to adopt $\tilde{\phi}$ defined by
%
\begin{equation}
\tilde{\phi} = \phi - \Omega_{\rm zamo} t.
\end{equation}
%
instead of $\phi$. Here, $\Omega_{\rm zamo}$ is the angular velocity of
the zero-angular-momentum
observers (ZAMO) staying at fixed values of $r_*$ and $\theta$ and rotating
in the $\phi$ direction keeping
vanishing  angular momentum. With this coordinate, stable simulations become
feasible. The interpretation is that the frame associated with the Boyer-Lindquist
coordinates propagates superluminally in the ergoregion, and the numerical evolution
equation cannot respect the causality. On the other hand, the ZAMO frame
are timelike everywhere and the causality is properly included in the evolution equation.
Since the ZAMO coordinates gradually become distorted, we pull back
the coordinates periodically. See \cite{Yoshino:2012} for more details.

\subsubsection{Pseudospectral method}

Our new code is based on the pseudospectral approach, where
the angular dependence of the function $\varphi$ is spectrally
decomposed. First, we decompose the coordinate $\phi$ as
%
\begin{equation}
\varphi =  \sum_{m=0,\pm1,\pm2,...} \sin^{|m|}\theta f^{(m)}(t,r,\theta) \rme^{\rmi m\phi},
\qquad f^{(-m)} = f^{(m)*}.
\label{Eq:decompose-phi-scalar}
\end{equation}
%
The latter condition is imposed in order to guarantee that $\varphi$ is a real number.
As for the $\theta$ direction, we use $y$ defined by
%
\begin{equation}
y=\cos\theta
\end{equation}
%
rather than $\theta$, and decompose the coordinate $y$ as
%
\begin{equation}
f^{(m)}(t,r_*,y)=\sum_{n=0,1,2,...} a_n^{(m)}(t,r_*)y^n.
\label{Eq:decompose-y-scalar}
\end{equation}
%
The expansion of this form is practically same as the expansion
by the spherical harmonics. Then, we solve the equation
for $a_n^{(m)}(t,r_*)$ by the finite difference method with the sixth order in the $r_*$
direction and the fourth-order Runge-Kutta method in the time direction.
In the numerical simulation, we introduce the cut-off parameter
$n_{\rm max}$ and $m_{\rm max}$, and require $a_n^{(m)}=0$ for $n>n_{\rm max}$
and $m>m_{\rm max}$.

In order to derive the equation
for $a_n^{(m)}$, we have to express $\sin\varphi$ also in the form
%
\begin{equation}
\sin\varphi=\sum_{m=0,\pm1,\pm2,...} \sin^{|m|}\theta s^{(m)}(t,r,\theta) \rme^{\rmi m\phi},
\qquad s^{(-m)} = s^{(m)*}
\label{Eq:decompose-phi-sinvarphi}
\end{equation}
%
with
%
\begin{equation}
s^{(m)}(t,r_*,y)=\sum_{n=0,1,2,...} \sigma_n^{(m)}(t,r_*)y^n.
\label{Eq:decompose-y-sinvarphi}
\end{equation}
%
The calculation of $\sigma_n^{(m)}$ can be reduced to the algebraic computation
using the Taylor series of $\sin\varphi$. The details of this method
and the equation for $a_n^{(m)}$ are presented in \ref{Appendix:A}.

We note one subtlety in this decomposition. 
Due to the reality condition $f^{(-m)}=f^{(m)*}$ in \eref{Eq:decompose-phi-scalar}, 
the distinction of the quantum numbers between $+m$ and $-m$ apparently loses a meaning. 
However, it is not the case because $f^{(m)}$ contains both positive frequency 
mode ($\propto \rme^{-\rmi\omega t}$) and the negative frequency mode ($\propto \rme^{\rmi\omega t}$) 
with $\omega\ge 0$. Then, the positive and negative frequency
modes are naturally reinterpreted as the real $+m$ and $-m$ modes, respectively.
Therefore, $f^{(m)}$ introduced in \eref{Eq:decompose-phi-scalar} represents 
the superposition of the $\pm m$ modes.  
In explaining the simulation results below, we often 
refer to the excitation of the $m=-1$ mode defined in this way.

Here, we discuss the features of this new code.
Since the convergence with respect to $m$ is relatively fast,
the simulation is not as heavy as the 3D code.
Also, the 3D code requires a small time step for a stable simulation,
$\Delta t/\Delta r_*=0.05$, while the pseudospectral code
allows relatively large time step, $\Delta t/\Delta r_*=0.25$,
although a small time step must be adopted for a calculation that requires
a high accuracy like the one in the previous section.
In the case where nonlinear self-interaction is present,
the calculation of $\sigma_n^{(m)}$ causes a large number of operations,
and the simulation time becomes much longer compared to the
massive Klein-Gordon case.
But in general,
the calculation of the pseudospectral code is much faster compared
to the 3D code.

The numerical stability is a little problematic. In the pseudospectral approach,
a stable simulation is available as long as the modes with
large $n$ remain small, and the results by the two codes
are consistent. But when violent dynamics causes a
huge excitation
of the modes with large $n$, the simulation easily crashes.
Although simulations of bosenovae could be done in some cases,
we found many cases where a bosenova causes a crash of the simulation.
However, since the simulation is faster compared to the 3D code,
it can be used as a tool to search for the system parameters that lead to
interesting phenomena. In this sense, having the two codes is fairly useful
for our investigations.

\subsection{Simulations}

Now, we present the results of several simulations including nonlinear self-interaction effects.
The setups are summarized in \Tref{Table:simulations}. 
In all simulations, we fix the black hole rotation parameter to be $a_*=0.99$.
We choose the initial condition to be (sum of) the quasibound state solutions
in the Klein-Gordon case (in the absence of nonlinear interaction). 
Then, the system parameter is specified by $M\mu$ and
the amplitudes of the modes that are labeled by the angular quantum numbers
$(\ell, m)$ and the radial quantum number $n_{\rm r}$.

%
\begin{table}[tbh]
\caption{\label{Table:simulations}Setup of the performed simulations in this paper.}
\begin{indented}
\item[]\begin{tabular}{@{}cccc}
\br
Simulations & Initial $(\ell, m, n_{\rm r})$ $[\textrm{Amplitude}]$ & ~~~~$M\mu$~~~~ & Bosenova? \\
\mr
(1a)  & $(1,1,0)$ $[0.40]$  & $0.30$ & No\\
(1b)  & $(1,1,0)$ $[0.45]$  & $0.30$ & Yes\\
(2)  & $(2,2,0)$ $[1.00]$  & $0.80$ & No (but see text)\\
(3)  & $(1,1,0)$ $[0.70]$ and $(2,2,0)$ $[0.01]$  & $0.40$ & Yes\\
(4)  & $(1,1,0)$ $[0.60]$ and $(1,1,1)$ $[0.20]$ &  $0.40$ & Yes\\
\br
\end{tabular}
\end{indented}
\end{table}
%

In \sref{Sec:Pure_lm1}, we present the simulations (1a) and (1b), 
where the initial condition
is chosen to be a pure mode with $\ell = m = 1$ and $n_{\rm r}=0$
for $M\mu=0.30$. 
The same situation was studied in our previous paper \cite{Yoshino:2012} but for $M\mu=0.40$.
Here, we are interested in the dependence on the
axion mass, especially how the condition for the bosenova occurrence
changes with the mass. 
In \sref{Sec:Pure_lm2}, we present the results for the case
(2) where the initial condition is represented by a pure mode with
$\ell = m = 2$ and $n_{\rm r}=0$.

In simulations (3) and (4),  we study what happens 
if the initial condition is given by superposition of two modes. 
In \sref{Sec:Superposition_lm1_lm2}, we consider the 
case (3) that initially the axion cloud is dominated by the $\ell = m = 1$
mode but a small contribution of the $\ell = m = 2$ mode is present. 
This situation is realistic because for the chosen mass, $M\mu=0.40$,
both the $\ell = m = 1$ and $2$ modes are unstable and the growth rate of
the $\ell = m = 2$ mode is much smaller than that of the $\ell = m = 1$ mode.
Here, we are interested in whether the $\ell=m=2$ mode can continue to grow or not
in the situation where the nonlinear effect of the $\ell = m = 1$ is strong.
In \sref{Sec:Superposition_nr0_nr1}, we consider the superposition of
the fundamental  and overtone modes ($n_{\rm r}=0$ and $1$, respectively) in the
$\ell = m = 1$ case and study to what extent the property of the bosenova is
changed. 

\subsubsection{Pure $\ell = m = 1$ mode}
\label{Sec:Pure_lm1}

We begin with the cases (1a) and (1b) where 
the simulation starts with a pure  mode with $\ell  = m = 1$ and  $n_{\rm r}=0$. 
The initial amplitude of the first peak is chosen to be $0.40$ and $0.45$
in the cases (1a) and (1b), respectively.

%
\begin{figure}[tbh]
\centering
\includegraphics[width=0.65\textwidth,bb= 0 0 360 252]{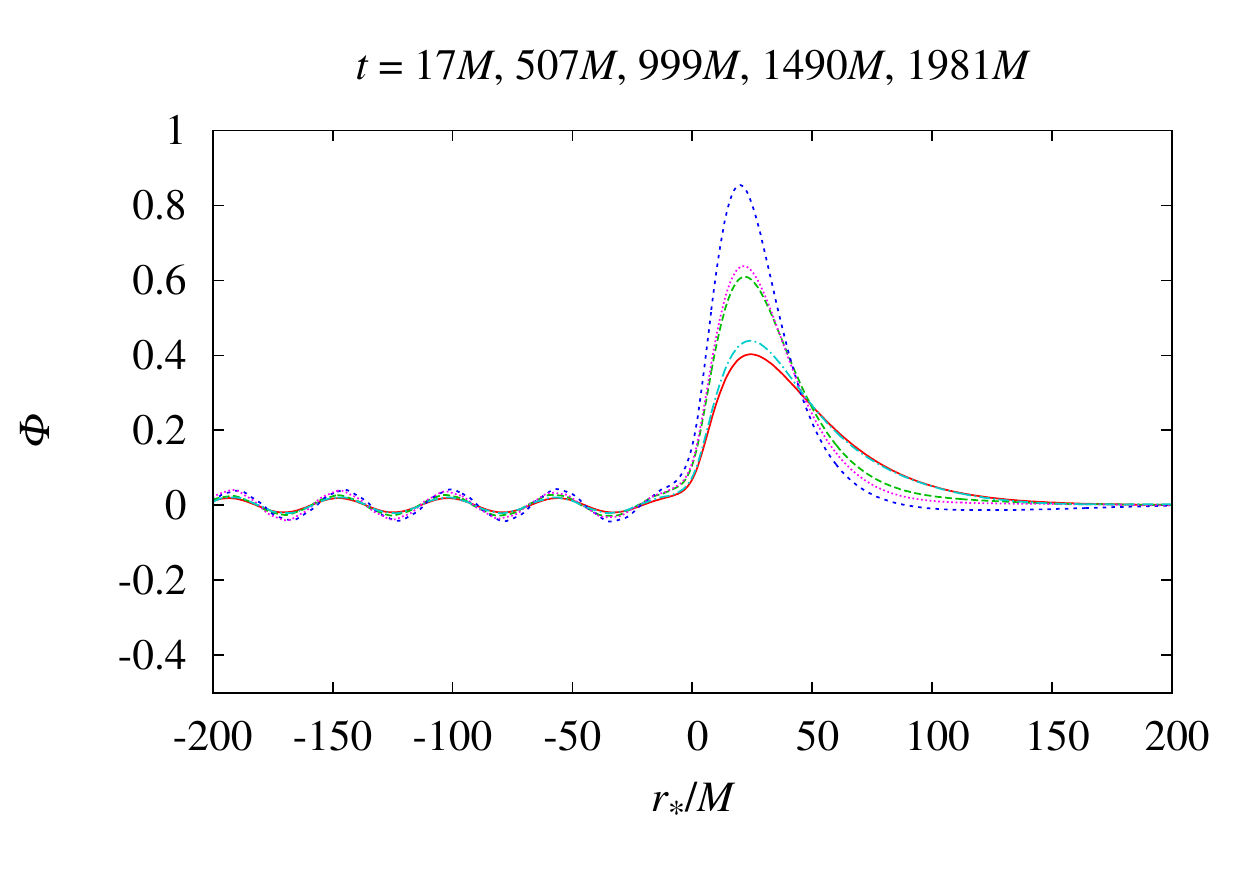}
\caption{Snapshots of the scalar field on the $\phi=0$ line in the equatorial plane
for $t=17M$ (solid line, red online),
$507M$ (long-dashed line, green online), $999M$ (short-dashed line, blue online),
$1490M$ (dotted line, purple online), and $1981M$ (dash-dotted line, sky blue online)
in the simulation (1a).
Time is chosen in order to depict the moments when the scalar
field peak passes the $\phi=0$ line.}
\label{Fig:Snapshots-scalar-amp04-lm1-a099}
\end{figure}
%

In the simulation (1a), the nonlinear effect is not so strong. 
\Fref{Fig:Snapshots-scalar-amp04-lm1-a099} shows the profile
of the scalar field on the $\phi=0$ line in the equatorial plane
at five moments when the scalar field peak passes the $\phi=0$ line.
During $0\lesssim t/M\lesssim 1000$, the axion cloud becomes
concentrated around the black hole and the peak value becomes larger.
But after that, the axion cloud recedes from the black hole,
and the profile at $t/M\approx 2000M$ resembles the initial profile.
Therefore, in this case, the nonlinear self-interaction effect just causes the radial
oscillation of the axion cloud  with the period of $\approx 2000M$.

The superradiant instability continues throughout the simulation (1a),
and the averaged 
rate of the energy extraction is given 
by $\langle \rmd E/\rmd t\rangle  \approx 4\times 10^{-4}(f_{\rm a}/M_{\rm pl})^2$
with the Planck mass $M_{\rm pl} \sim 10^{19}~\mathrm{GeV}$. This value 
is larger by a factor of $\sim 3$
than that of the Klein-Gordon field with the same
amplitude due to the nonlinear self-interaction effect. 
The energy extraction is expected to change the
axion cloud energy from that of  (1a) to that of (1b), which are
$E/M\approx 2.6\times 10^3(f_{\rm a}/M_{\rm pl})^2$ and $\approx 3.3\times 10^3(f_{\rm a}/M_{\rm pl})^2$, respectively. 
The required time for the growth from (1a) to (1b) is estimated to be
$\lesssim 1.7\times 10^6 M$.

%
\begin{figure}[tbh]
\centering
\includegraphics[width=0.40\textwidth,bb=0 0 250 230]{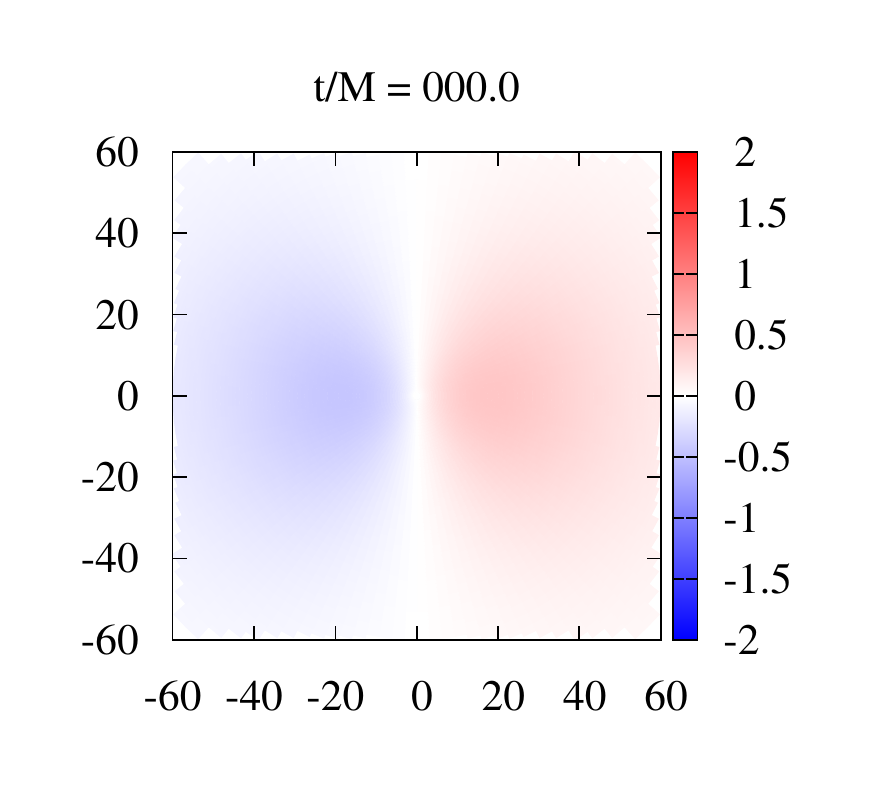}
\includegraphics[width=0.40\textwidth,bb=0 0 250 230]{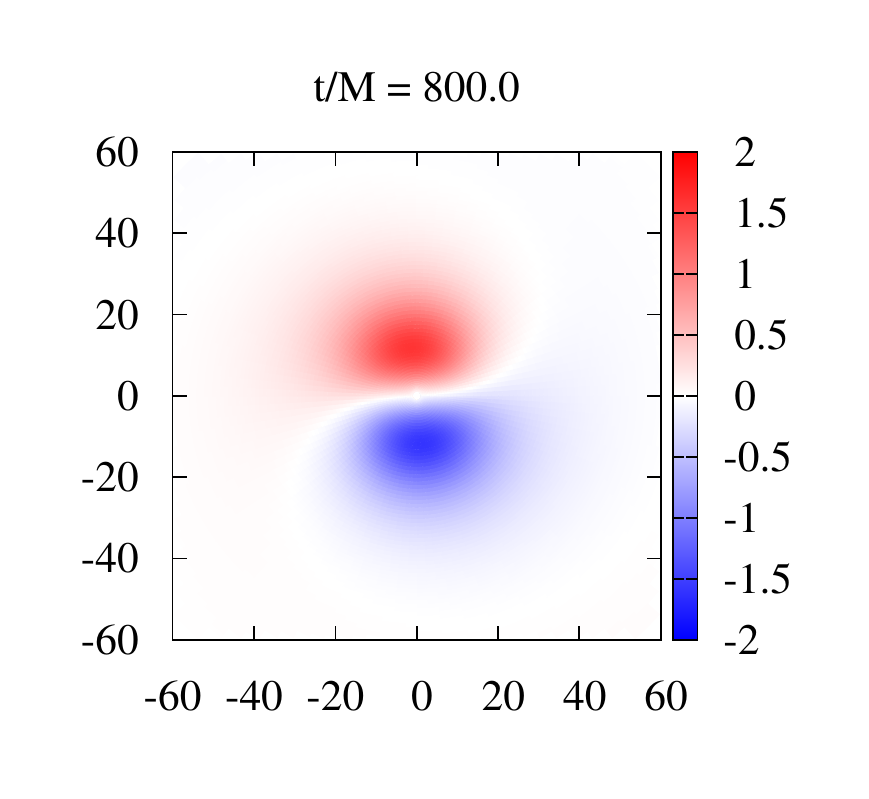}
\includegraphics[width=0.40\textwidth,bb=0 0 250 230]{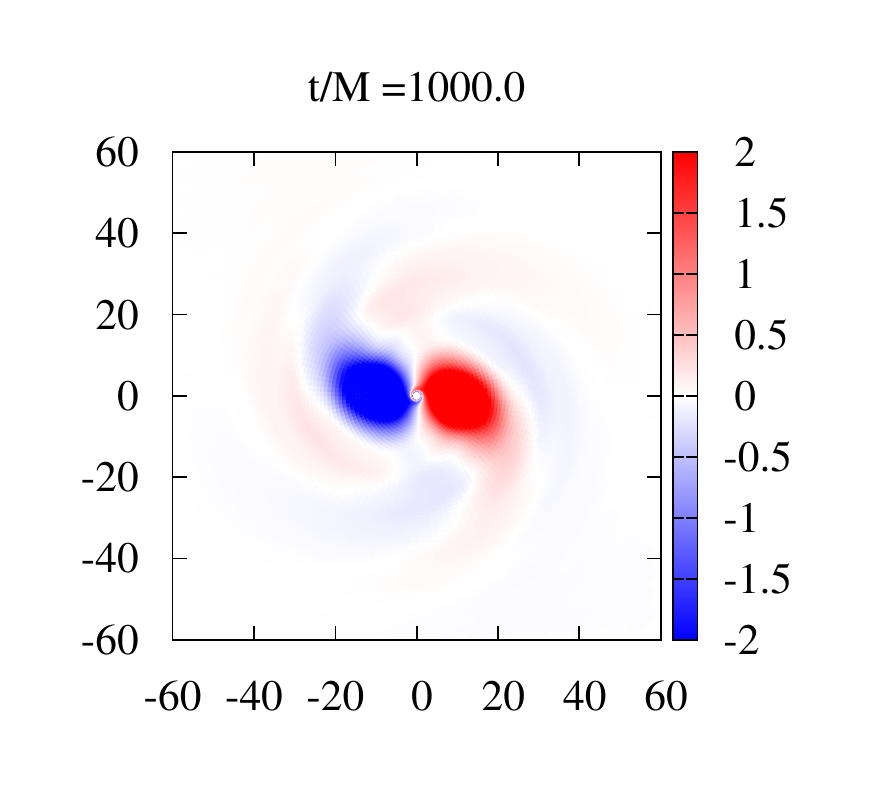}
\includegraphics[width=0.40\textwidth,bb=0 0 250 230]{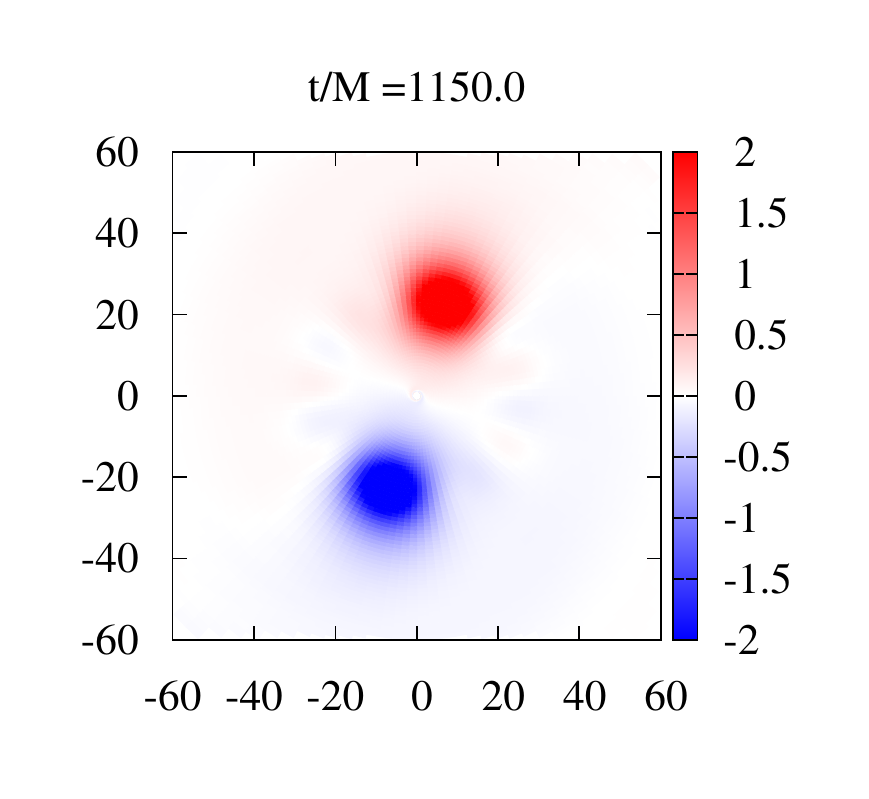}
\includegraphics[width=0.40\textwidth,bb=0 0 250 230]{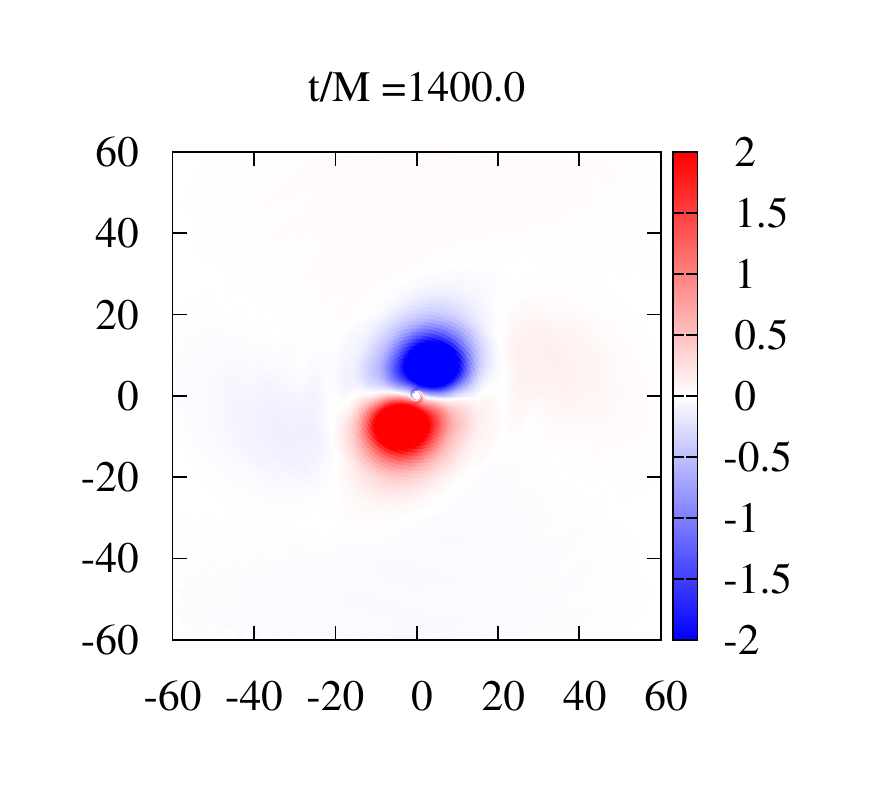}
\includegraphics[width=0.40\textwidth,bb=0 0 250 230]{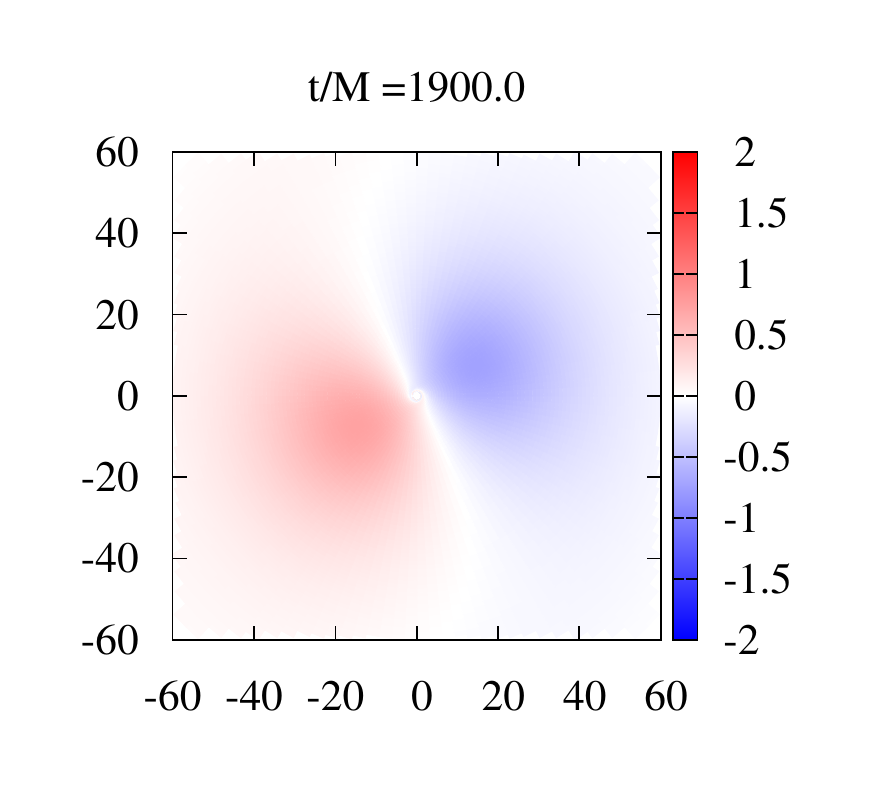}

\caption{Snapshots of the scalar field in the strongly nonlinear, $\ell = m = 1$ case [simulation (1b)]
at $t=0$ (top left), $800M$ (top right), $1000M$ (middle left), $1150M$ (middle right), 
$1400M$ (bottom left), and $1900M$ (bottom right) 
in the equatorial plane in the simulation (1b).}
\label{Fig:Snapshots-2D-a099-L1M1}
\end{figure}
%

In the simulation (1b), the field behaviour is fairly different from the case of (1a). 
\Fref{Fig:Snapshots-2D-a099-L1M1} shows the snapshots of the scalar
field value in the equatorial plane at $t=0$ (top left), $800M$ (top right), 
$1000M$ (middle left), $1150M$ (middle right), 
$1400M$ (bottom left), and $1900M$ (bottom right). 
The axion cloud gradually becomes concentrated around the black hole ($t=800M$),
and around $t=1000M$, the axion cloud becomes strongly distorted
by the excitation of various modes, especially the $m=-1$ mode that
falls into the black hole. 
After that, the axion cloud recedes from the black hole for a while ($t=1150M$).
The axion cloud in this intermediate state consists of two compact solitonlike objects
slowly rotating around the black hole. 
These solitons approach the black hole again around $t=1400M$, exciting
various modes. After that the axion cloud settles down to a state
somewhat similar to the initial state but still continuing the excitation of the $m=-1$ mode
($t=1900M$). This is the bosenova collapse in the case $M\mu=0.30$.

%
\begin{figure}[tbh]
\centering
\includegraphics[width=0.95\textwidth,bb= 0 0 407 171]{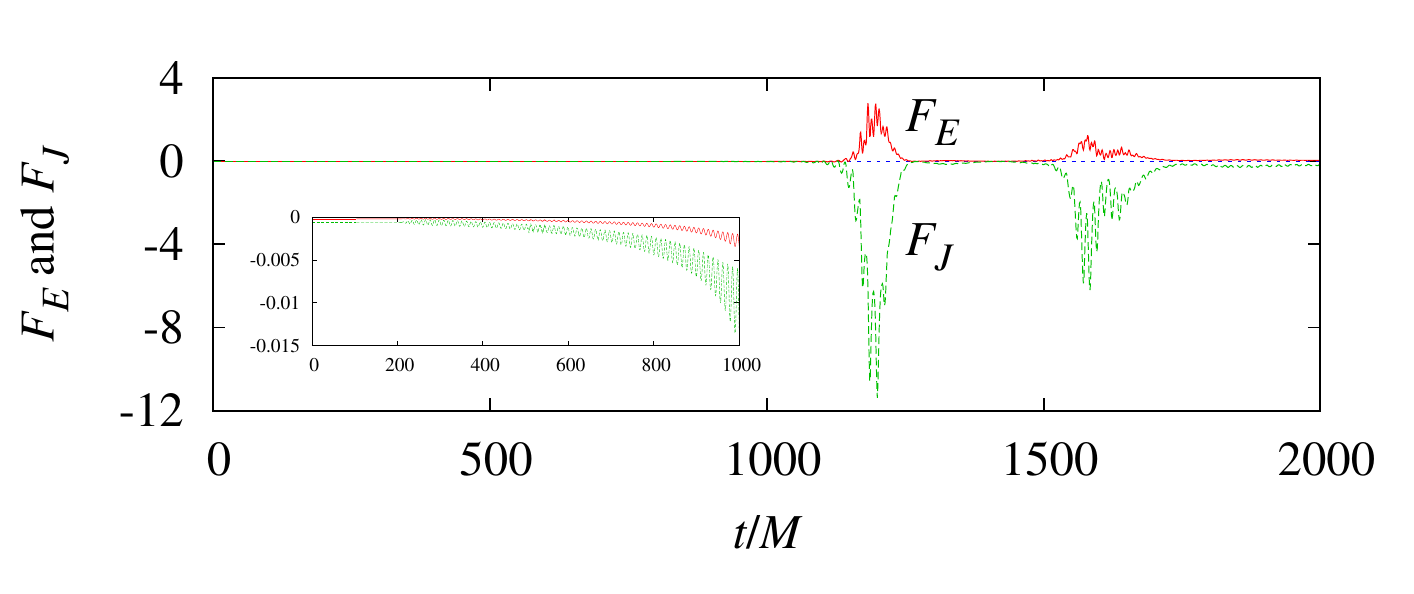}
\caption{Energy flux (solid line, red online) and angular momentum flux 
(dashed line, green online) towards the horizon observed
at $r_*=-200M$ as functions of time in the simulation (1b). The inset enlarges the period $0\le t/M\le 1000$.}
\label{Fig:FEFJ-a099-L1M1}
\end{figure}
%

\Fref{Fig:FEFJ-a099-L1M1} shows the energy and angular momentum fluxes
towards the horizon observed at $r_*=-200M$ as functions of time. 
The inset enlarges the period $0\le t/M\le 1000$. In this period,
both the energy and angular momentum fluxes are negative, and the energy is extracted
from the black hole. During $1000\le t/M\le 2000$,  
there are two periods when large positive energy rapidly falls into the
black hole, corresponding to the high concentration of the axion cloud
at $t/M\approx 1000$ and $1400$. The bosenova collapse
is characterized by the excitation of the $m=-1$ mode,
which carries positive energy and negative angular momentum to the black hole terminating
the superradiant instability. 
During this period, about 5.9\% of the total energy falls into the black hole.

%
\begin{figure}[tbh]
\centering
\includegraphics[width=0.6\textwidth,bb=0 0 360 252]{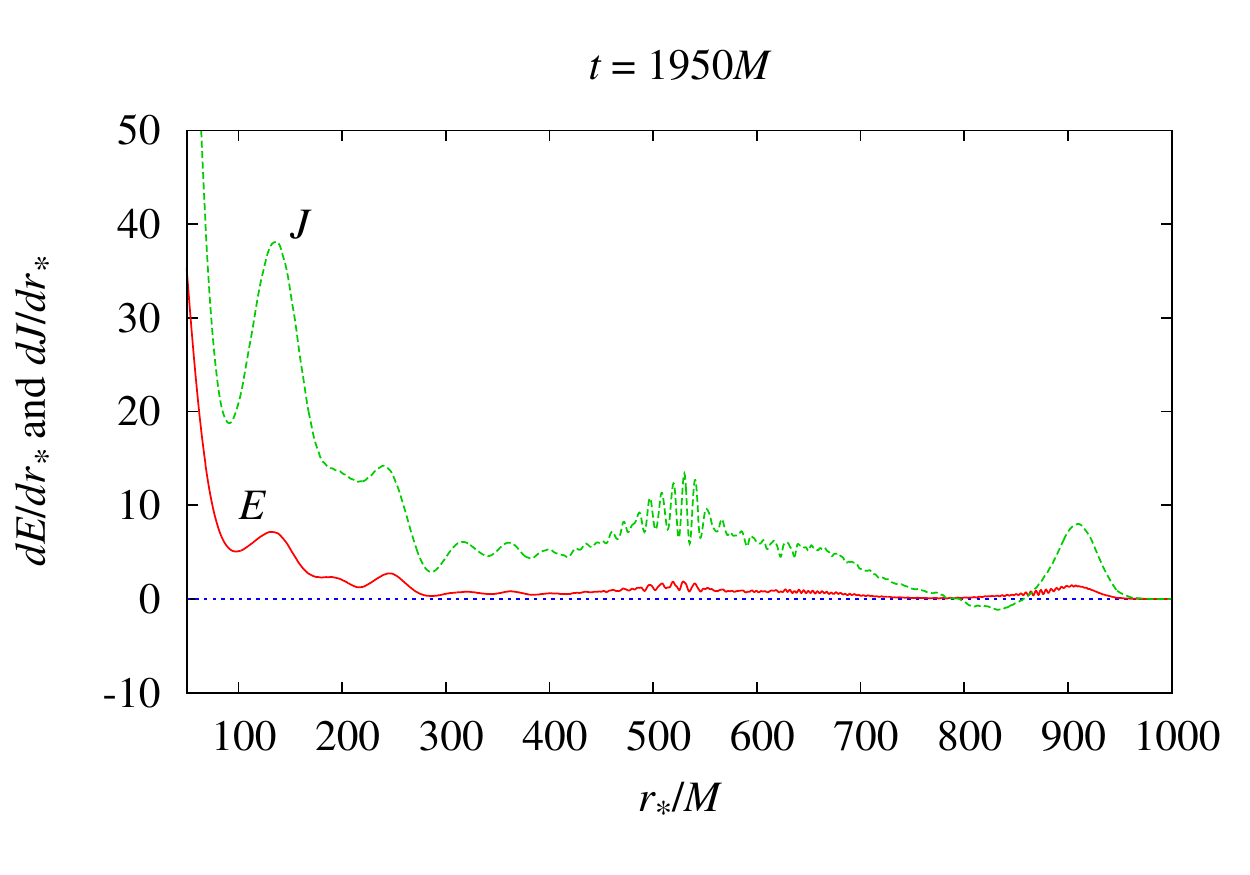}
\caption{Energy density (solid line, red online) and angular momentum density 
(dashed line, green online) with respect to 
the tortoise coordinate $r_*$
in the distant region at $t=1950M$ in the simulation (1b).}
\label{Fig:dEdrdJdr-a099-L1M1-t1950}
\end{figure}
%

Next, let us look at the scattering of the scalar field to the distant region.
Before the bosenova, the field configuration is practically
confined in the region $r_*\lesssim 200M$, and it is approximately zero 
beyond this region due to the exponential decay.
But after the bosenova, the axion cloud begins to spread out to the distant
domain. \Fref{Fig:dEdrdJdr-a099-L1M1-t1950}
shows the energy and angular momentum densities with respect
to the tortoise coordinate $r_*$ in the distant region at $t=1950M$.
About 13.4\% of the total energy distributes in the domain $r_*\ge 300M$.

%
\begin{figure}[tbh]
\centering
\includegraphics[width=0.45\textwidth,bb=0 0 360 252]{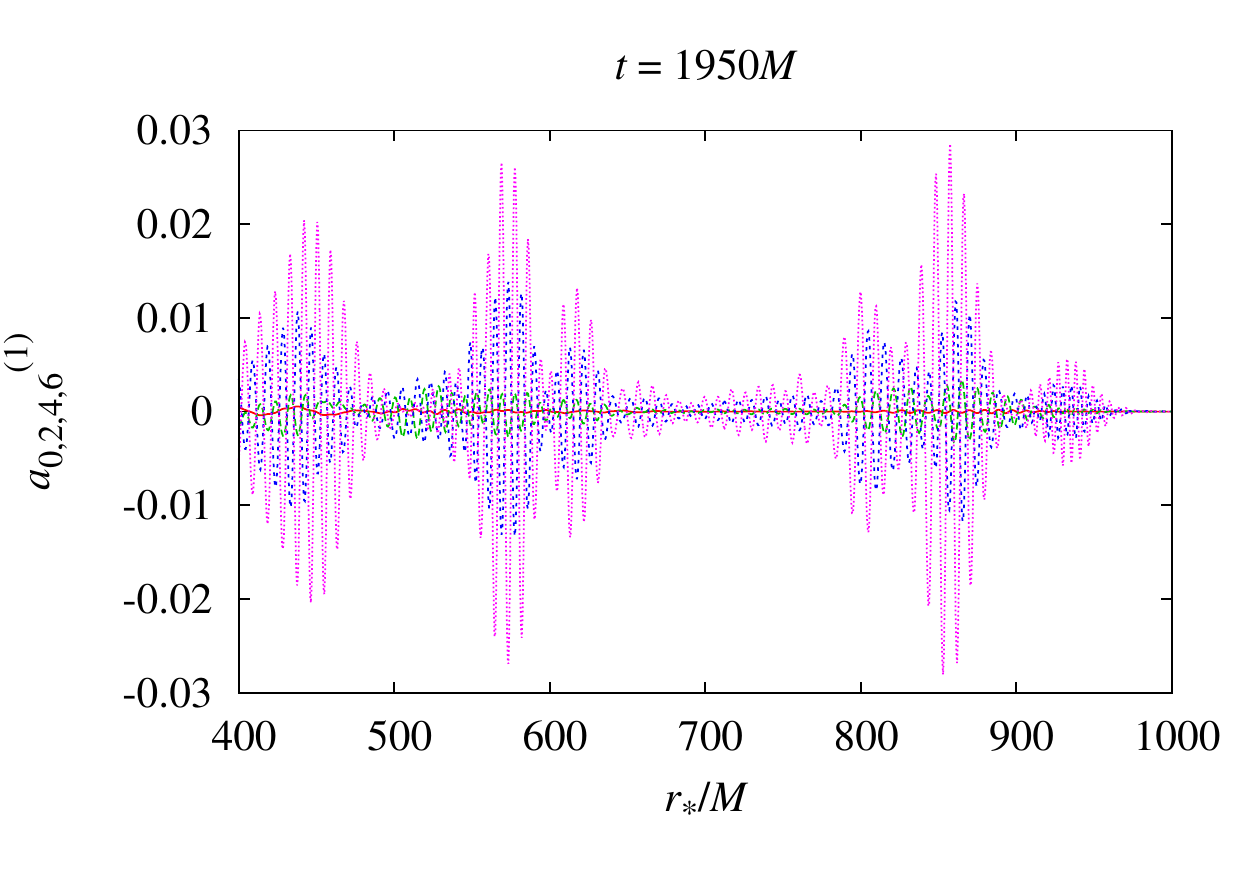}
\includegraphics[width=0.45\textwidth,bb=0 0 360 252]{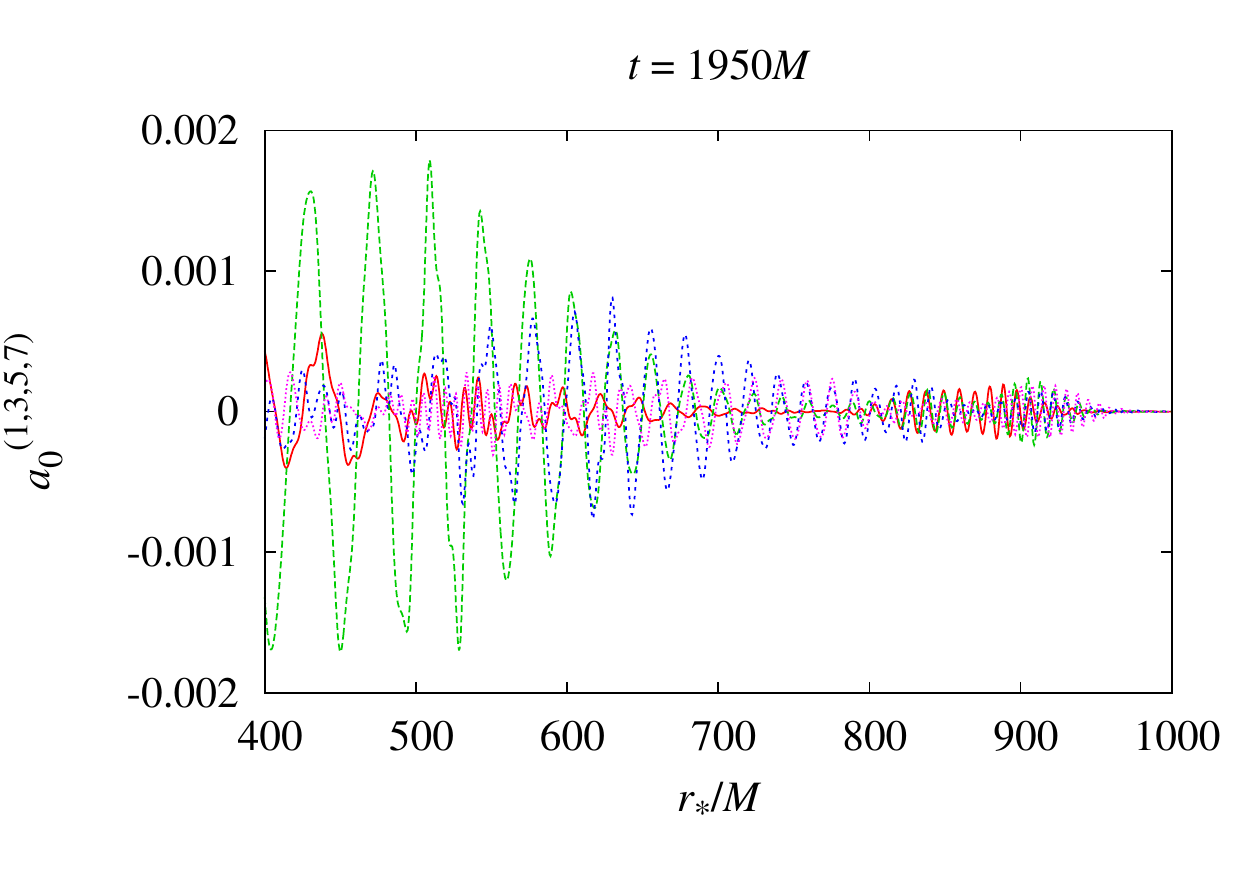}
\caption{Left panel: Snapshots of the real part of the 
scalar field components $a_n^{(m)}$ for $m=1$ and $n=0$ (solid line, red online), 
$2$ (long-dashed line, green online), $4$ (short-dashed line, blue online), 
and $6$ (dotted line, purple online) as functions of the tortoise coordinate $r_*$
in the distant region at $t=1950M$ in the simulation (1b).
Right panel: The same as the left panel but 
for $m=1$ (solid line, red online), $3$ (long-dashed line, green online), 
$5$ (short-dashed line, blue online), and $7$ (dotted line, purple online) and $n=0$. }
\label{Fig:snapshot-L1M1-mu03-a099-t1950}
\end{figure}
%

Whether these scattered fields to the far region are gravitationally 
bounded or not is of interest because it affects the evolution of the
axion cloud in a longer time scale. The exact separation
of bounded and unbounded modes
requires mode decomposition both in time and radial directions, 
and we have not calculated it. But 
we can understand
the gross feature by looking at the waveform because 
bounded modes have frequencies smaller than the mass $\mu$ and
decay exponentially at large $r_*$, while unbounded modes 
oscillate with frequency $\omega$ larger than $\mu$
and decay slowly. 
\Fref{Fig:snapshot-L1M1-mu03-a099-t1950}
shows snapshots of $a_{n}^{(m)}$ for $m=1$ and $n=0,2,4,6$ (left panel)
and for $m=1,3,5,7$ and $n=0$ (right panel) at $t=1950M$  as functions of 
the tortoise coordinate $r_*$. Clearly the major component of the $m=1$ mode oscillates
with frequencies larger than $\mu$. It is also found that  
the modes with $m\ge 3$ include frequencies larger than $\mu$,
although some waves have frequencies close to $\mu$.
Therefore, fairly large part of the fields scattered to the 
distant region is unbounded, and they propagate towards infinity. 
This is in contrast to the case of $M\mu=0.40$ studied in \cite{Yoshino:2012},
where only a small portion of the field is scattered to infinity and most 
part of the scattered fields remains bounded. 
In this sense, the bosenova explosion becomes more violent
as $M\mu$ is decreased.

In a more realistic situation, the bosenova is expected to happen when 
the energy amount close the case (1a). 
From this, the condition
for the bosenova occurrence is estimated to be $E_a/M\approx 3\times 10^3(f_{\rm a}/M_{\rm pl})^2$
for $M\mu=0.30$.
This energy amount is larger than that for $M\mu=0.40$ by
a factor of two. The axion cloud energy at the bosenova occurrence
becomes larger as $M\mu$ is decreased.
This is consistent with the naive expectation that the bosenova happens 
when the peak field amplitude becomes of the order $f_{\rm a}$, 
because in that case, we obtain the estimate 
$E/M \sim\mu^2 \varphi^2 V /M \propto f_{\rm a}^2/(\mu M)$.

\subsubsection{Pure $\ell = m = 2$ mode}
\label{Sec:Pure_lm2}

For $a_*=0.99$, the fundamental $\ell = m = 2$ mode has the fastest growth rate
for $0.448\lesssim M\mu \lesssim 0.928$. 
In this section, we choose $M\mu=0.80$ and simulate the 
evolution of the axion cloud starting from 
the fundamental $\ell = m = 2$ mode whose first peak value is $1.00$.

%
\begin{figure}[tbh]
\centering
\includegraphics[width=0.45\textwidth,bb= 0 0 250 230]{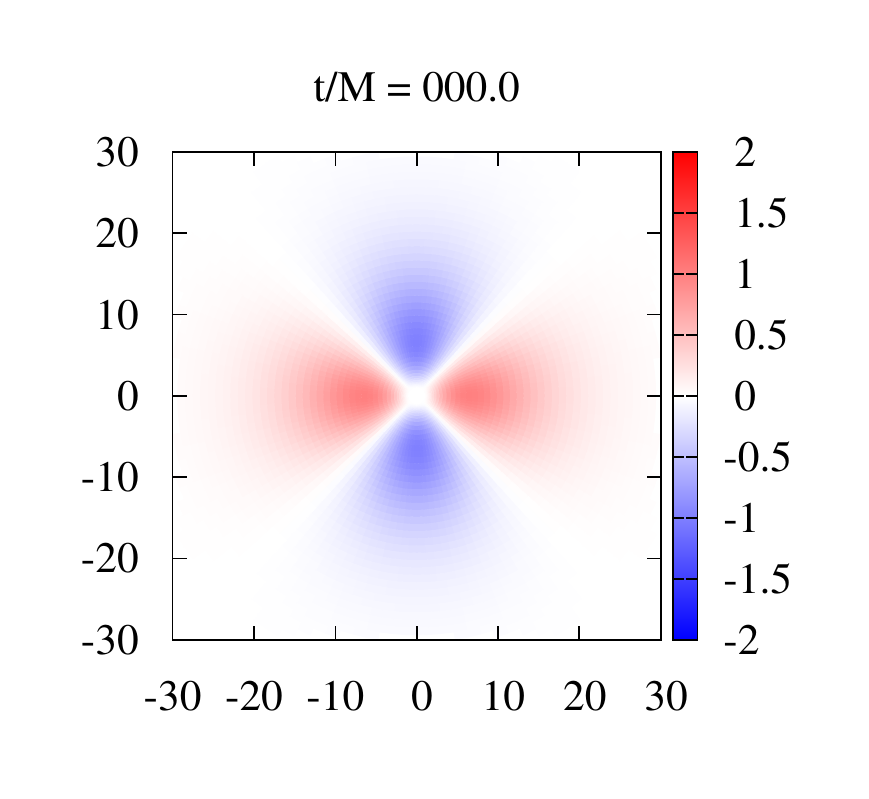}
\includegraphics[width=0.45\textwidth,bb= 0 0 250 230]{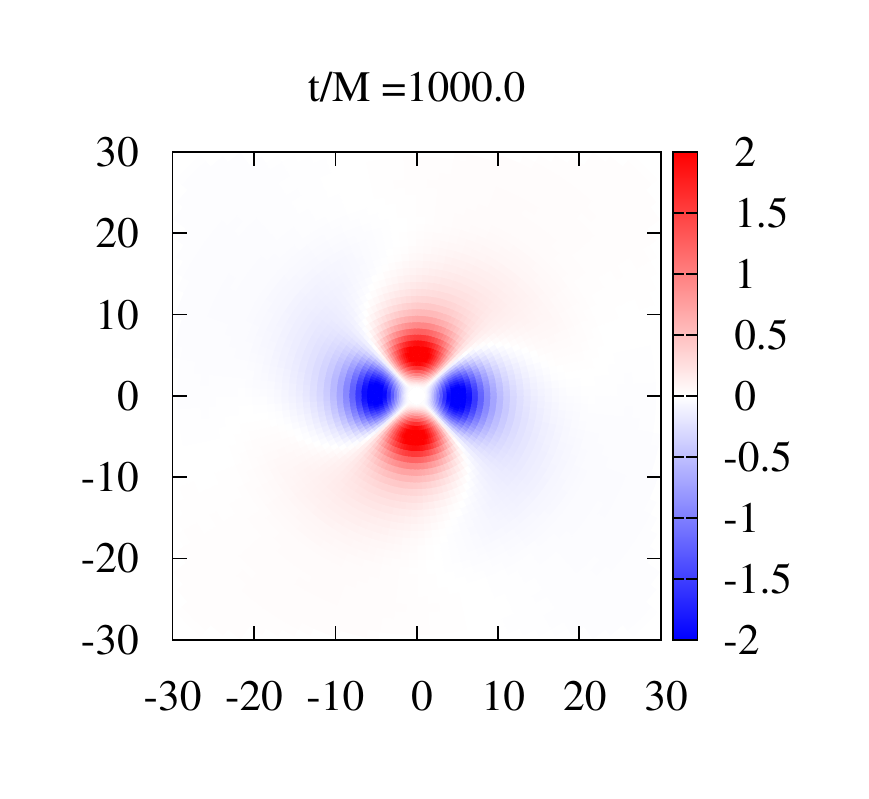}
\caption{Snapshots of the scalar field value in the equatorial plane
 in the $\ell = m = 2$ case [simulation (2)]
at $t=0$ (left) and $1000M$ (right).}
\label{Fig:Kerr099-L2M2-0000-1000}
\end{figure}
%

\Fref{Fig:Kerr099-L2M2-0000-1000} shows the snapshots
of the field value on the equatorial plane at $t/M=0$ and $1000$. 
Because we consider the $\ell = m = 2$ fundamental mode, there are two local
maxima and two local minima. Due to the nonlinear interaction, 
the axion cloud becomes closer to the black hole during
the evolution. This is a kind of relaxation process, and the axion cloud remains concentrated 
around the black hole after that.  
The local maxima keep fairly large values that are between 
two and three. The typical configuration is depicted in the right panel
of \fref{Fig:Kerr099-L2M2-0000-1000}. 
This phenomenon is in contrast to the $\ell = m = 1$ case.
In the simulation (1a) in the previous section, the axion cloud
starting from the pure $\ell = m = 1$ mode oscillates in the radial direction. 
After a mild concentration near the black hole,
the axion cloud recedes from the black hole, and the configuration at $t\approx 2000M$
is very close to the initial configuration. In the $\ell = m = 2$ case,
the axion cloud never resembles the initial configuration
and keeps high concentration around the black hole.

%
\begin{figure}[tbh]
\centering
\includegraphics[width=0.9\textwidth,bb=0 0 407 152]{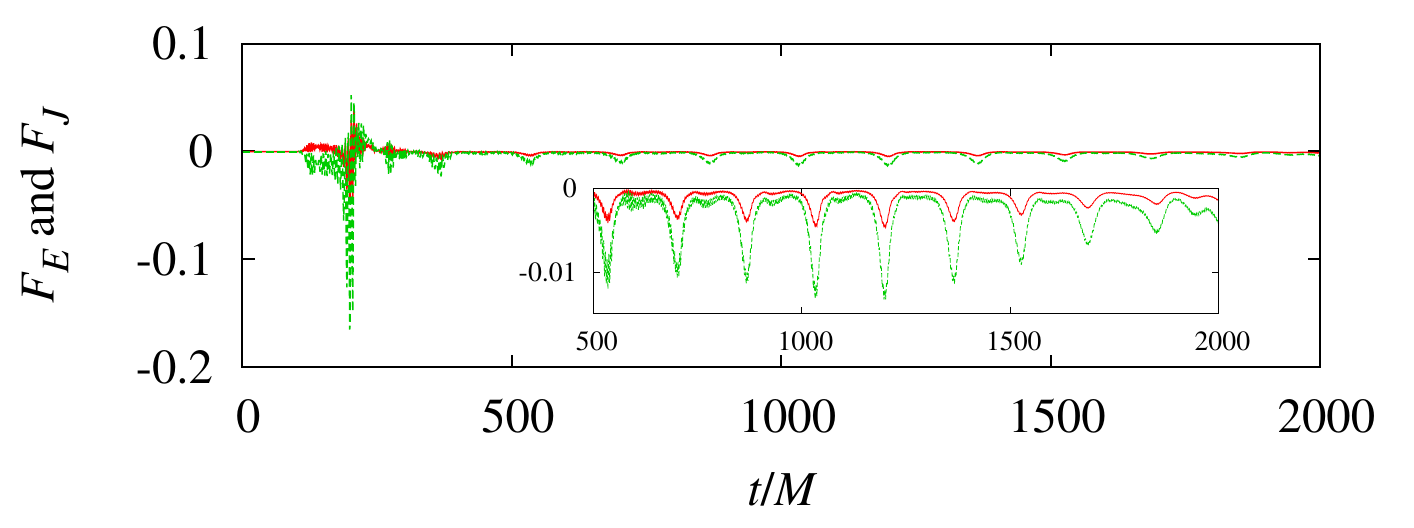}
\caption{Energy flux (solid line, red online) and angular momentum flux 
(dashed line, green online) towards the horizon observed
at $r_*=-100M$ as functions of time in the simulation (2). 
The inset enlarges the period $500\le t/M\le 2000$.}
\label{Fig:FEFJ-a099-L2M2-amp10-combine}
\end{figure}
%

\Fref{Fig:FEFJ-a099-L2M2-amp10-combine} shows the
energy and angular momentum fluxes towards the horizon 
observed at $r_*/M=-100$. The fluxes during $100\lesssim t/M\lesssim 300$
are generated by the relaxation process. After this relaxation process, 
the energy flux is always negative and the energy extraction continues.
Periodically, the absolute value of the energy flux becomes large. 
This is because the axion cloud shows a small radial oscillation.
When the axion cloud becomes closer to the black hole,
the amplitude of the field oscillation increases to $\sim 3$,
and at that moment large amount of energy flux falls into the black hole.
When the axion cloud becomes a bit far from the black hole,
the amplitude of the field oscillation decreases to $\sim 2$ and the
absolute value of the energy flux also becomes small.

%
\begin{figure}[tbh]
\centering
\includegraphics[width=0.6\textwidth,bb= 0 0 360 252]{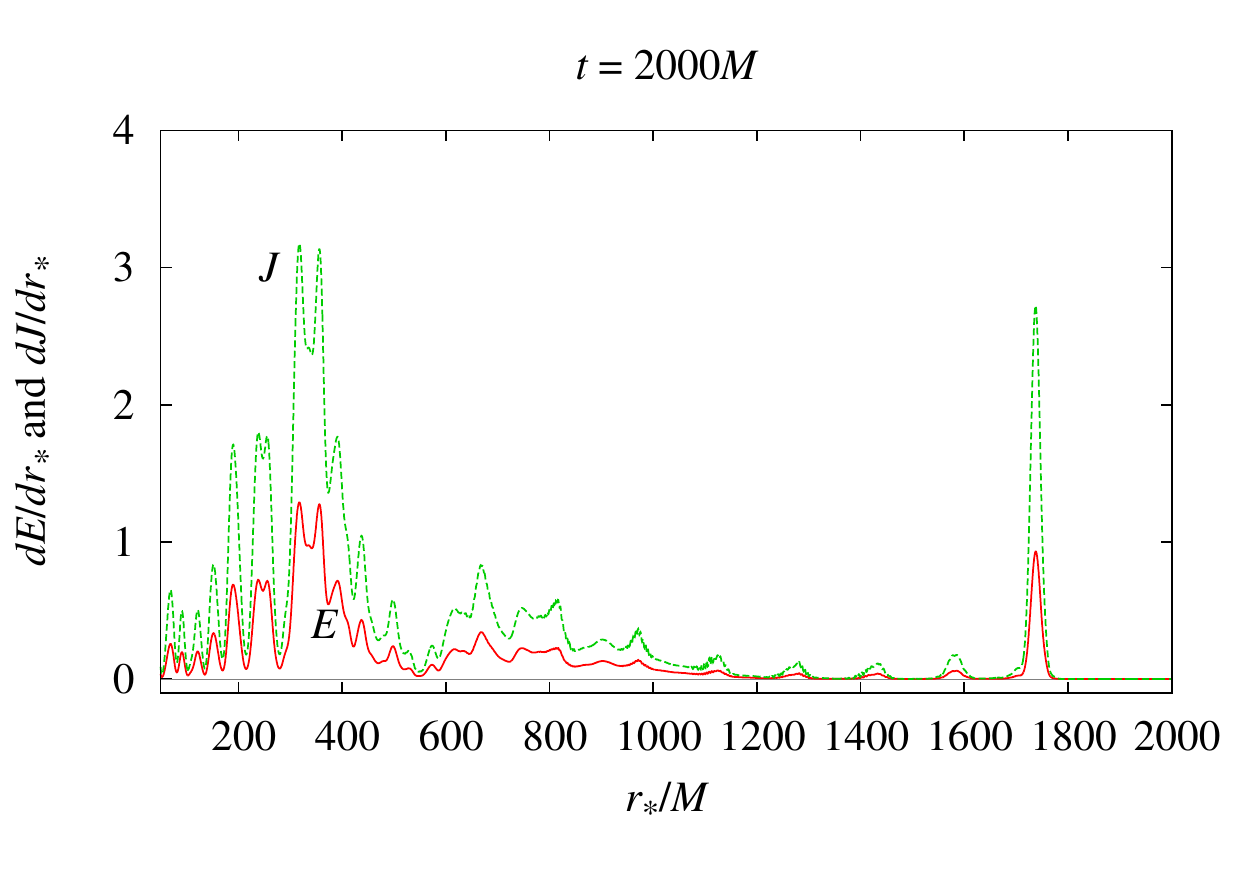}
\caption{Energy density (solid line, red online) and angular momentum density 
(dashed line, green online) with respect to 
the tortoise coordinate $r_*$
in the far region $50\le r_*/M\le 2000$ at $t=2000M$ in the simulation (2).}
\label{Fig:dEdrdJdr-a099-L2M2-amp10-t2000}
\end{figure}
%

This behaviour of the energy flux towards the horizon is analogous to 
the weakly nonlinear case in the $\ell = m = 1$ case.
But an important difference appears in the field behaviour 
at the distant region. \Fref{Fig:dEdrdJdr-a099-L2M2-amp10-t2000}
shows a snapshot of the energy and angular momentum densities 
at $t/M=2000$ at far region, $50\le r_*/M\le 2000$. 
About 15\% of the total energy distributes in this domain.
This is in contrast to the $\ell = m = 1$ case,
where the amount of the energy 
scattered to the distant region is very small (see figure~7 of \cite{Yoshino:2012}). 
Therefore, in spite of the fact that the bosenova collapse (i.e., the positive energy infall 
towards the horizon) does not happen, the scattering
of the field to the distant region occurs continuously. 
As a result, the axion cloud 
in the domain $r_*/M\le 50$ continuously 
loses its energy. 

%
\begin{figure}[tbh]
\centering
\includegraphics[width=0.6\textwidth,bb= 0 0 360 252]{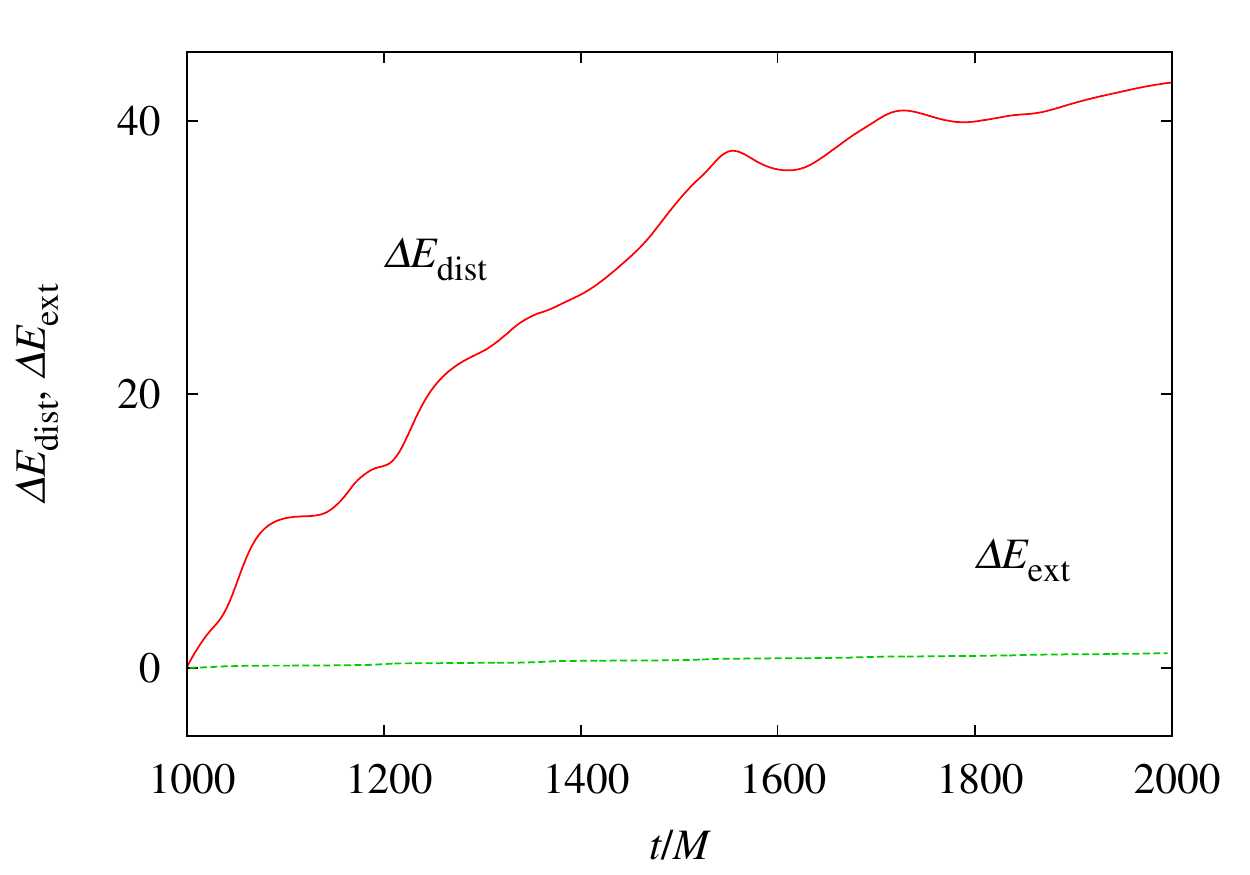}
\caption{The increse in the energy in the distant region $\Delta E_{\rm dist}$
 (solid line, red online)
and the extracted energy (dashed line, green online) 
from the black hole $\Delta E_{\rm ext}$ from time $1000M$ to $t$
in the simulation (2). See text for details.}
\label{Fig:distantE-extractedE}
\end{figure}
%

To see this fact more clearly, let us define the following quantities:
\numparts
\begin{eqnarray}
\Delta E_{\rm dist} &:= E_{\rm dist}(t)-E_{\rm dist}(1000M), \\
\Delta E_{\rm ext} &:= \int_{1000M}^{t} \left| F_E\right| \rmd t.
\end{eqnarray}
\endnumparts
Here, $E_{\rm dist} (t)$ denotes the energy amount in the domain
$r_*\ge 50M$, and therefore, $\Delta E_{\rm dist}$ 
indicates the energy amount that is supplied from the central region $r_*\le 50M$
between time $1000M$ and $t$. On the other hand,
$\Delta E_{\rm ext} $ indicates the amount of the extracted
energy from the black hole between $1000M$ and $t$. 
\Fref{Fig:distantE-extractedE} shows the behaviour
of these two quantities as functions of time. 
We see that although some fluctuation exists, 
the distant energy has tendency to increase in time,
and its amount is much larger than the extracted energy. 

The important question is whether the scattered field at the distant region
is gravitationally bounded or not. In order to derive the answer,
one must decompose the modes 
in both time and radial directions, 
and we have not calculated it. Here, we discuss the gross feature
from figures \ref{Fig:distantE-extractedE} and \ref{Fig:dEdrdJdr-a099-L2M2-amp10-t2000}. In  
\fref{Fig:distantE-extractedE}, the value of $\Delta E_{\rm dist}$
sometimes decreases in time, and this means that the
scattered fields include bounded modes.
But since the gross tendency is that $\Delta E_{\rm dist}$ increases in time,
the scattered fields also include the nonnegligible fraction of unbounded modes.
\Fref{Fig:dEdrdJdr-a099-L2M2-amp10-t2000} also 
indicates that some energy generated after the relaxation process
propagates towards the very far region.

Let us discuss the implication of this simulation. 
Since the main part of the axion cloud 
loses its energy while the energy extraction
from the black hole continues in this simulation, the setup here is 
unrealistic in the sense that it does not approximate the
end point of the superradiant instability. 
In a realistic situation, when the amount of the axion cloud energy is small,
the nonlinear effect is not important and the axion cloud grows by the
superradiant instability. As the nonlinear effect becomes important,
the scattering towards the far region gradually occurs, while the extraction 
of the energy from the black hole continues. 
At some energy, the scattering rate of energy to the distant region
and the rate of the energy extraction from the black hole
would balance, and the final state would 
not be the bosenova collapse, but rather a steady outflow of the field
from the axion cloud. This steady state should be realized 
before the axion cloud energy reaches
the energy amount in the simulation here, that is, 
$E/M\approx 1.9\times 10^3(f_{\rm a}/M_{\rm pl})^2$.

Note that we have performed several simulations by changing
the initial amplitude. If we choose an artificially large initial amplitude such as 1.2,
the phenomena analogous to the bosenova in the $\ell = m = 1$ case
occurs. There, the excitation of the $m=-2$ mode
is observed and it continues for the period of $\sim 400M$. 
However, we consider this ``bosenova collapse'' would not happen
in a realistic case, because
the axion cloud energy cannot reach the required energy to cause the bosenova
due to the energy loss by the scattering of the field to the distant region.

\subsubsection{Superposition of $\ell = m = 1$ and $2$ modes}
\label{Sec:Superposition_lm1_lm2}

In this section, we study the case that the axion cloud consists 
of the superposition of the fundamental $\ell=m=1$ and $\ell = m = 2$ modes
initially.
Here, the primary component of the axion cloud is the $\ell = m = 1$ mode,
and the $\ell = m = 2$ mode is added as a perturbation. 
To be more specific, we choose the case of $a_*=0.99$ and $M\mu=0.40$,
and assume that the initial peak value of the $\ell = m = 1$ mode
to be $0.70$. If the $\ell = m = 2$ mode is absent, the system is same
as the one simulated in our previous paper~\cite{Yoshino:2012}. 
To this system, we add the $\ell = m = 2$ mode whose 
first peak value is $0.01$. 
Our primary interest here is what happens to  the $\ell = m = 2$ mode 
under the influence of the nonlinear effect of the $\ell = m = 1$ mode. 
To see this, it is useful to calculate the decomposition
\begin{equation}
\Phi = \sum_{\ell m} b_{\ell m}(t,r_*)Y_{\ell m}(\theta,\varphi) \quad b_{\ell(-m)} = b_{\ell m}^{*}.
\end{equation}
Then, $b_{22}$ approximately gives the $\ell = m = 2$ component of the 
axion cloud.

%
\begin{figure}[tbh]
\centering
\includegraphics[width=0.6\textwidth,bb= 0 0 360 252]{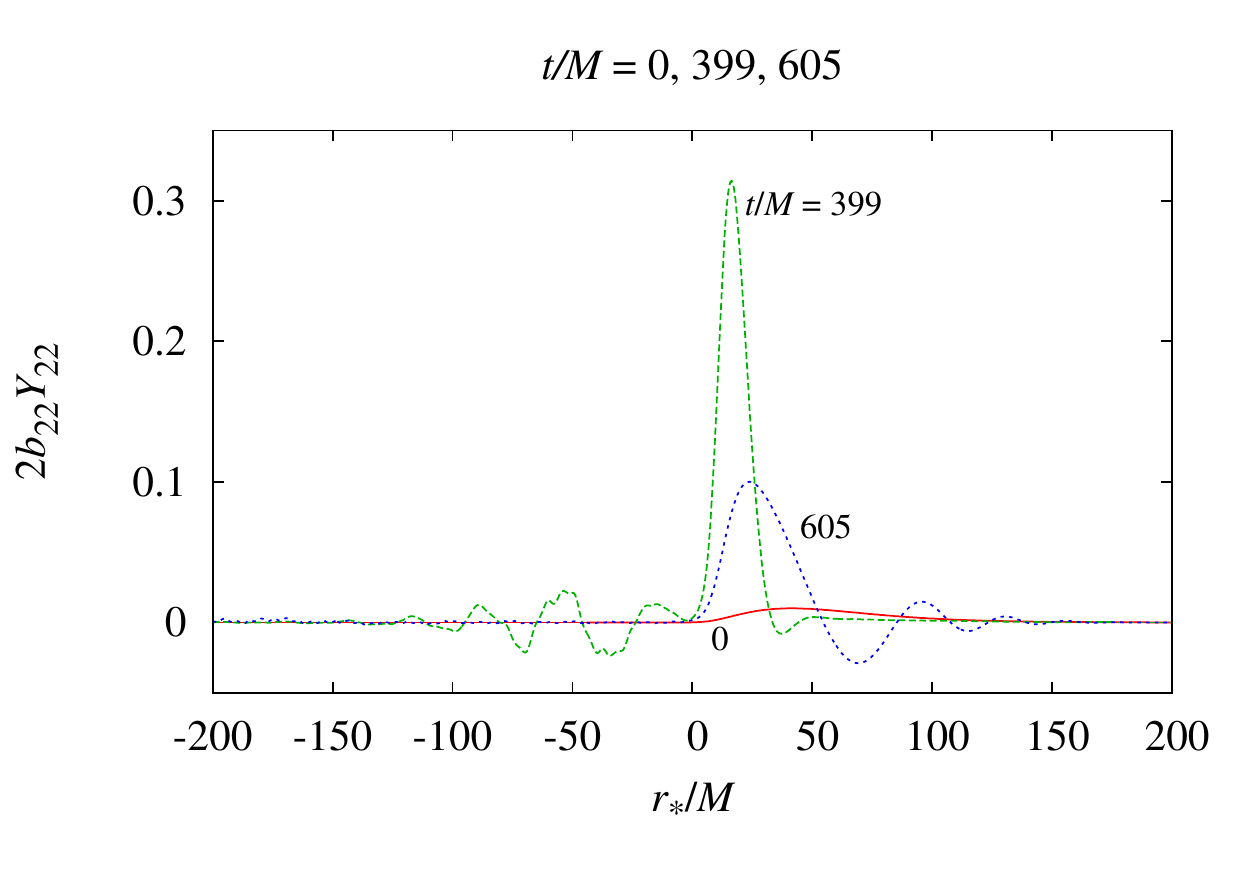}
\caption{Snapshots of the real part of the component $2b_{\ell m}Y_{\ell m}$ for
$\ell = m = 2$ on the $\phi=0$ line in the equatorial plane at $t/M=0$ (solid line, red online), 
$399$ (long-dashed line, green online), and $605$ (short-dashed line, blue online) in the simulation (3). 
At $t/M=399$, the field is concentrated around the black hole and very high peak value.
After that, the field settles to a configuration that resembles an overtone mode
($t/M=605$).}
\label{Fig:Snapshots-L2M2-caseL1M1pL2M2}
\end{figure}
%

\Fref{Fig:Snapshots-L2M2-caseL1M1pL2M2} shows snapshots of the
real part of $2b_{22}Y_{22}$ on the $\phi=0$ line in the equatorial plane
at $t/M=0$, $399$, and $605$. 
Although the added $\ell = m = 2$ mode is small initially, 
during the bosenova, 
it grows to become order one analogously to the resonance by a forced oscillation 
($t=399M$). 
The profile of the $\ell = m = 2$ mode becomes highly concentrated around the black hole,
and generated ingoing waves carry
positive energy flux to the horizon.
As the bosenova ceases,
the $\ell = m = 2$ mode also settles down, 
but the mode function is very different 
compared to the initial state ($t=605M$). 
The $\ell = m = 2$ mode after the bosenova has a larger frequency compared to the
initial state  and spreads over a distant region, 
causing several oscillations in the radial profile of the mode function. 
In this phase,  infalling waves to the horizon include both the $m=\pm 2$ modes,
and the net flux carried by the $\ell = m = 2$ mode remains positive
at least up to $t=1000M$.

As observed above, the evolution of the $\ell = m = 2$ mode 
is quite complicated. Although
the superradiant instability by the $\ell = m = 2$ fundamental mode
no longer occurs after the bosenova, 
the $m=2$ mode survives showing complicated dynamics. 
With the simulation results up to $t=1000M$, 
it is very difficult to predict the final fate of the $\ell = m = 2$ mode
in a long-term evolution.

%
\begin{figure}[tbh]
\centering
\includegraphics[width=0.6\textwidth,bb= 0 0 360 252]{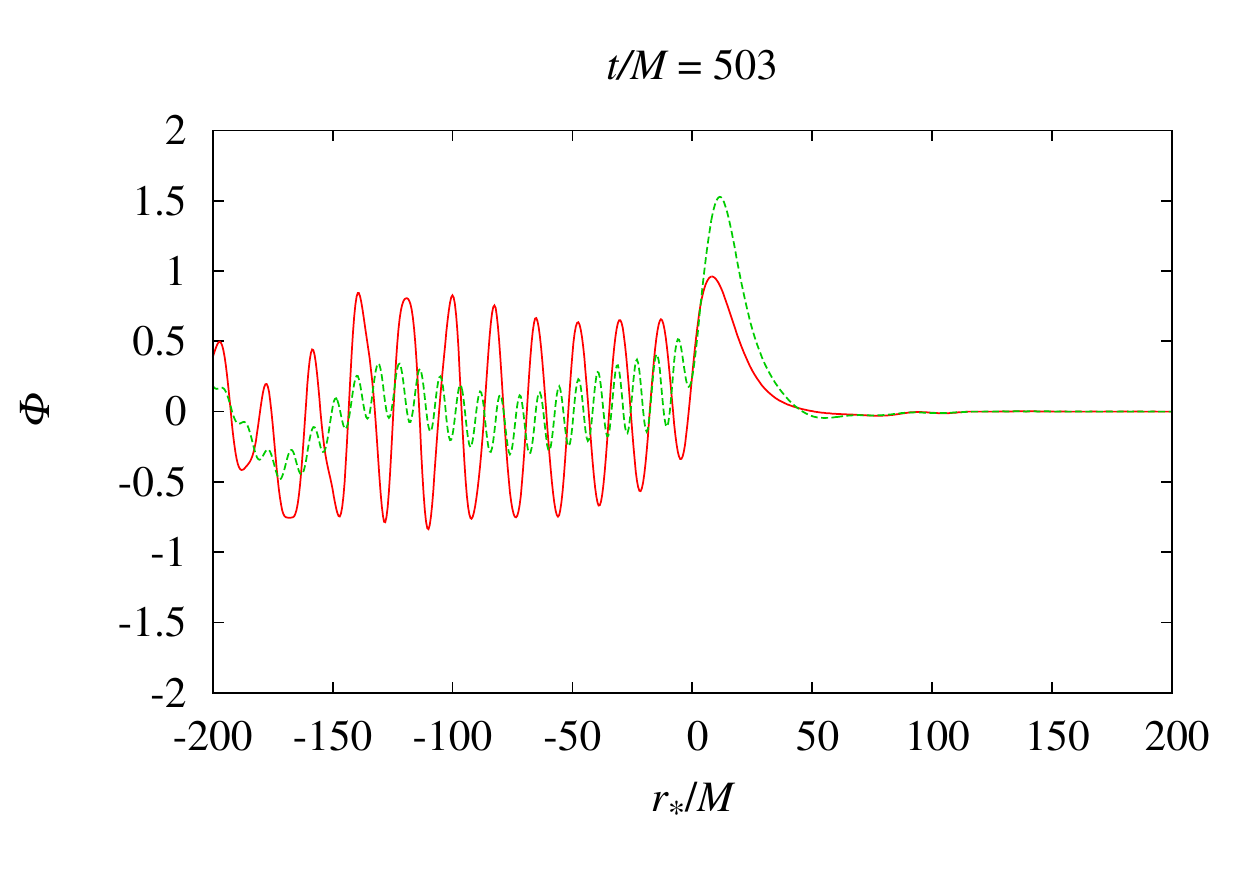}
\caption{Snapshots of the scalar field $\Phi$ on the $\phi=0$ line in the 
equatorial plane at $t=503M$ in the cases that the small $\ell = m = 2$ is present (solid line, red online)
 [simulation (3)]
and absent (dashed line, green online) \cite{Yoshino:2012}.}
\label{Fig:Snapshots-Phi-caseL1M1pL2M2}
\end{figure}
%

Another interesting phenomena observed in the simulation is that
the property of the mode excitation during the bosenova is rather changed 
by the presence of the small $\ell = m = 2$ mode. 
\Fref{Fig:Snapshots-Phi-caseL1M1pL2M2} shows a snapshot
of the field value $\Phi$ on the line $\phi=0$ in the equatorial plane
at $t=503M$. The cases that the $\ell = m = 2$ mode is present 
and absent are shown by the solid line and the dashed line,
respectively. When the small $\ell = m = 2$ mode is added, 
the excited mode has larger amplitudes and the typical frequency
becomes smaller.

%
\begin{figure}[tbh]
\centering
\includegraphics[width=0.45\textwidth,bb= 0 0 360 252]{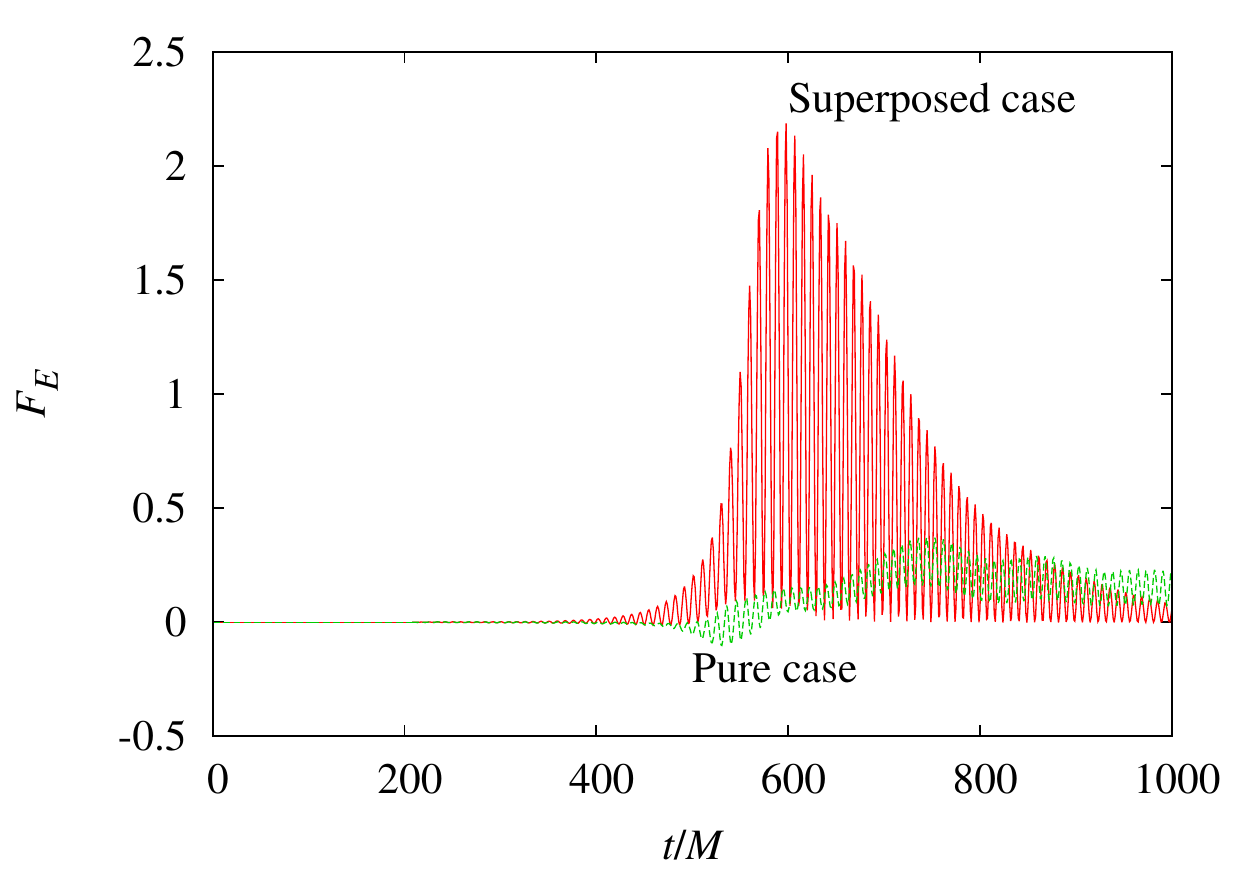}
\includegraphics[width=0.45\textwidth,bb= 0 0 360 252]{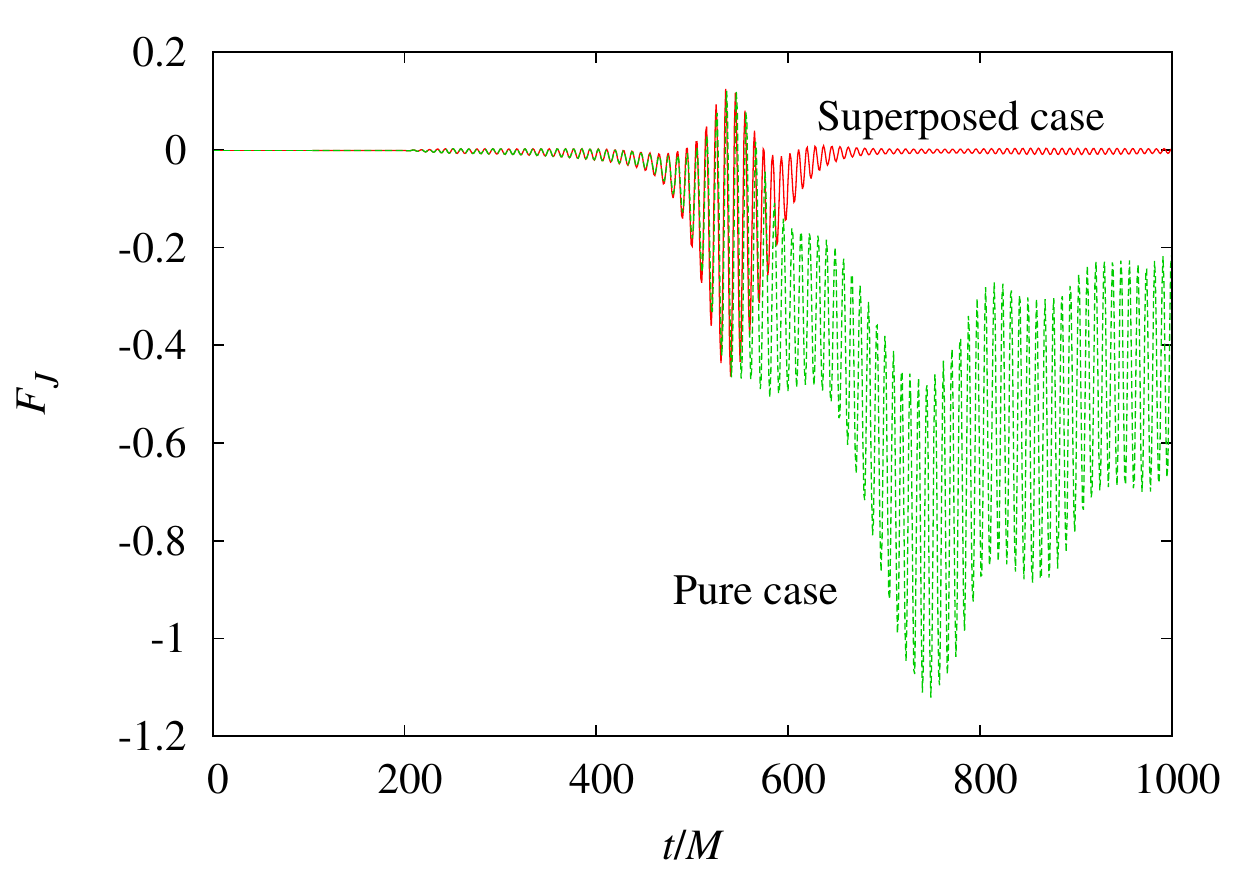}
\caption{The energy flux $F_E$ (left panel) and the angular momentum flux 
$F_J$ (right panel) towards the horizon observed at $r_*=-200M$
as functions of time 
for the cases that the small $\ell=m=2$ is present [simulation (3)]
(solid line, indicated by ``Superposed case'')
and absent \cite{Yoshino:2012} (dashed line, indicated
by ``Pure case'').}
\label{Fig:FEFJ-a099-L1M1pL2M2}
\end{figure}
%

Corresponding to this phenomena, the energy and angular momentum fluxes towards the
horizon also change as shown in \fref{Fig:FEFJ-a099-L1M1pL2M2}.
In this figure, we compare 
the energy flux (left panel) and the angular momentum flux (right panel)
towards the horizon observed at $r_*/M=-200$ in the two cases that
a small $\ell = m = 2$ mode is present (solid line) and absent (dashed line). 
The energy flux becomes much larger while the absolute value of the angular momentum flux
becomes smaller when the small $\ell = m = 2$ mode is added.
Therefore, the evolution of the energy and angular momentum of the axion cloud
is strongly affected by the mixture of additional modes to the axion cloud
even if they are small.

\subsubsection{Superposition of $n_{\rm r}=0$ and $1$ modes}
\label{Sec:Superposition_nr0_nr1}

Here, we consider the case in which the initial state is given by the
superposition of the fundamental mode ($n_{\rm r}=0$) and the overtone mode ($n_{\rm r}=1$)
for $\ell = m = 1$.  As the system parameters, we choose $a_*=0.99$
and $M\mu=0.40$, and prepare 
the initial data by adding the two modes $n_{\rm r}=0$ and $n_{\rm r}=1$ with
the first peak values $0.60$ and $0.20$, respectively.
When only the fundamental mode $n_{\rm r}=0$ is present with
the peak value $0.60$, the bosenova does not happen \cite{Yoshino:2012}. 
We are interested in what is changed if the $n_{\rm r}=1$ mode is superposed
to this situation.

%
\begin{figure}[tbh]
\centering
\includegraphics[width=0.4\textwidth,bb=0 0 250 230]{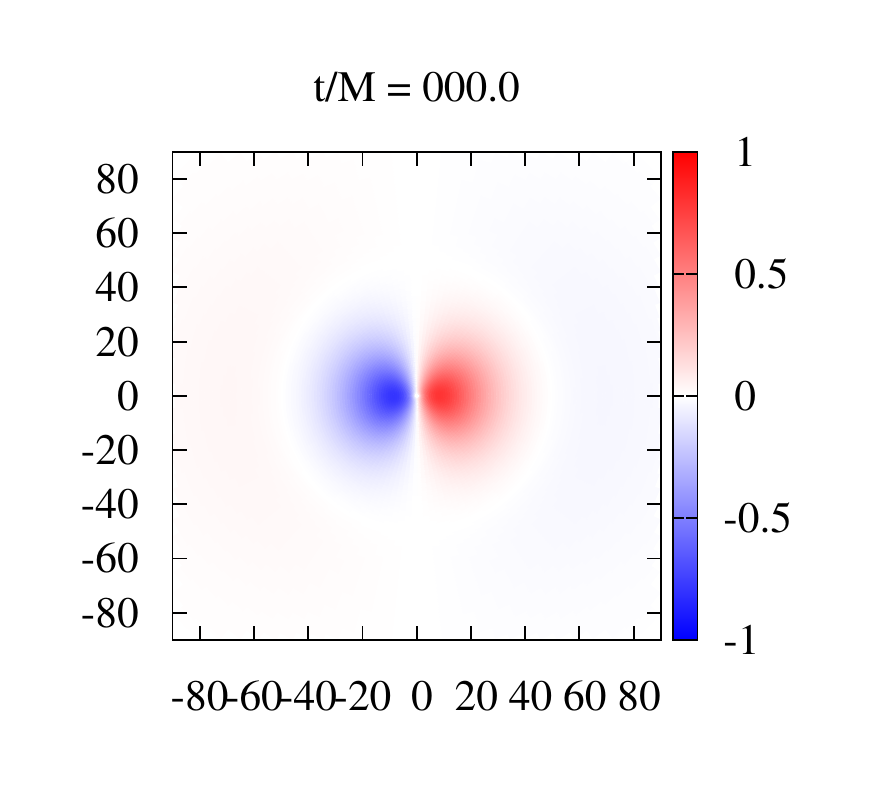}
\includegraphics[width=0.4\textwidth,bb=0 0 250 230]{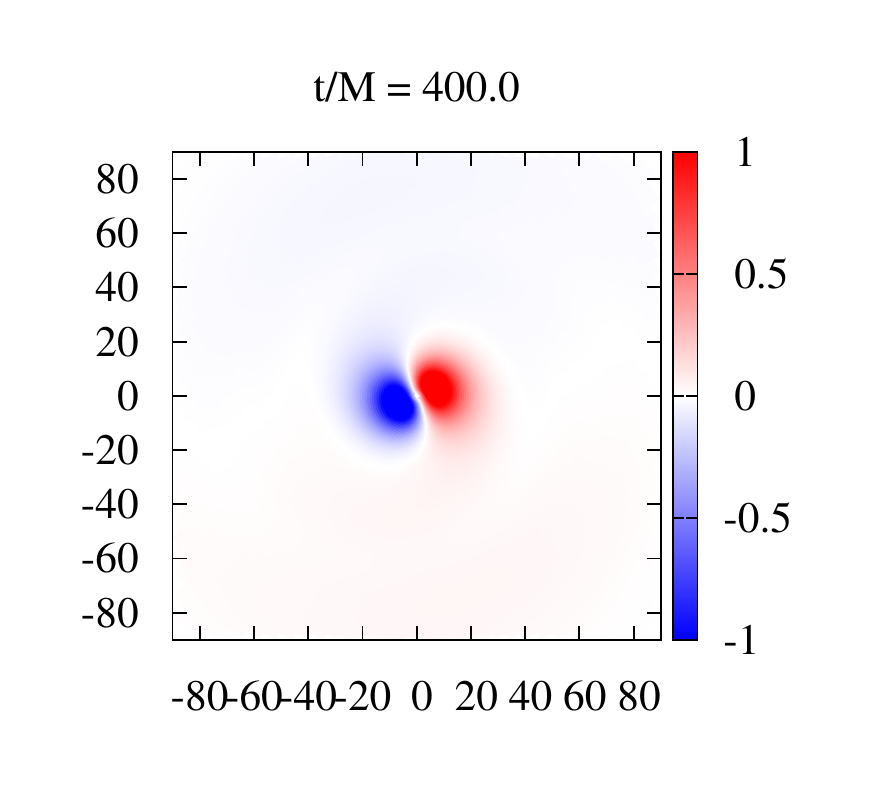}
\includegraphics[width=0.4\textwidth,bb=0 0 250 230]{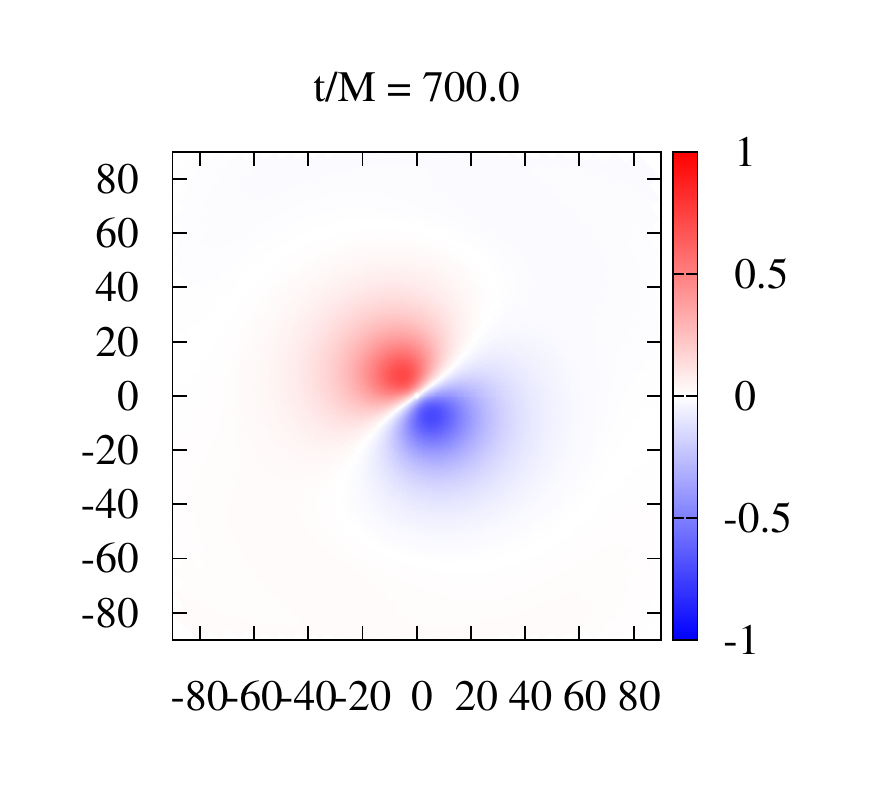}
\caption{Snapshots of the scalar field in the case that the two modes
$n_{\rm r}=0$ and $n_{\rm r}=1$ are superposed  [simulation (4)]
at $t=0$ (top left), $400M$ (top right) and  $700M$ (bottom)
in the equatorial plane.}
\label{Fig:Kerr099-L1M1-nr0nr1-t0t400t700}
\end{figure}
%

\Fref{Fig:Kerr099-L1M1-nr0nr1-t0t400t700} shows
the field configuration in the equatorial plane at $t/M=0$, $400$,
and $700$. In our initial data, the inner local maximum (minimum) exists at 
$r_*/M\approx 12.5$  and $\phi=0$ ($\pi$) and the
outer local minimum (maximum) exists at
$r_*/M\approx 77.5$  and $\phi=0$ ($\pi$) 
in the equatorial plane.  
During the evolution, it was found that 
the outer local maximum/minimum outside is absorbed to the
inner local maximum/minimum ($t/M=400$). 
In other words, the merger of the two maxima/minima
happens. Then, the system resembles
an axion cloud with just one fundamental mode ($t/M=700$). 
Therefore, a kind of conversion from the overtone mode
to the fundamental mode occurs due to the nonlinear self-interaction.

%
\begin{figure}[tbh]
\centering
\includegraphics[width=0.6\textwidth,bb=0 0 360 252]{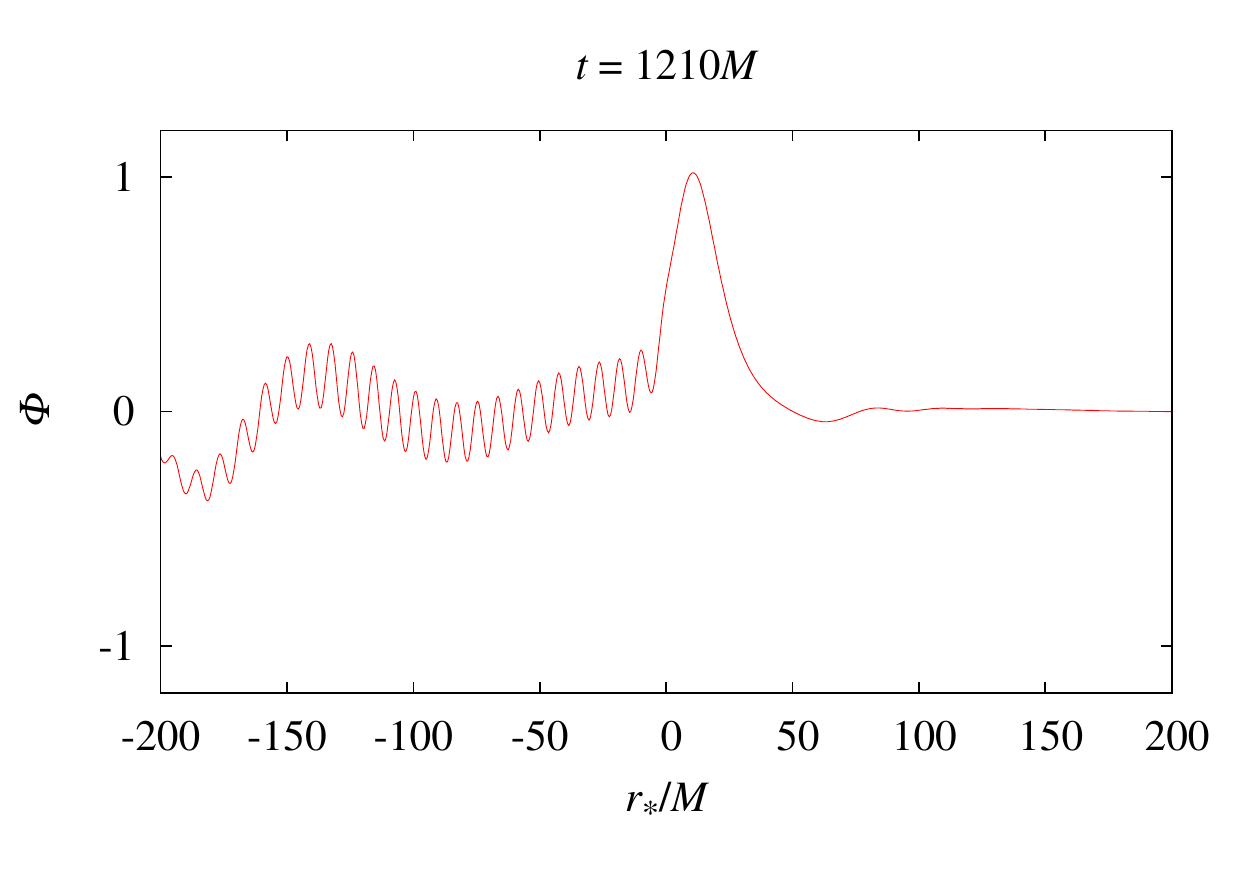}
\caption{A snapshot of the scalar field in the case that the two modes
$n_{\rm r}=0$ and $n_{\rm r}=1$ are superposed [simulation (4)]  as a function of $r_*/M$
at $t=1210M$ on the $\phi=0$ line
in the equatorial plane.}
\label{Fig:snapshot-L1M1-nr0nr1-mu04-a099-t1210}
\end{figure}
%

The evolution after this merger is very analogous to the case of
pure fundamental mode with $\ell = m = 1$ with initial peak value $0.70$
simulated in \cite{Yoshino:2012}. 
After high concentration around the black hole, various modes
are excited. In particular, the $m=-1$ mode is the primary mode
that falls into the horizon. \Fref{Fig:snapshot-L1M1-nr0nr1-mu04-a099-t1210} 
depicts the field configuration
at $t=1210M$ on the $\phi=0$ line in the equatorial plane. 
The excitation of the $m=-1$ mode is clearly observed.

To summarize, the existence of the overtone mode with $n_{\rm r}=1$
does not much affect the dynamics of the axion cloud which is
dominated by the fundamental mode with $n_{\rm r}=0$, because
the overtone mode is converted 
to the fundamental mode due to the nonlinear self-interaction effect. 
Our results suggest that when the nonlinear effect is strong,
it would be sufficient to consider the dynamics of the
fundamental mode.

\subsection{Summary of this section}

In this section, we numerically studied the phenomena caused by the nonlinear
interaction for various setups. For the $\ell = m = 1$ mode, 
we studied the dependence of the bosenova phenomena on 
the value of $M\mu$. In the case of $M\mu=0.30$ simulated in (1b), 
the axion bosenova happens when the axion cloud acquires
a larger energy compared to the case of $M\mu=0.40$ studied
in \cite{Yoshino:2013}. The bosenova phenomena becomes
more violent for a smaller $M\mu$, causing stronger excitation of 
gravitationally unbounded modes.
The excitation of the $m=-1$ mode
is commonly observed for both $M\mu=0.30$ and $0.40$. The $m=-1$ mode
falls into the black hole carrying positive energy, and thus,
it terminates the superradiant instability.

In the $\ell = m = 2$ case, we found the 
evidence against the bosenova occurrence. 
In the simulation (2), 
no phenomena analogous to the bosenova happens, and instead, 
we observed the continuous energy extraction from the black hole 
and generation of a steady outflow
to the distant region, simultaneously. 
For a chosen parameter in the simulation (2),  the
outflow of the scalar field energy was larger than the
rate of the energy extraction. 
From these results, we suspect that 
the growth of a realistic axion cloud formed by the superradiant instability
would stop when the energy extraction rate and energy outflow rate
balance with each other. Note that if we choose a large initial
amplitude by hand, the phenomena analogous
to the bosenova was observed. In particular, we observed the excitation 
of the $m=-2$ mode,
which is an analogous to the $m=-1$ mode excitation in the $\ell = m = 1$ case.
However,  in a realistic situation, the axion cloud would not be able to acquire the required 
energy to cause a bosenova collapse due to the generation of
a steady outflow.

In the simulations (3) and (4), we studied the cases where two modes
are superposed. In the simulation (3), we added the $\ell = m = 2$ fundamental mode
as a perturbation to the axion cloud in the $\ell = m = 1$ mode.
During the bosenova, the $\ell = m = 2$ mode also carries a positive energy flux 
to the horizon, but the forced oscillation by the bosenova 
increases the energy of the $\ell = m = 2$ mode. 
The mode function after the bosenova is very different from the
initial state, and spreads over a larger region.
It was also  observed that the added $\ell = m = 2$ mode strongly affects
the properties of the bosenova, such as the values of the
energy/angular momentum fluxes towards the horizon.
On the other hand, the simulation (4)
indicates that the addition of the overtone mode $n_{\rm r}=1$ to the 
 fundamental $n_{\rm r}=0$ mode does not strongly affect the 
 properties of the bosenova, because the overtone mode is
 converted into the fundamental mode by the
 nonlinear self-interaction.

%
%
\section{Gravitational wave emission}

\label{Sec:V}

One of the important phenomena caused by an axion cloud around a black hole
is the gravitational wave emission. In our previous research,
we calculated the amplitudes of continuous gravitational waves emitted
by a Klein-Gordon scalar cloud ignoring the nonlinear self-interaction~\cite{Yoshino:2013},
and discussed observational constraints on axion model parameters~\cite{Yoshino:2014}.
The neglect of the nonlinear self-interaction in this calculation may, however,
result in an estimation largely different from reality, because the axion bosenova
may emit burst-type gravitational waves with amplitudes much larger than those of
continuous emissions, and besides, the self-interaction may cause
fluctuation/modulation in the amplitude/frequency of
continuous waves emitted in the superradiant phase, which has
significant influences on the wave search in the data analysis.
Thus, more elaborate calculations of gravitational waves taking account of
the nonlinear self-interaction are highly desired.
In this section, we describe the current status of our research
in this direction.

\subsection{Formulation and numerical method}

We first explain the basic strategy of the calculation.
In the previous section, we have numerically determined the behaviour of the
axion scalar field in the test field approximation.  Using this numerical solution,
the energy-momentum tensor $T_{\mu\nu}$ can be calculated
using \eref{Eq:Energy-momentum-tensor-scalar}.
Then, the amplitudes of gravitational waves can be determined
by solving the linearized Einstein equations with the source term $T_{\mu\nu}$
in the Kerr background. Throughout the present paper,
we ignore the radiation reaction to the dynamics of the
scalar field and the black hole parameters.

Various approaches can be taken to determine
the metric perturbation for a given source term.
In the present paper, we choose the approach to directly solve
the Teukolsky equation \cite{Teukolsky:1973} in the time domain. In this approach,
the perturbed Einstein equations in the Kerr background are reduced
to a set of PDEs for the perturbation of the spin coefficients and
the Newman-Penrose quantities $\psi_{0}$ and $\psi_{4}$
constructed from the Weyl curvature tensor. This set of PDEs
can be further reduced to decoupled second-order
master wave equations for two gauge-invariant complex functions ${}_{\pm 2}\Psi$
that are algebraically related to $\psi_{0}$ and $\psi_{4}$, known as the Teukolsky equations,
and the relation between the two Teukolsky functions ${}_{\pm2}\Psi$.
With the help of this relation and the original PDEs, we can completely determine
all Newman-Penrose quantities as well as  the metric perturbation from either of these two functions
\cite{Wald:1973,Cohen:1974,Cohen:1975,Chrzanowski:1975,Wald:1978,Chandrasekhar.S1983}.

The explicit form of the Teukolsky equation for a gravitational perturbation \cite{Teukolsky:1973} is given by
%
\begin{eqnarray}
\fl
\left[\frac{(r^2+a^2)^2}{\Delta}-a^2\sin^2\theta\right]
\frac{\partial^2{}_{s}\Psi}{\partial t^2}
+\frac{4Mar}{\Delta}\frac{\partial^2{}_{s}\Psi}{\partial t\partial\phi}
+\left[\frac{a^2}{\Delta}-\frac{1}{\sin^2\theta}\right]
\frac{\partial^2{}_{s}\Psi}{\partial\phi^2}
\nonumber\\
\fl
-\Delta^{-s}\frac{\partial }{\partial r}
\left(\Delta^{s+1}\frac{\partial{}_{s}\Psi}{\partial r}\right)
-\frac{1}{\sin\theta}\frac{\partial}{\partial\theta}
\left(\sin\theta\frac{\partial{}_{s}\Psi}{\partial\theta}\right)
-2s\left[
\frac{a(r-M)}{\Delta}+\frac{\rmi\cos\theta}{\sin^2\theta}
\right]\frac{\partial{}_{s}\Psi}{\partial\phi}
\nonumber\\
-2s\left[
\frac{M(r^2-a^2)}\Delta{}-r-\rmi a\cos\theta
\right]
\frac{\partial{}_{s}\Psi}{\partial t}
+(s^2\cot^2\theta-s){}_{s}\Psi = 4\pi\Sigma T
\label{Eq:Teukolsky-original}
\end{eqnarray}
%
with $s=+2$ or $-2$.
The right-hand side is the source term, and
$T$ is given in terms of components of the energy-momentum tensor
that generates the gravitational waves.
In our case, we choose $s=-2$ for a technical reason in numerical calculations.
For this choice of $s$,
%
\begin{equation}
T=({2}/{\rho^4}) T_4,
\label{Eq:T-definition}
\end{equation}
%
with
%
\begin{eqnarray}
\fl
T_4=(\hat{\Delta} + 3\gamma-\gamma^*+4\mu + \mu^*)
    [(\delta^*-2\tau^*+2\alpha)T_{nm^*}
        -(\hat{\Delta}+2\gamma-2\gamma^*+\mu^*)T_{m^*m^*}]
\nonumber\\
\fl
+(\delta^*-\tau^*+\beta^*+3\alpha+4\pi)
[(\hat{\Delta}+2\gamma+2\mu^*)T_{nm^*}-(\delta^*-\tau^*+2\beta^*+2\alpha)T_{nn}],
\label{Eq:T4-definition}
\end{eqnarray}
%
where $T_{nn}=T_{\mu\nu}n^\mu n^\nu$, and so on.
Here, $n^\mu$ and $m^\mu$ are the Kinnersley null tetrads,
%
\numparts
\begin{eqnarray}
l^\mu = \frac{1}{\Delta}\left(r^2+a^2, \ 1, \ 0, \ a \right),
\\
n^\mu = \frac{1}{2\Sigma}\left(r^2+a^2, \ -\Delta, \ 0, \ a\right),
\\
m^\mu = \frac{1}{\sqrt{2}(r+\rmi a\cos\theta)}
\left(\rmi a\sin\theta, \ 0, \ 1, \ \frac{\rmi}{\sin\theta}\right),
\end{eqnarray}
\endnumparts
%
$\hat{\Delta}$ and $\delta$ are the derivative operators,
$\hat{\Delta} := n^\mu\nabla_\mu$ and $\delta := m^\mu\nabla_\mu$, and
$\alpha$, $\beta$, $\gamma$, $\mu$, and $\pi$ are non-vanishing Newman-Penrose
variables for the spin coefficients \cite{Newman:1961}.

The Teukolsky equation has terms with divergent coefficients proportional
to $1/\sin^2\theta$,
which may cause a problem in numerical simulations. In order to resolve this, we
spectrally decompose ${}_{-2}\Psi$ as
%
\begin{equation}
{}_{-2}\Psi = \sum_{\tilde{m}=0,\pm 1, \pm 2,\cdots}\frac{\Delta^2}{r}\psi^{(\tilde{m})}(t,r,\theta)
\sin^{|\tilde{m}-2|}\frac{\theta}{2}\cos^{|\tilde{m}+2|}\frac{\theta}{2}
\rme^{\rmi\tilde{m}\phi}.
\end{equation}
%
Note that the this expression is consistent with the behaviour of the
spin-weighted spheroidal harmonics if $\psi^{(\tilde{m})}$ is finite and regular
at the two poles. The extra factor $\Delta^2/r$ is chosen so that the
amplitudes of $\psi^{(\tilde{m})}$ for ingoing waves to the horizon
and for outgoing waves to infinity  become constant.
If the source term is similarly decomposed as
%
\begin{equation}
T= \sum_{\tilde{m}=0,\pm1,\pm2,...}T^{(\tilde{m})}\rme^{\rmi\tilde{m}\phi},
\label{T-summation-m}
\end{equation}
%
we have the equation
%
\begin{eqnarray}
\fl
-\left[(r^2+a^2)^2-\Delta a^2(1-y^2)\right]\ddot{\psi}^{(\tilde{m})}+{(r^2+a^2)^2}\psi^{(\tilde{m})}_{,r_*r_*}
\nonumber\\
\fl
-{4}
\left[M(r^2-a^2)-\Delta(r+\rmi ay)+\rmi\tilde{m}Mar\right]\dot{\psi}^{(\tilde{m})}
+{2(r^2+a^2)}
\left[\Delta_{,r}-\frac{\Delta}{r}+\frac{r\Delta}{r^2+a^2}\right]
\psi_{,r_*}^{(\tilde{m})}
\nonumber\\
\fl
+{\Delta}
\left\{
(1-y^2)\psi^{(\tilde{m})}_{,yy}
-\left[|\tilde{m}-2|-|\tilde{m}+2|+(|\tilde{m}-2|+|\tilde{m}+2|+2)y\right]\psi^{(\tilde{m})}_{,y}
\right\}
\nonumber\\
\fl
+\left\{
{\tilde{m}^2a^2}-{4\rmi\tilde{m}a(r-M)}
+{\Delta}\left(2-\frac{3}{r}\Delta_{,r}+\frac{2\Delta}{r^2}\right)
\right.
\nonumber\\
\left.\qquad\qquad\qquad
-\frac{\Delta}{2}[(\tilde{m}^2-4)+|\tilde{m}^2-4|+|\tilde{m}-2|+|\tilde{m}+2|]
\right\}\psi^{(\tilde{m})}
\nonumber\\
=
-4\pi\frac{2^{\frac{|\tilde{m}-2|+|\tilde{m}+2|}{2}}\Sigma r}{\Delta (1-y)^{|\tilde{m}-2|/2}(1+y)^{|\tilde{m}+2|/2}}
T^{(\tilde{m})}.
\label{Eq:Teukolsky-modify}
\end{eqnarray}
%
The left-hand side of this equation is explicitly regular. Instead, the divergent coefficient
appears in the right-hand side. The regularity of this source term
is discussed below.

Now, let us discuss the source term in more detail.
It is convenient to adopt the decay constant $f_{\rm a}$ as the unit of the scalar
field $\Phi$ as done in \eref{Eq:normalized-scalar},
and we use $\varphi=\Phi/f_{\rm a}$ below. This means that
the Teukolsky variable is normalized by $f_{\rm a}^2$.
For the energy-momentum tensor
of the scalar field \eref{Eq:Energy-momentum-tensor-scalar}, each component in
\eref{Eq:T4-definition} becomes
%
\begin{equation}
\fl
T_{nm^*}=(n^\mu\partial_\mu\varphi)(m^{*\nu}\partial_\nu\varphi),
\quad
T_{m^*m^*}=(m^{*\mu}\partial_\mu\varphi)^2,
\quad
T_{nn}=(n^\mu\partial_\mu\varphi)^2.
\label{Eq:T-components}
\end{equation}
%
The important point is that since $n^\mu$ and $m^\mu$ are null and orthogonal to each other,
there is no contribution from the potential term $\hat{U}(\varphi)$.
We substitute \eref{Eq:decompose-phi-scalar} into \eref{Eq:T-components}
and then, into \eref{Eq:T4-definition}. Since $T_4$ is quadratic in the scalar field $\varphi$,
we label one of this by $m_1$ and the other by $m_2$. Then, $T$
can be expressed in the form
%
\begin{eqnarray}
T=\sum_{m_1,m_2}T^{(m_1,m_2)} \rme^{\rmi (m_1+m_2)\phi}.
\label{Eq:T-decompoision-m1m2}
\end{eqnarray}
%
This relation implies that the coupling 
between the $ m_1$ mode and the $m_2$ mode of the scalar field 
generates gravitational waves in the $\tilde{m}=m_1+m_2$ mode, denoted by $\psi^{(m_1,m_2)}$.
Comparing this expression with \eref{T-summation-m},
we find that gravitational waves in the $\tilde{m}$ mode
can be expressed as the sum of the contributions from all pairs of modes 
$(m_1,m_2)$  satisfying the condition $\tilde{m}=m_1 + m_2$.

We have to note the subtlety in the meaning of the mode number $\tilde{m}$.
As noted in the previous section, the mode function $f^{(m)}$ of the scalar field 
in \eref{Eq:decompose-phi-scalar} includes both positive and
negative frequency modes, which can be reinterpreted as the true 
$+m$ and $-m$ modes, respectively. As a result,  
the Teukolsky function $\psi^{(m_1,m_2)}$ also contains both 
positive and negative frequency modes, which are naturally
reinterpreted as the $+\tilde{m}$ and $-\tilde{m}$ modes, respectively.
Therefore, in order to specify the true $+\tilde{m}$ mode,
we have to decompose positive and negative frequency
modes in $\psi^{(m_1,m_2)}$ and $\psi^{(-m_1,-m_2)}$, and 
superpose the positive frequency mode in $\psi^{(m_1,m_2)}$
and the negative frequency mode in $\psi^{(-m_1,-m_2)}$. 
In this paper, we do not calculate this decomposition and resummation,
because the negative $m$ mode of the scalar field gives a minor contribution
to the axion cloud in the Schwarzschild case considered below.
But this procedure would be required in the more realistic Kerr case.

Since the denominator of the right-hand side of \eref{Eq:Teukolsky-modify}
becomes zero in general at $y=\pm 1$, its regularity has to be checked and
a regular expression must be implemented in the numerical code.
In the Kerr case, we checked the regularity at the poles using {\small MATHEMATICA},
although we do not present the explicit results here because it is very tedious.
In the Schwarzschild case, we can find a relatively simple expression,
and it is presented in \ref{Appendix:B}.

Similarly to the pseudospectral approach in the scalar case,
we further decompose $\psi^{(\tilde{m})}$ as
%
\begin{equation}
\psi^{(\tilde{m})}(t,r_*,y) = \sum_{\tilde{n}=0,1,2,...}\alpha^{(\tilde{m})}_{\tilde{n}}(t,r_*) y^{\tilde{n}}.
\label{Eq:psi-y-decomposition}
\end{equation}
%
By substituting this expression into the left-hand side of \eref{Eq:Teukolsky-modify},
and the corresponding series expressions for $f^{(m_1)}$ and $f^{(m_2)}$ with respect $y$,
\eref{Eq:decompose-y-scalar}, into the right-hand side, and
by equating the coefficients of each $y^n$,  we obtain a sequence of $(1+1)$-dimensional 
PDEs  for $\alpha_{\tilde{n}}^{(\tilde{m})}$ with sources expressed 
in terms of known functions of $a_n^{(m)}$.
More detailed formulas are shown in \ref{Appendix:B}.

In simulations, we first calculate the functions $a_n^{(m)}$ for the scalar field.
Then, for fixed $m_1$ and $m_2$, we solve the functions
$\alpha_{\tilde{n}}^{(m_1+m_2)}$
by using the data $a_n^{(m_1)}$ and $a_n^{(m_2)}$ for the source term.
We use the sixth-order finite differencing in spatial direction
and the fourth-order Runge-Kutta method in time direction. In this method,
a numerical instability develops around $0\lesssim r_*\lesssim 100M$. In order to stabilize
the numerical data, we added a stabilization term $\delta(r_*)\dot{\psi}^{(\tilde{m})}_{,r_*r_*}$
to the left-hand side of \eref{Eq:Teukolsky-modify},
where
%
\numparts
\begin{equation}
\delta(r_*) =
\epsilon_1\exp\left[-\left(\frac{r_*-r_*^{(1)}}{w_1}\right)^2\right]
+\epsilon_2\exp\left[-\left(\frac{r_*-r_*^{(2)}}{w_2}\right)^2\right],
\end{equation}
with
\begin{equation}
\fl
\epsilon_1=0.015, \quad \epsilon_2=0.005, \quad
r_*^{(1)} = 20M, \quad r_*^{(2)} = 70M, \quad w_1=w_2=30M.
\end{equation}
\endnumparts
%
Because adding such a term is artificial, there remains a room for improvement.
Here, we postpone this issue as a future research, and show the results
obtained under this method. But we would like to point out that
because generated gravitational waves propagates towards the horizon or infinity,
the period during which
the stabilization term affects the gravitational wave form is relatively small.
Note that in the scalar case in the previous section, it was not
necessary to add such a stabilization term. This problem is specific to
the Teukolsky equation for a gravitational perturbation.

\subsection{Numerical results for the Schwarzschild case}

At this point, we have not yet developed a code for a Kerr black hole case
because the source term  is very complicated.
In this paper, as a preliminary research, we present the results for the
Schwarzschild case.  Because the energy extraction does not
happen, we cannot discuss the generation of gravitational waves
at bosenova as a result of the superradiant instability.
However, in the case of Klein-Gordon field without self-interaction
in the Schwarzschild background,
there exist long-lived quasibound states with small decaying rates:
For  $M\mu= 0.30$, the decay rate of the $\ell=m=1$ fundamental mode
is $M\omega_{\rm I}\approx -0.9456\times 10^{-5}$.
Then, we can prepare a Klein-Gordon quasibound state with a certain amplitude
as the initial condition by hand, and evolve the scalar field by switching on the
nonlinear self-interaction. The phenomena in this setup captures some of the features
of the axion bosenova in the more realistic Kerr-background case within the time scale
of few thousand $M$, and therefore,
we can understand the gross feature of gravitational waves emitted during the bosenova.

In what follows, we show the numerical results for the three cases: the Klein-Gordon case
(without nonlinear self-interaction), the mildly nonlinear case, and the
strongly nonlinear case, one by one. The mass of the scalar field is fixed to be $M\mu=0.30$,
and we use the scalar field in the quasibound state in the $\ell = m = 1$ fundamental mode
as the source for gravitational waves.

Note that if the black hole is a nonrotating Schwarzschild black hole
and the scalar field is symmetric with respect to the equatorial plane
as assumed above,
$\psi^{(-\tilde{m})}$ is related to $\psi^{(\tilde{m})}$ as
$\psi^{(-\tilde{m})}(t,r,y)=\psi^{(\tilde{m})*}(t,r,-y)$.
Therefore, it is sufficient to consider the positive number of $\tilde{m}$.

\subsubsection{Klein-Gordon case}

We first briefly discuss  gravitational wave emissions from the Klein-Gordon
quasibound state. In this case,
the spectral component for gravitational waves,
$\alpha_{\tilde{n}}^{(\tilde{m})}$ with $(\tilde{n},\tilde{m})=(0,2)$,
is generated from the scalar field $a_{n}^{(m)}$ with $(n,m)=(0,\pm 1)$.
No excitation of higher $\tilde{n}$ modes has been found, which is consistent with
our previous perturbative study \cite{Yoshino:2013}.

%
\begin{figure}[tbh]
\centering
\includegraphics[width=0.45\textwidth,bb=0 0 360 252]{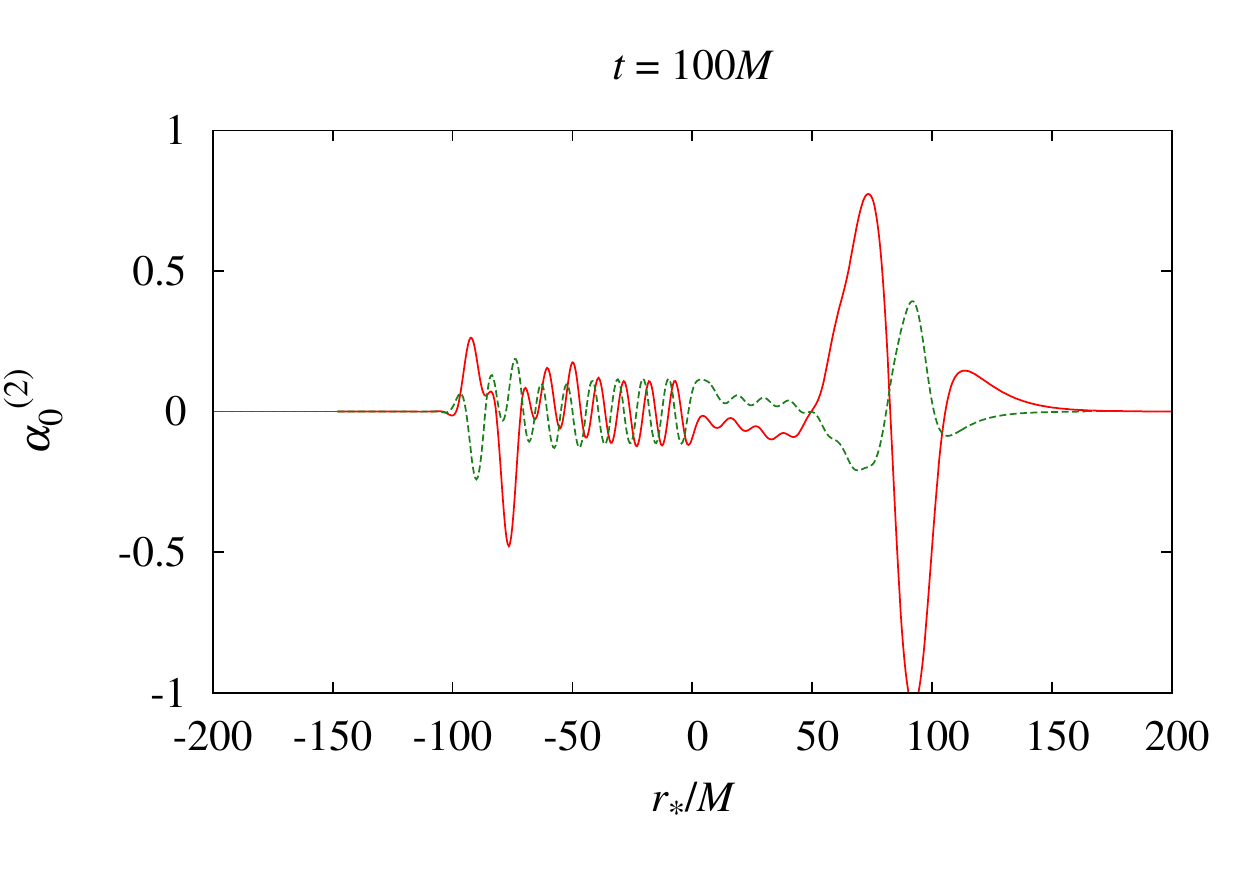}
\includegraphics[width=0.45\textwidth,bb=0 0 360 252]{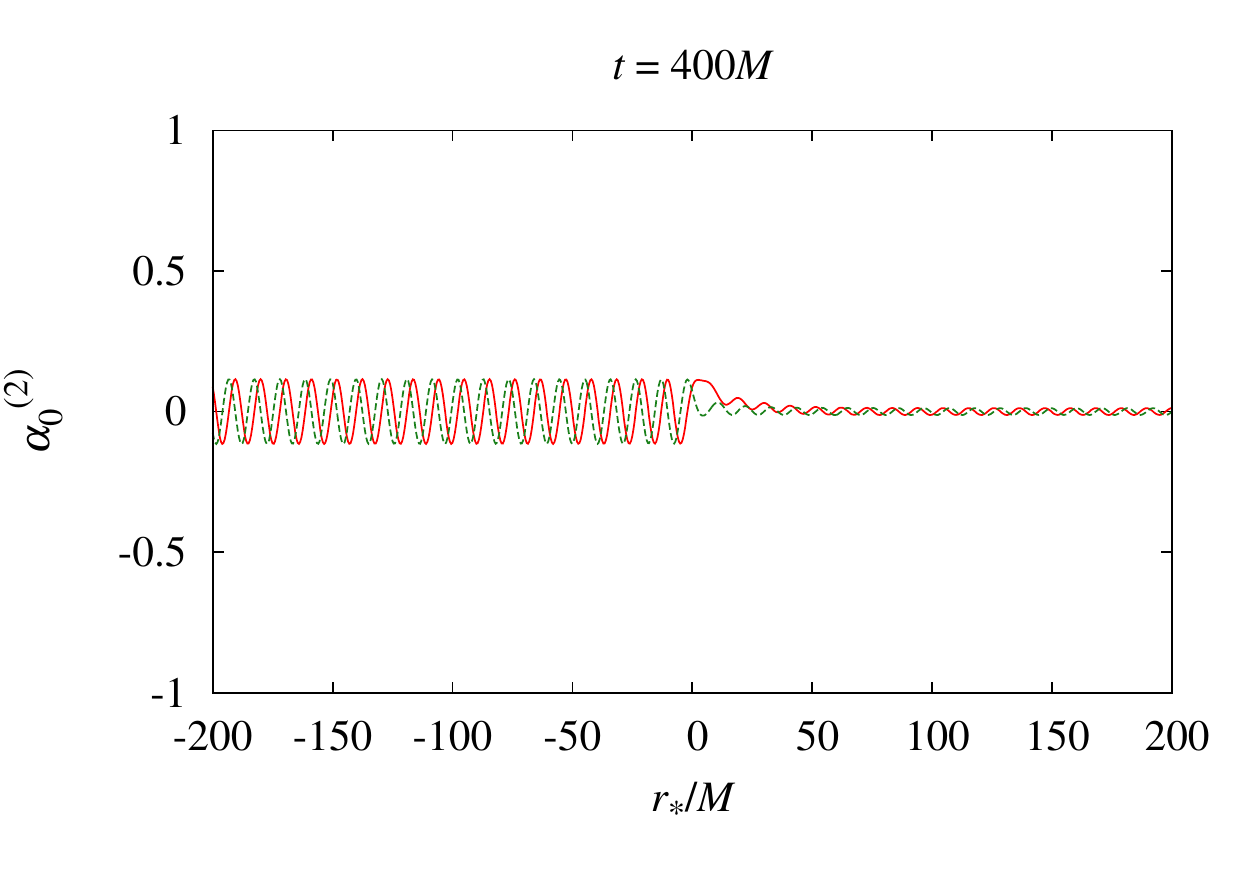}
\caption{Snapshots of the real part (solid line, red online) and the imaginary part
(long dashed line, green online) for $\alpha_{0}^{(2)}$ ($\psi^{(\tilde{m})}$ for $\tilde{m}=2$
in the equatorial plane) at time $t=100M$ (Left panel) and $t=400M$ (Right panel) in the
Klein-Gordon case.}
\label{Fig:Snapshots-GW-KleinGordon}
\end{figure}
%

\Fref{Fig:Snapshots-GW-KleinGordon} shows snapshots of
the real and imaginary parts of $\alpha_{\tilde{n}}^{(\tilde{m})}$ with $(\tilde{n},\tilde{m})
=(0,2)$ for $t=100M$ (Left panel) and $t=400M$ (Right panel).
We started with the initial data $\psi^{(\tilde{m})} = \dot{\psi}^{(\tilde{m})} = 0$.
Since they are not realistic initial data in the sense that they do not reflect continuous waves
emitted before $t=0$, a spurious gravitational wave burst can be seen
in the initial stage.
But they are soon radiated away, and the generation of continuous waves are observed
after a sufficient time.

\subsubsection{Mildly nonlinear case}

Now, we consider the case in which the scalar field follows the sine-Gordon equation.
Here, we consider the mildly nonlinear case,
where the initial amplitude of the scalar field oscillation is chosen as
$\varphi_{\rm peak}=0.40$.

%
\begin{figure}[tbh]
\centering
\includegraphics[width=0.65\textwidth,bb=0 0 360 252]{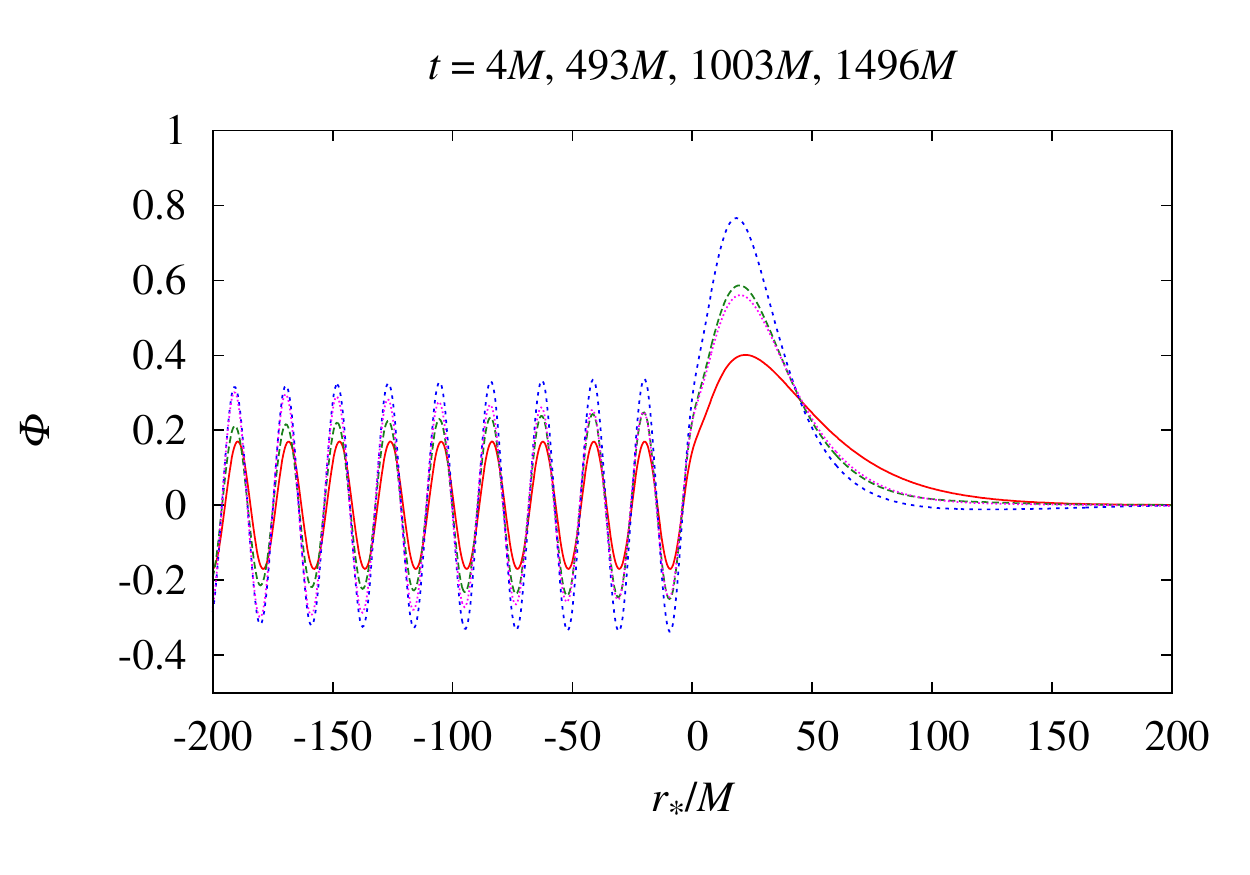}
\caption{A snapshot of the scalar field on the $\phi=0$ line in the equatorial plane
for $t=4M$ (solid line, red online),
$493M$ (long-dashed line, green online), $1003M$ (short-dashed line, blue online),
and $1496M$ (dotted line, purple online). 
Time is chosen in order to depict the moment when the scalar
field peak passes the $\phi=0$ line.}
\label{Fig:Snapshots-scalar-amp04-n0m1}
\end{figure}
%

Before looking at generated gravitational waves, let us inspect the
scalar field behaviour. \Fref{Fig:Snapshots-scalar-amp04-n0m1}
shows the snapshots of the scalar field on the line $\phi=0$ in the
equatorial plane at time $t=4M$ (solid line),
$493M$ (long-dashed line), $1003M$ (short-dashed line),
and $1496M$ (dotted line). Here, we selected the moments at which
the scalar field peak passes the $\phi=0$ line. Although
the field amplitude changes by a factor due to self-interaction,
the system does not experience a violent phenomena analogous to the bosenova.
In fact, the amplitude increases up to $t\approx 1000M$,
but it decreases after that. As a result, the scalar field
distribution is very similar for $t=493M$ and $1496M$. Therefore,
the situation is similar to the weakly nonlinear case discussed in the
rotating background in \cite{Yoshino:2012}. In the pseudospectral approach,
the field primarily consists of $a_n^{(m)}$ with $(n,m)=(0,1)$, and other components
are scarcely excited.

Although the behaviour of the scalar field caused by the nonlinear self-interaction
is not very dynamic, interesting phenomena appear in the generated gravitational waves.
Below, we show just the gravitational wave forms generated by $m_1=m_2=\pm 1$ mode of the
scalar field, because the higher $m$ modes remain to be small.
Therefore, we consider the gravitational waves in the $\tilde{m} = 2$ mode excited
by the $m_1=m_2=1$ mode of the scalar field. In the mildly nonlinear case,
gravitational waves in the $\tilde{m}=0$ mode excited by
the $m_1=-m_2=\pm 1$ modes are scarcely observed.

%
\begin{figure}[tbh]
\centering
\includegraphics[width=0.45\textwidth,bb=0 0 360 252]{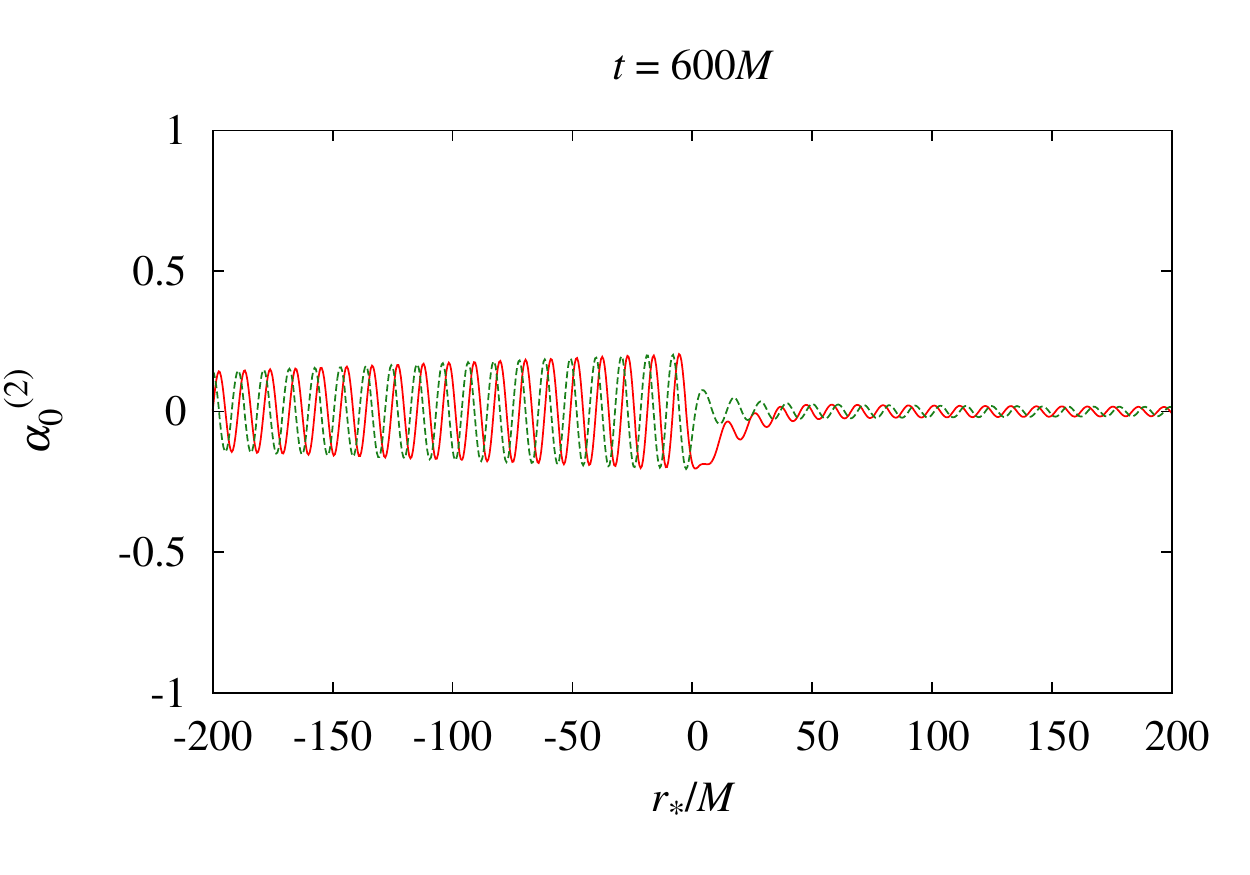}
\includegraphics[width=0.45\textwidth,bb=0 0 360 252]{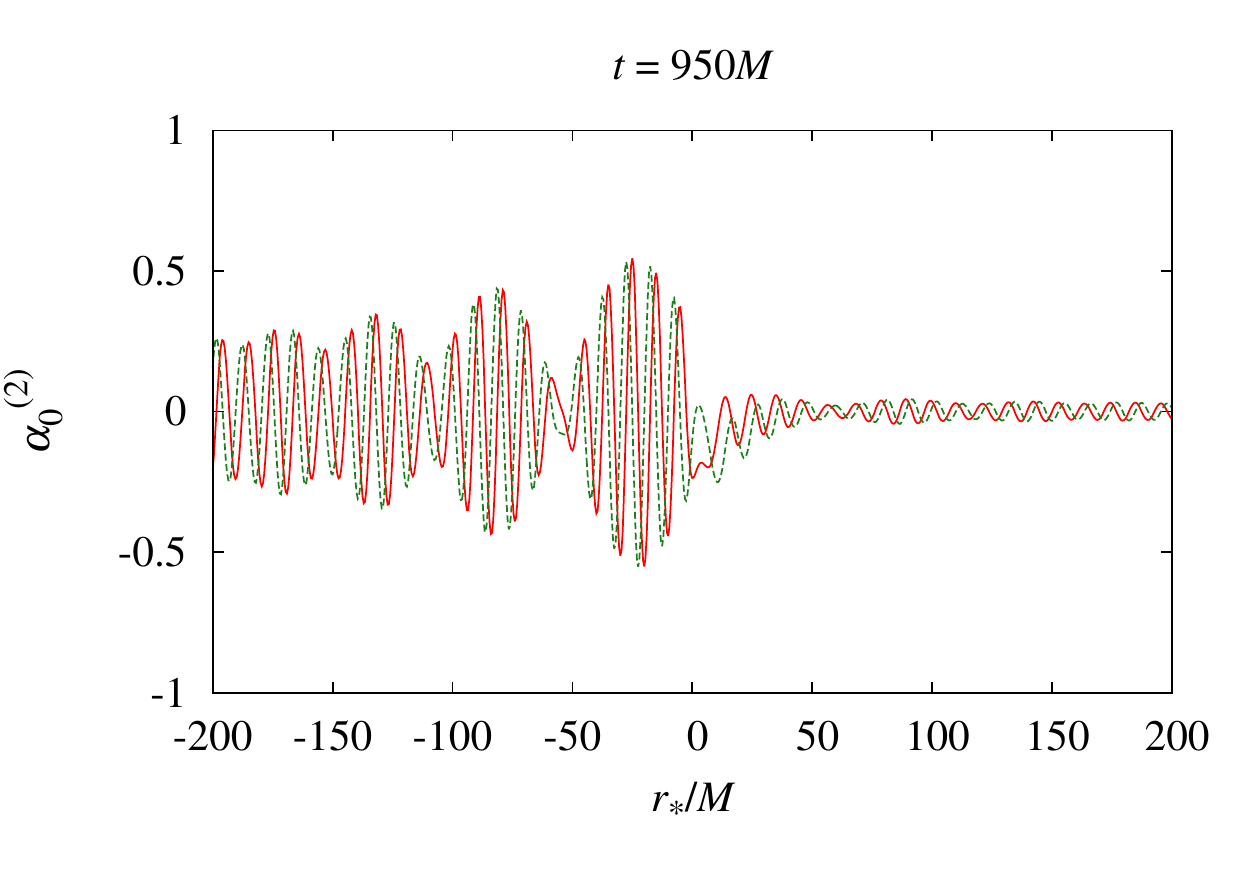}
\includegraphics[width=0.45\textwidth,bb=0 0 360 252]{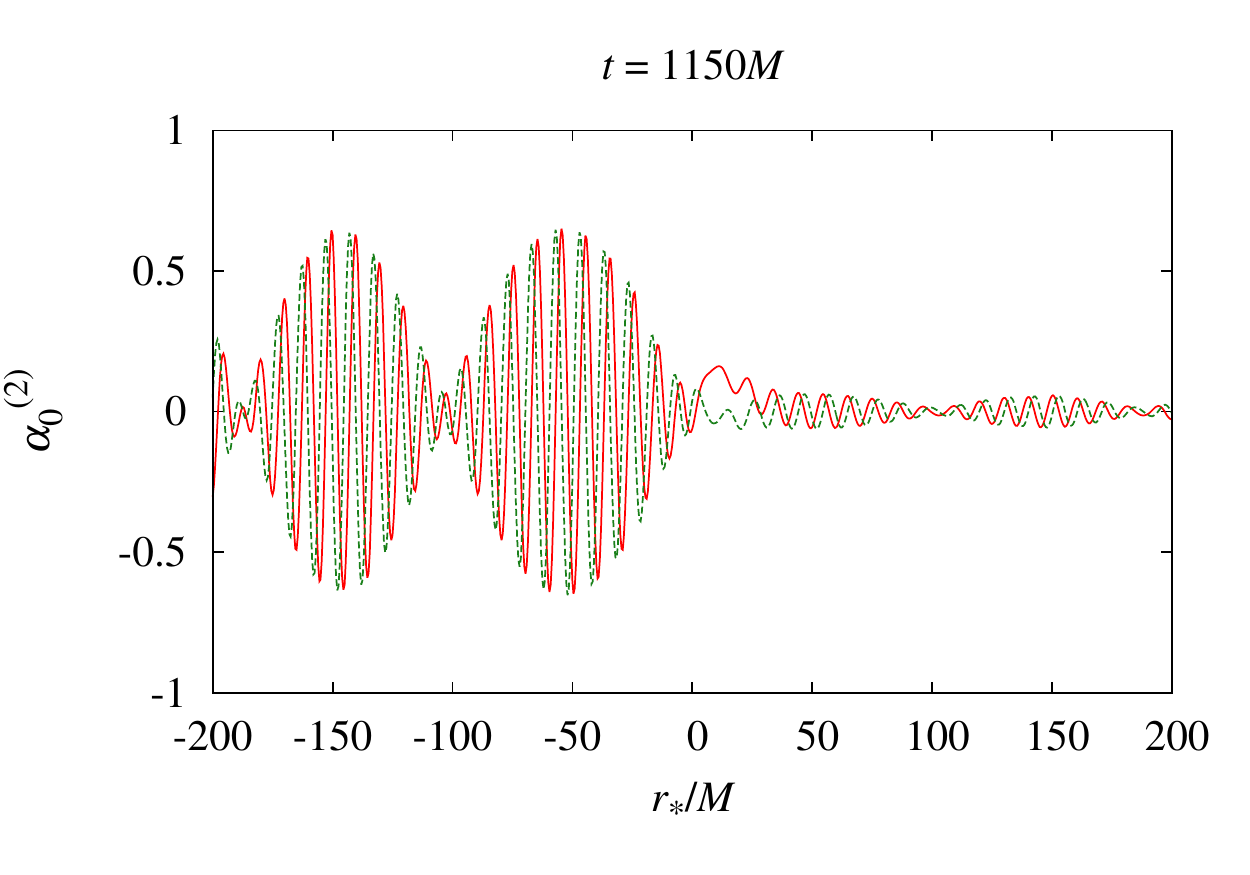}
\includegraphics[width=0.45\textwidth,bb=0 0 360 252]{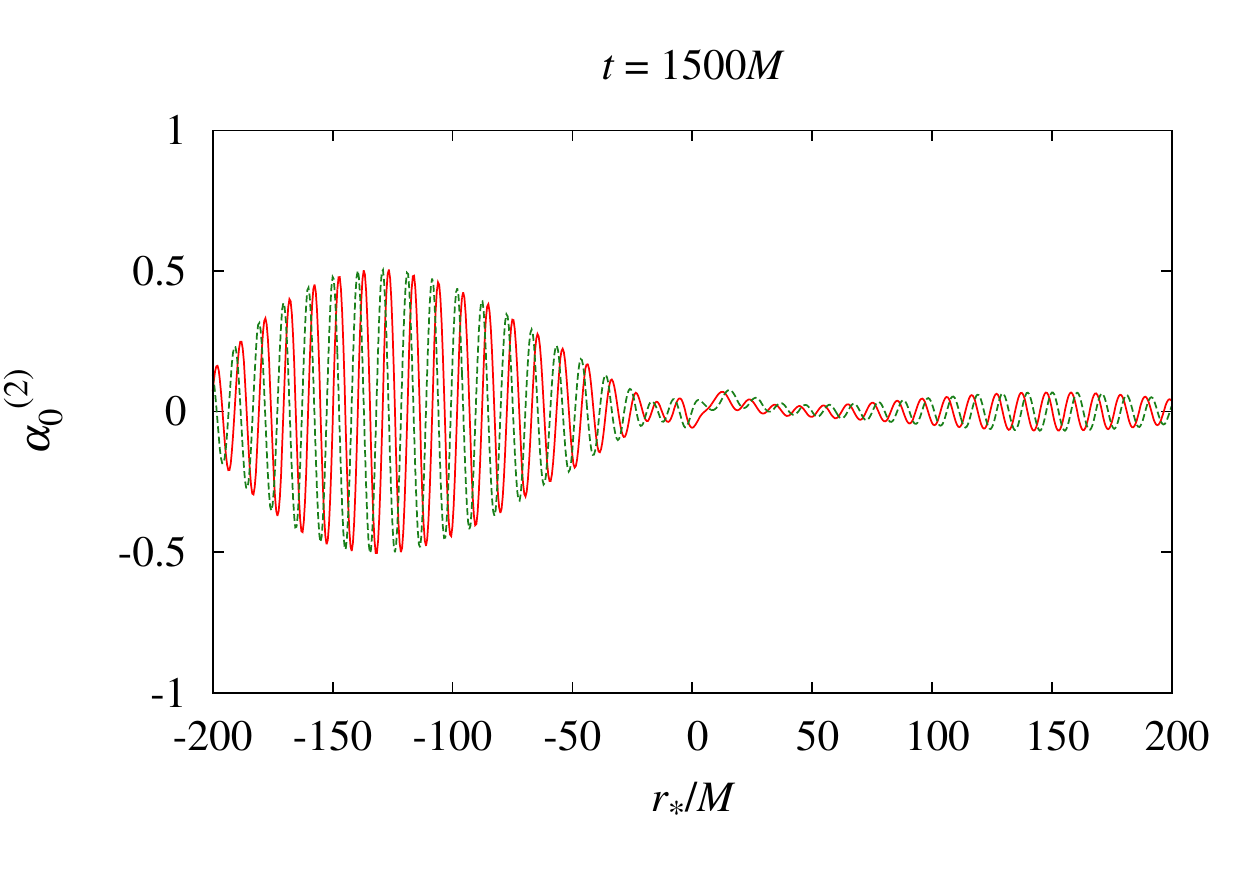}
\caption{Snapshots of the real  (solid line, red online) and imaginary (long dashed line, green online)
part of spectral component $\alpha_{0}^{(2)}$
of the Teukolsky variable for a gravitational perturbation
at time $t=600M$ (top left panel), $t=950M$ (top right panel),
$t=1150M$ (bottom left panel) and $t=1500M$ (bottom right panel) in the
mildly nonlinear case.}
\label{Fig:Snapshots-GW-amp04-n0m2}
\end{figure}
%

\Fref{Fig:Snapshots-GW-amp04-n0m2} shows snapshots of
the real and imaginary parts of $\alpha_{\tilde{n}}^{(\tilde{m})}$ with $(\tilde{n},\tilde{m})
=(0,2)$ for $t=600M$ (top left panel), $t=950M$ (top right panel),
$t=1150M$ (bottom left panel), and $t=1500M$ (bottom right panel).
Note that this quantity agrees with the value of $\psi^{(\tilde{m})}$ in the equatorial plane
($y=0$).
Similarly to the Klein-Gordon case, continuous waves appear after the radiation of
spurious initial burst. But in this case, its amplitude increases in time.
Around $t=900M$, the ``beating'' pattern appears in the radiated gravitational wave,
and it becomes more prominent as time goes.
The ``beat frequency'' changes in time and becomes smaller as time goes,
as can be understood by comparing the three panels for $t=950M$, $1150M$, and $1500M$.

%
\begin{figure}[tbh]
\centering
\includegraphics[width=0.8\textwidth,bb= 0 0 350 155]{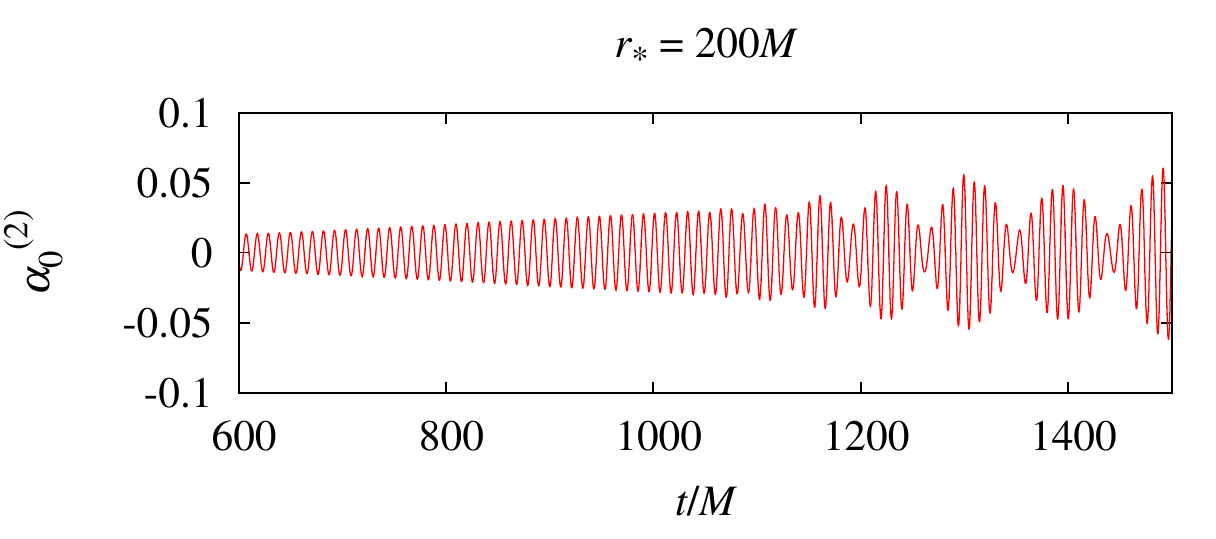}
\caption{Behaviour of real part of the spectral component $\alpha_0^{(2)}$
of the Teukolsky variable for a gravitational perturbation
as a function of time observed at the point $r_*=200M$ in the mildly nonlinear case.}
\label{Fig:Timeevolution-GW-amp04-n0m2}
\end{figure}
%

\Fref{Fig:Timeevolution-GW-amp04-n0m2} shows the value of
$\alpha_{\tilde{n}}^{(\tilde{m})}$ for $(\tilde{n},\tilde{m})=(0,2)$ as a function of time
observed at the fixed position $r_*=200M$. After a linear growth in the amplitude,
the beating phenomena are observed.

%
\begin{figure}[tbh]
\centering
\includegraphics[width=0.65\textwidth,bb=0 0 360 252]{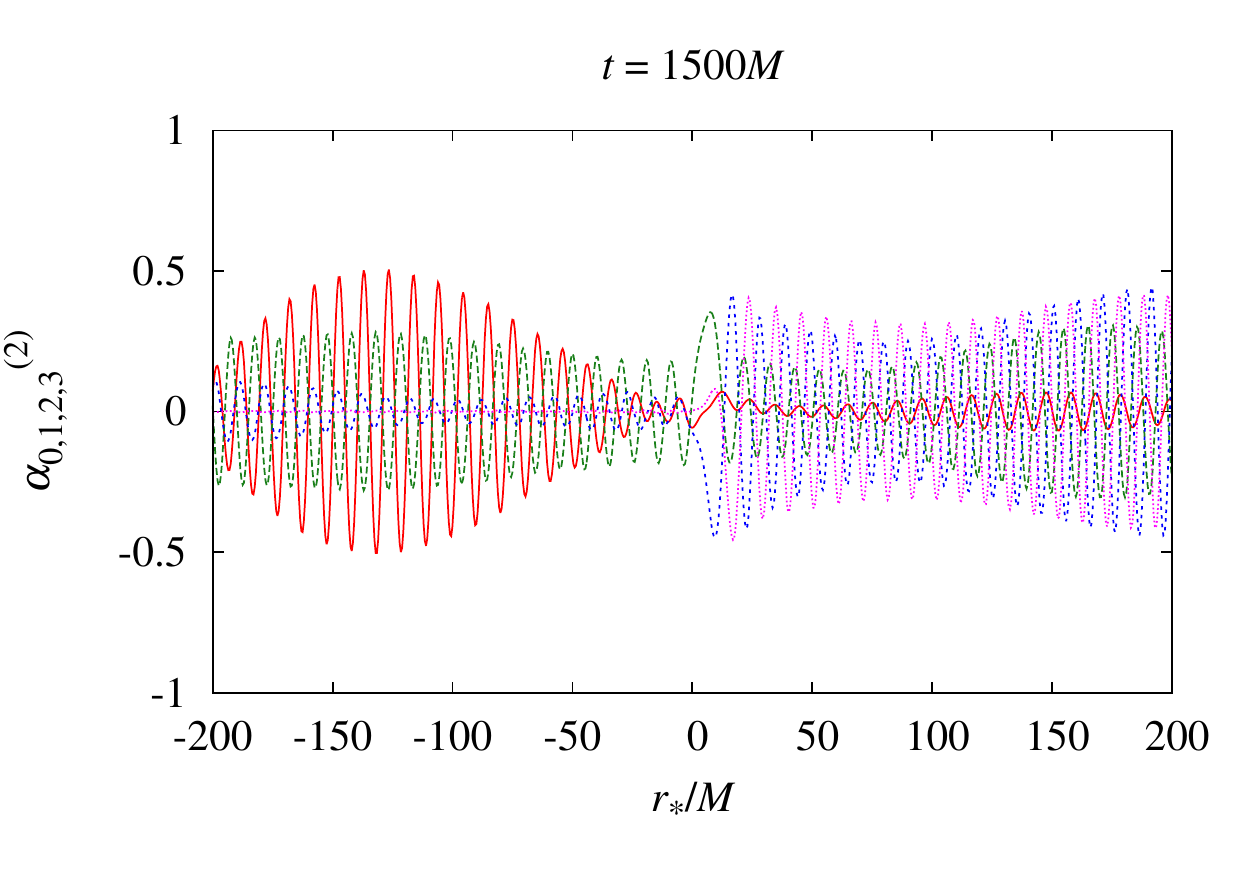}
\caption{A snapshot of the real part of the spectral component
$\alpha_{\tilde{n}}^{(\tilde{m})}$ of Teukolsky variable for a gravitational perturbation
for fixed $\tilde{m}=2$ and $\tilde{n}=0$ (solid line, red online),
$1$ (long-dashed line, green online), $2$ (short-dashed line, blue online),
and $3$ (dotted line, purple online) at time $t=1500M$ in the
mildly nonlinear case.}
\label{Fig:Snapshots-GW-amp04-n0n1n2n3m2}
\end{figure}
%

\Fref{Fig:Snapshots-GW-amp04-n0n1n2n3m2} shows a snapshot
for $t=1500M$, but for $\alpha_{\tilde{n}}^{(\tilde{m})}$ with $\tilde{m}=2$ and
$\tilde{n}=0$ (solid line),
$1$ (long-dashed line), $2$ (short-dashed line),
and $3$ (dotted line). In contrast to the case of the Klein-Gordon field,
the higher $\tilde{n}$ modes are also excited. This means that
gravitational waves in the modes with $\tilde{\ell}>2$
are excited due to nonlinear self-interaction of the scalar field when they are expanded by the spin-weighted spheroidal harmonics ${}_{-2}Y^{\tilde m}_{\tilde \ell}$.

To summarize, in the mildly nonlinear case, several modifications from the Klein-Gordon case
appear in the gravitational wave forms although the scalar field
is not highly dynamical. After a growth in the amplitude
of monochromatic waves, beating pattern appears at this point.
The beating frequency changes in time: In our simulation time scale, it
becomes smaller as $t$ is increased. Also,
gravitational waves in the higher $\tilde{\ell}$ modes are excited.

\subsubsection{Strongly nonlinear case}

%
\begin{figure}[tbh]
\centering
\includegraphics[width=0.45\textwidth,bb= 0 0 250 230]{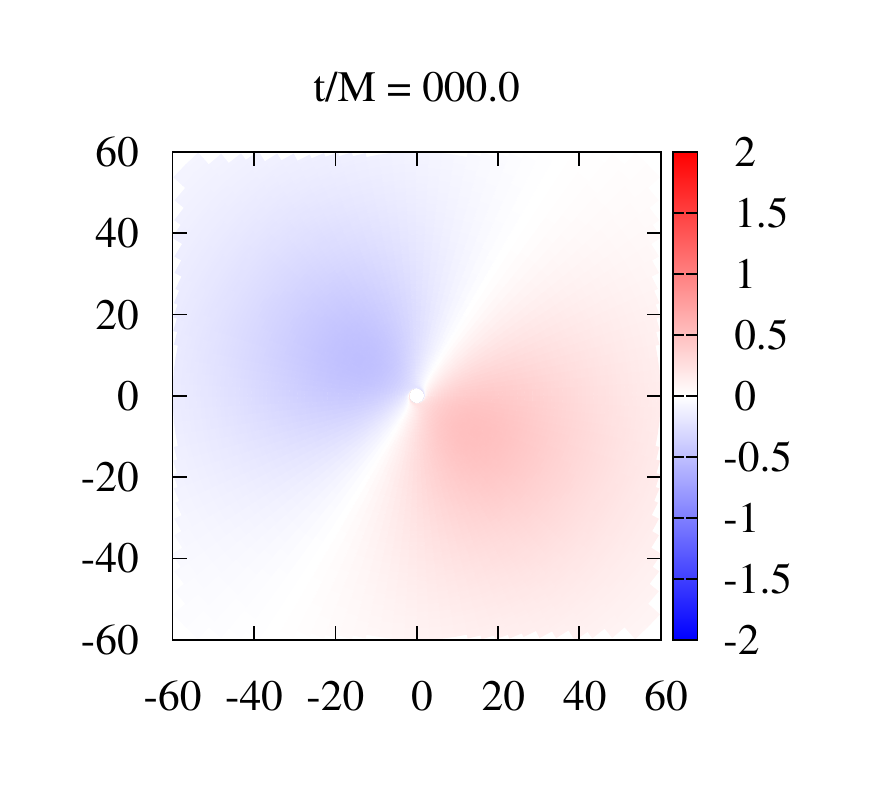}
\includegraphics[width=0.45\textwidth,bb= 0 0 250 230]{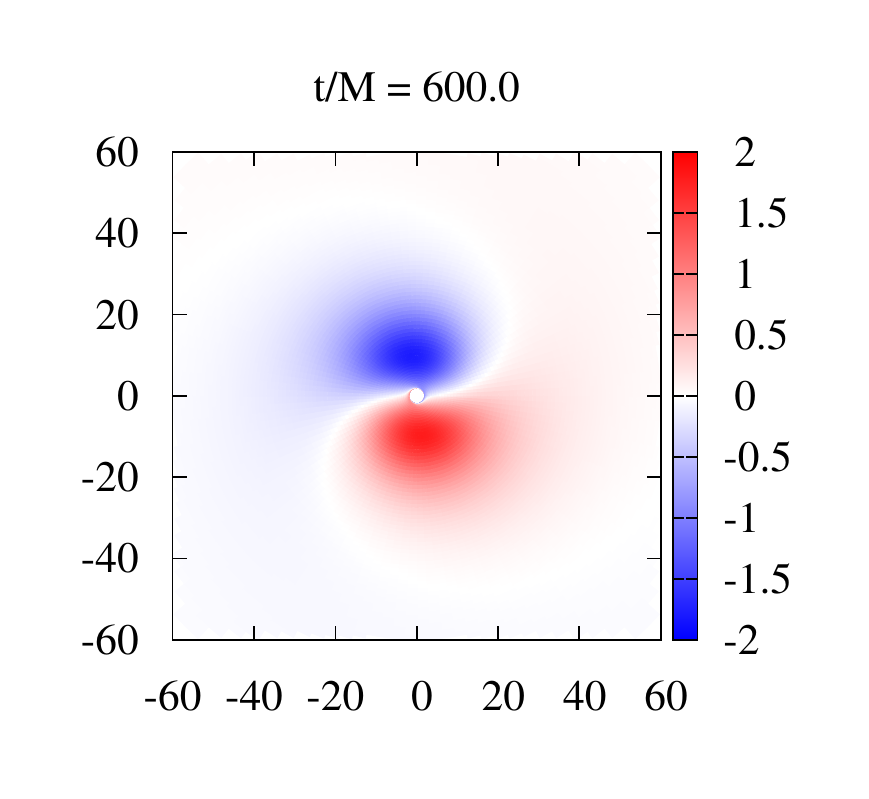}
\includegraphics[width=0.45\textwidth,bb= 0 0 250 230]{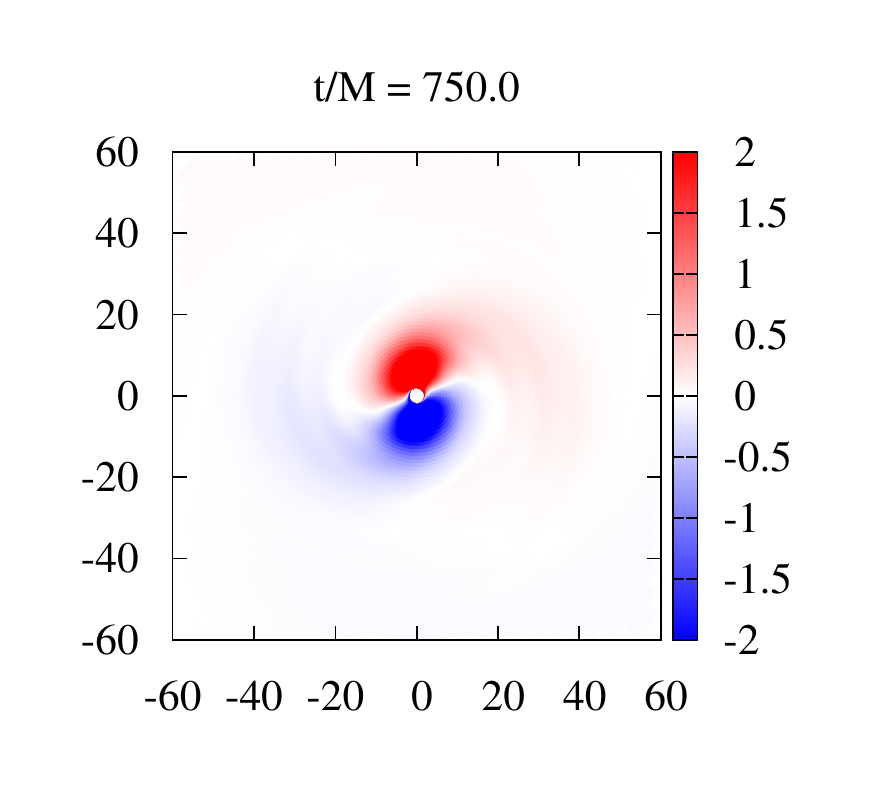}
\includegraphics[width=0.45\textwidth,bb= 0 0 250 230]{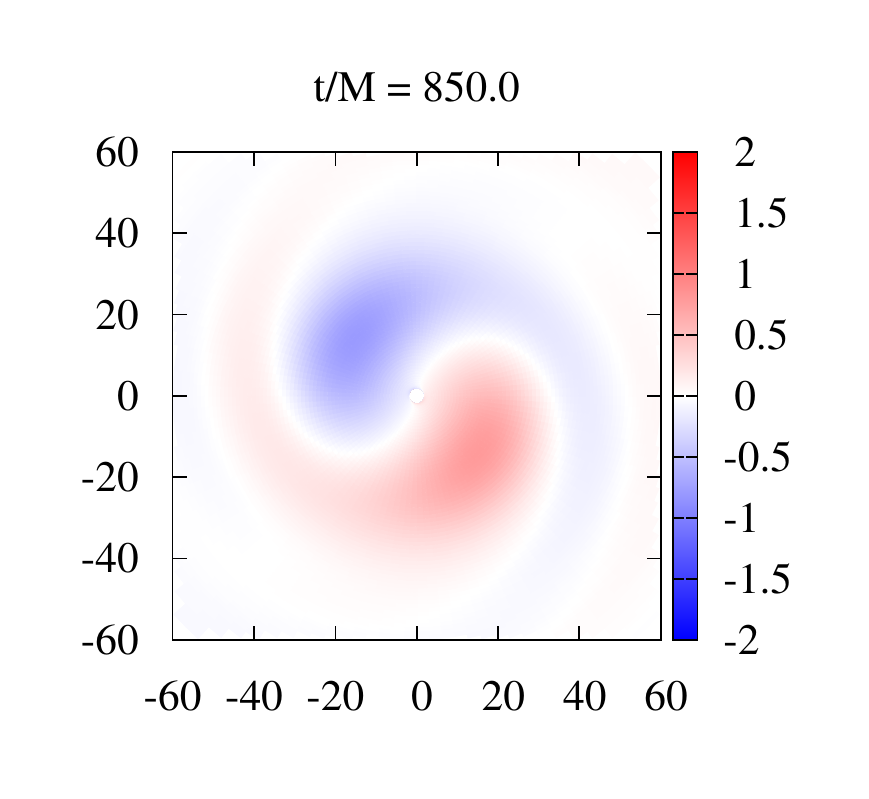}
\caption{Snapshots of the scalar field in the strongly nonlinear case
at $t=0$ (top left), $600M$ (top right), $750M$ (bottom left), and $850M$ (bottom right) 
in the equatorial plane.}
\label{Fig:Snapshots-scalar-2D}
\end{figure}
%
%
\begin{figure}[tbh]
\centering
\includegraphics[width=0.65\textwidth,bb=0 0 360 252]{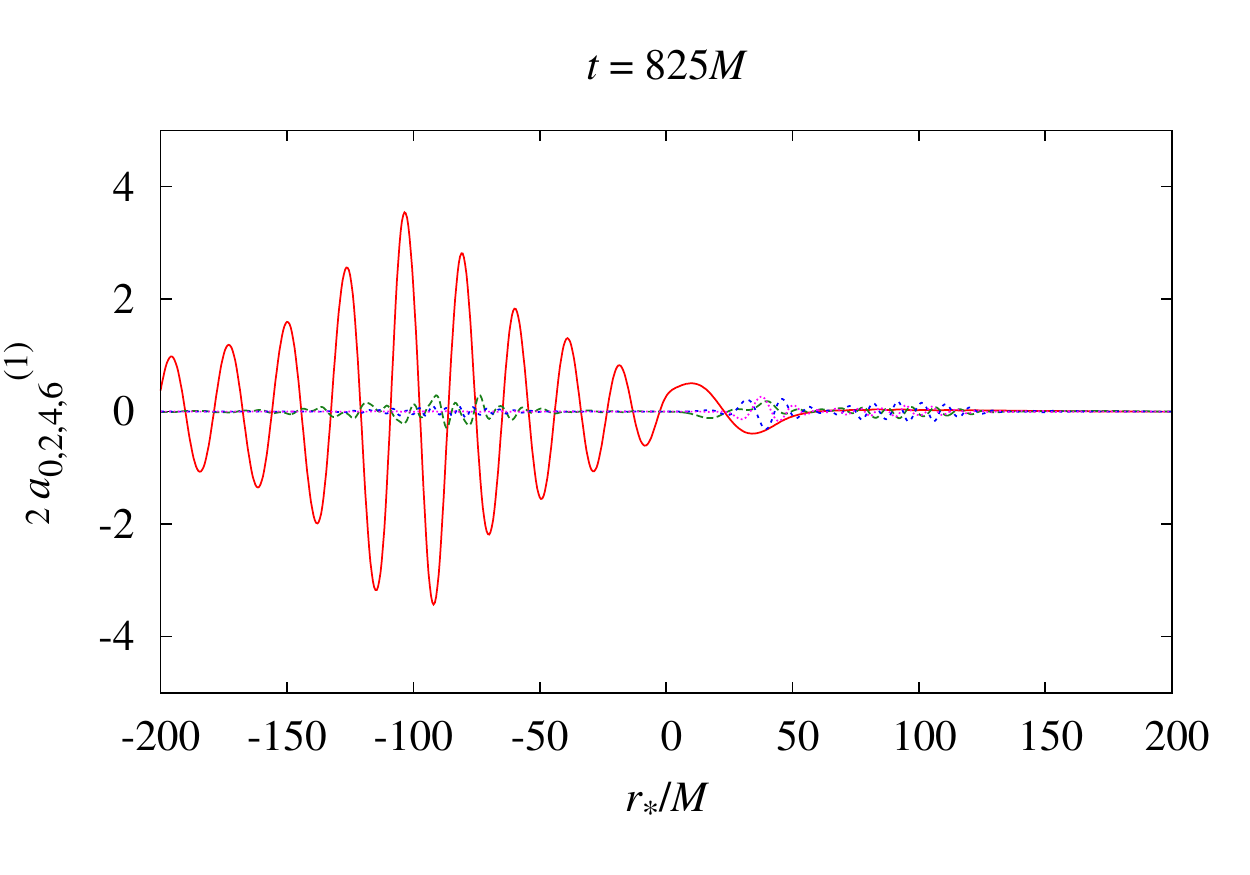}
\caption{A snapshot of the twice of the real part of the
spectral components $a_{n}^{(m)}$ for the scalar field
with $m=1$ and $n=0$ (solid line, red online), $2$ (long-dashed line, green online), 
$4$ (short-dashed line, blue online), and $6$
 (dotted line, purple online) at time $t=825M$. 
 This depicts the moment at which a large amount of the scalar
 field energy falls into the black hole.}
\label{Fig:Snapshots-scalar-amp05-n0n2n4n6m1}
\end{figure}
%

Now, we turn our attention to the case of the strongly nonlinear case.
Here, we choose the initial amplitude of the scalar field oscillation to be
$\varphi_{\rm peak}=0.5$.  \Fref{Fig:Snapshots-scalar-amp05-n0n2n4n6m1} shows
the scalar field configuration in the equatorial plane.
The scalar field becomes highly concentrated towards the black hole
due to nonlinear self-interaction rapidly,
and after several turns, the concentration is destroyed, and
the field configuration changes to a disperse spiral pattern.
This phenomenon is quite analogous to the bosenova observed in the rotating case
in our previous work:
 Compare this figure with figure~8 of \cite{Yoshino:2012}.
\Fref{Fig:Snapshots-scalar-amp05-n0n2n4n6m1} depicts twice of
the real part of the spectral component $2a_n^{(m)}$ of the scalar field
with $m=1$ and $n=0$ (solid line), $2$ (long-dashed line),
$4$ (short-dashed line), and $6$ (dotted line). In this case, the scalar field in the $m=1$
gets amplified and falls into the black hole, and no excitation of the $m=-1$ mode is observed.
This is a slight difference from the realistic bosenova in the
rotating black hole case \cite{Yoshino:2012}.

%
\begin{figure}[tbh]
\centering
\includegraphics[width=0.45\textwidth,bb= 0 0 360 252]{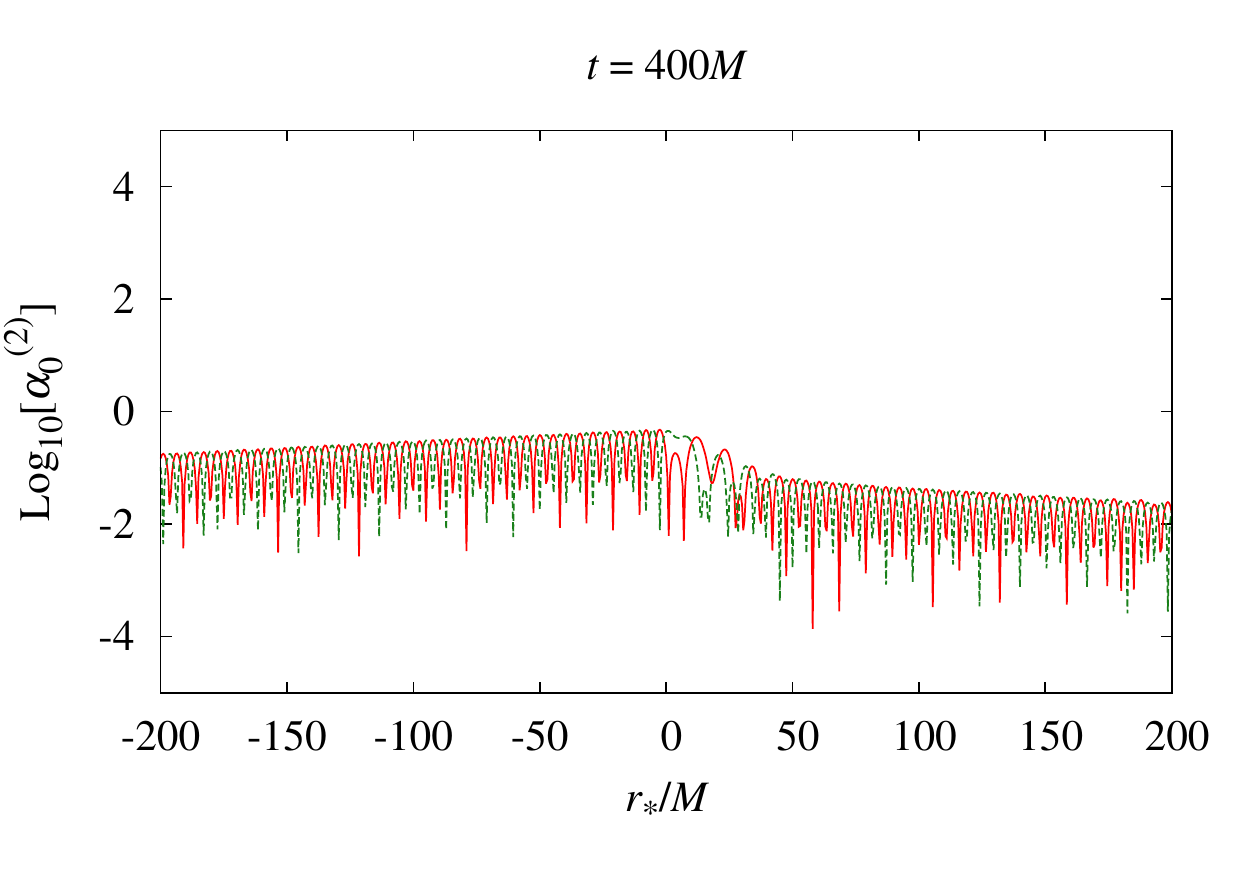}
\includegraphics[width=0.45\textwidth,bb= 0 0 360 252]{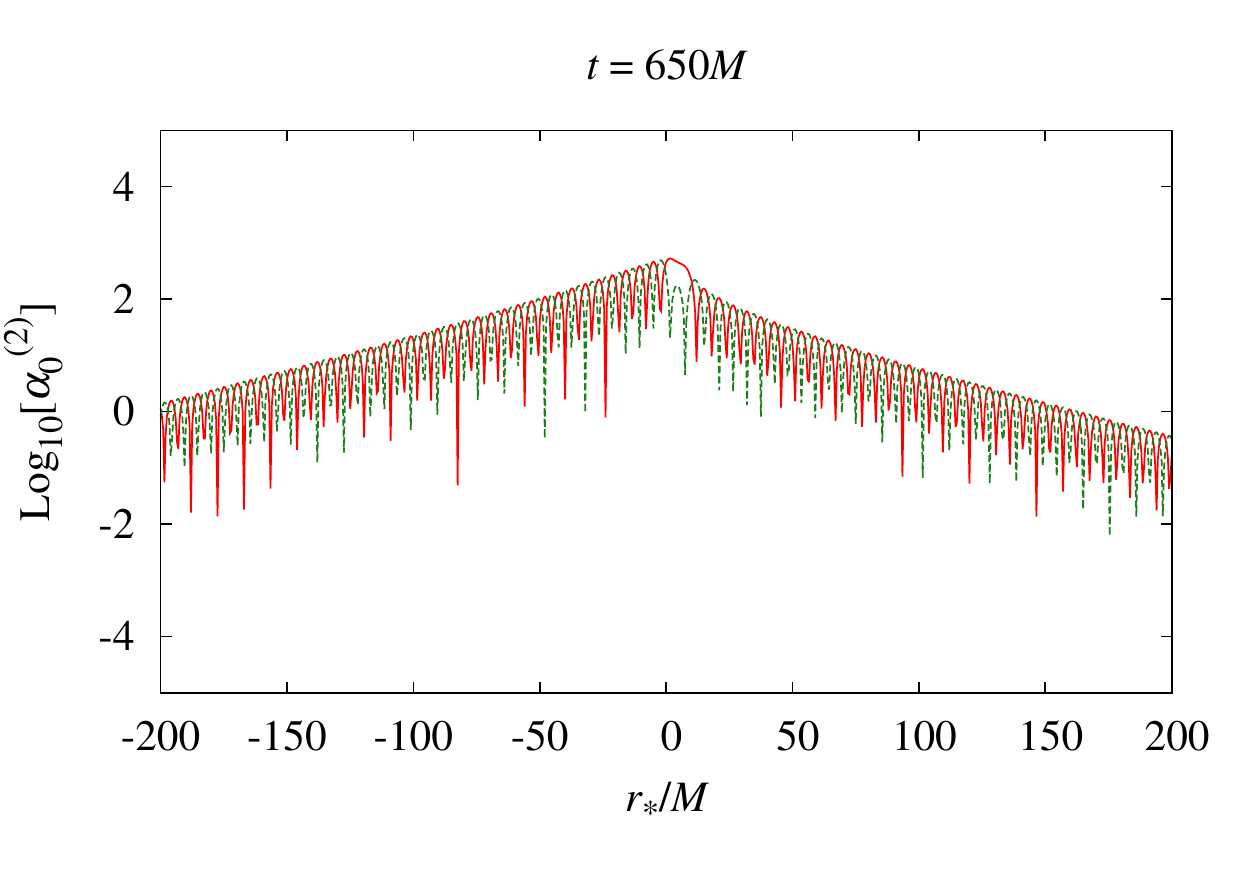}
\includegraphics[width=0.45\textwidth,bb= 0 0 360 252]{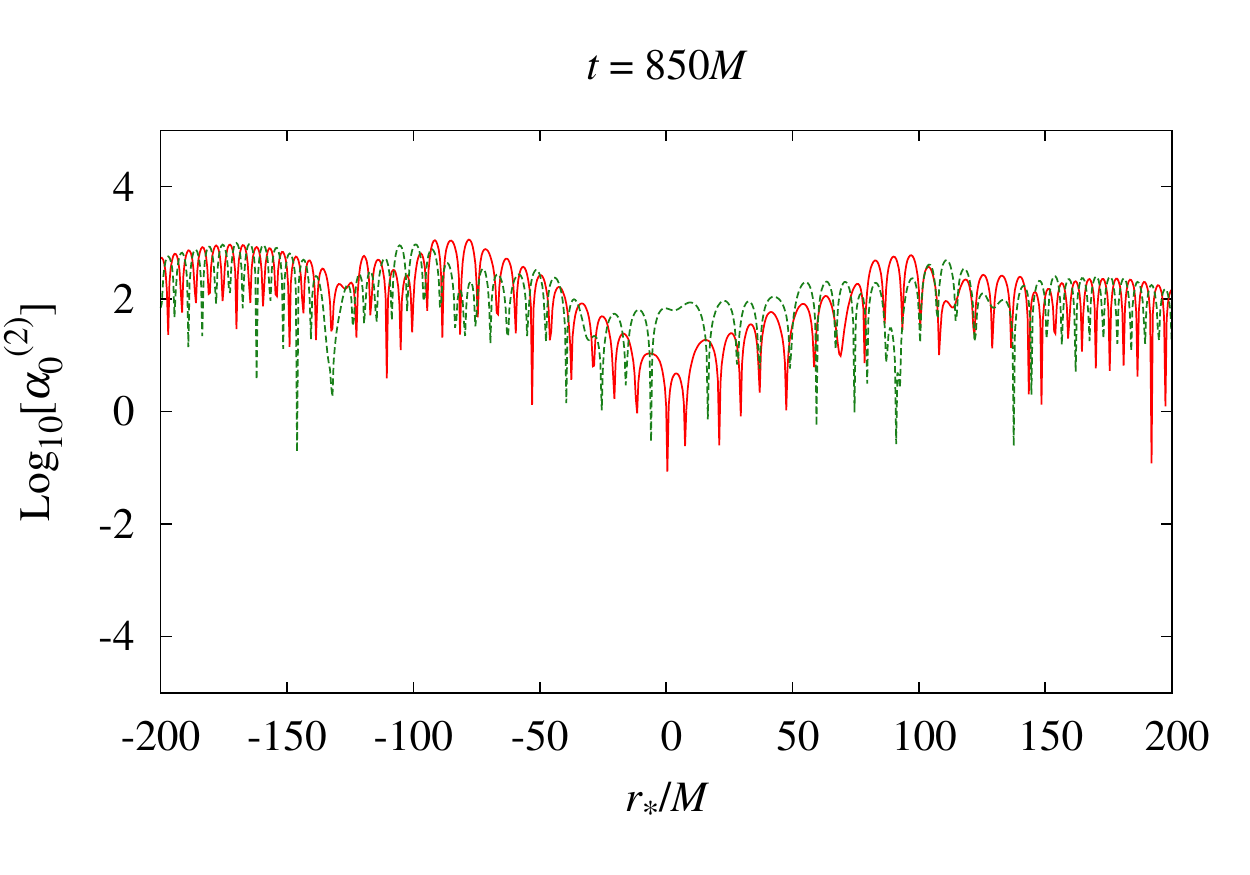}
\includegraphics[width=0.45\textwidth,bb= 0 0 360 252]{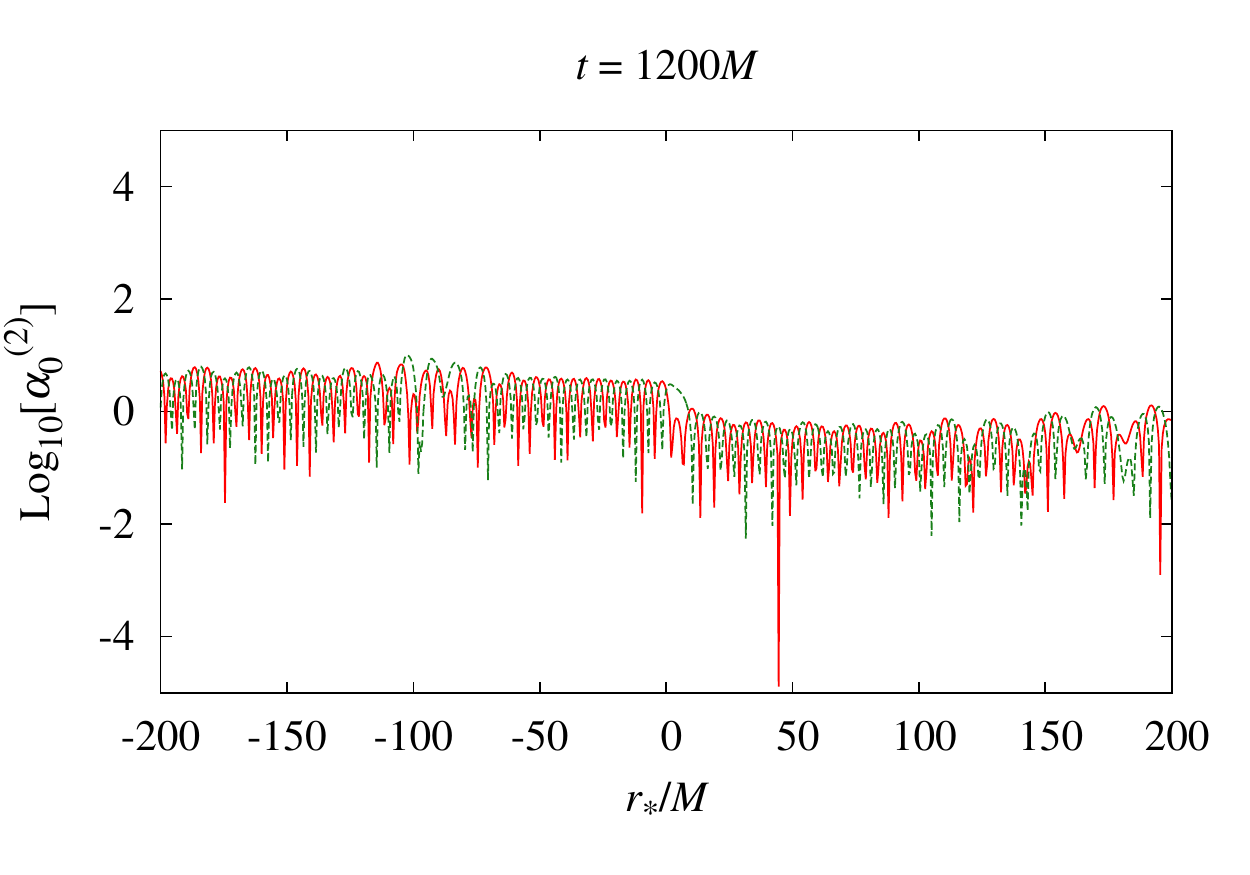}
\caption{Snapshots of the real part (solid line, red online) and imaginary part 
(long dashed line, green online)
of spectral component $\alpha_{0}^{(2)}$
of the Teukolsky variable for a gravitational perturbation in the logarithmic scale
at time $t=400M$ (top left panel), $t=650M$ (top right panel),
$t=850M$ (bottom left panel) and $t=1200M$ (bottom right panel) in the
strongly nonlinear case.}
\label{Fig:Snapshots-GW-amp05-n0m2}
\end{figure}
%

Now, we study gravitational waves.
Similarly to the mildly nonlinear case, we study just the gravitational wave forms
generated by the $m_1=m_2=\pm 1$ modes of the scalar field.
But in this case, we consider both the $\tilde{m} = 2$ mode excited
by the $m_1=m_2=1$ mode and the $\tilde{m}=0$ mode excited by
the $m_1=-m_2=\pm 1$ modes.
\Fref{Fig:Snapshots-GW-amp05-n0m2} shows snapshots of
the real and imaginary parts of $\alpha_{\tilde{n}}^{(\tilde{m})}$ with $(\tilde{n},\tilde{m})
=(0,2)$ for $t=400M$ (top left panel), $t=650M$ (top right panel),
$t=850M$ (bottom left panel), and $t=1200M$ (bottom right panel).
Because the gravitational wave amplitude changes significantly in time,
we show these quantities in the logarithmic scale.
Similarly to the mildly nonlinear case, continuous waves appear
and their amplitude increases in time (top left).
In this case the beating does not appear, and
the amplitude grows exponentially in time to become $10^4$
times larger than the initial amplitude (top right).
The bottom left panel for $t=850M$
shows the wave forms after the ``bosenova''.
After the amplitude stops growing, the gravitational waves
with a nontrivial waveform are radiated. Until the occurrence of the bosenova,
the gravitational wave frequencies are close to twice the axion mass,
$\tilde{\omega}\approx 2\mu$. But after the bosenova, the
gravitational wave frequencies are a few times smaller than $2\mu$.
The gravitational wave amplitude approximately shows exponential decay in time.
After that there appears another phase during which the gravitational waves
are generated with the frequency $\approx 2\mu$, but keeping a relatively larger
amplitude compared to the initial amplitude (compare the top left and bottom right panels).

%
\begin{figure}[tbh]
\centering
\includegraphics[width=0.9\textwidth,bb= 0 0 360 155]{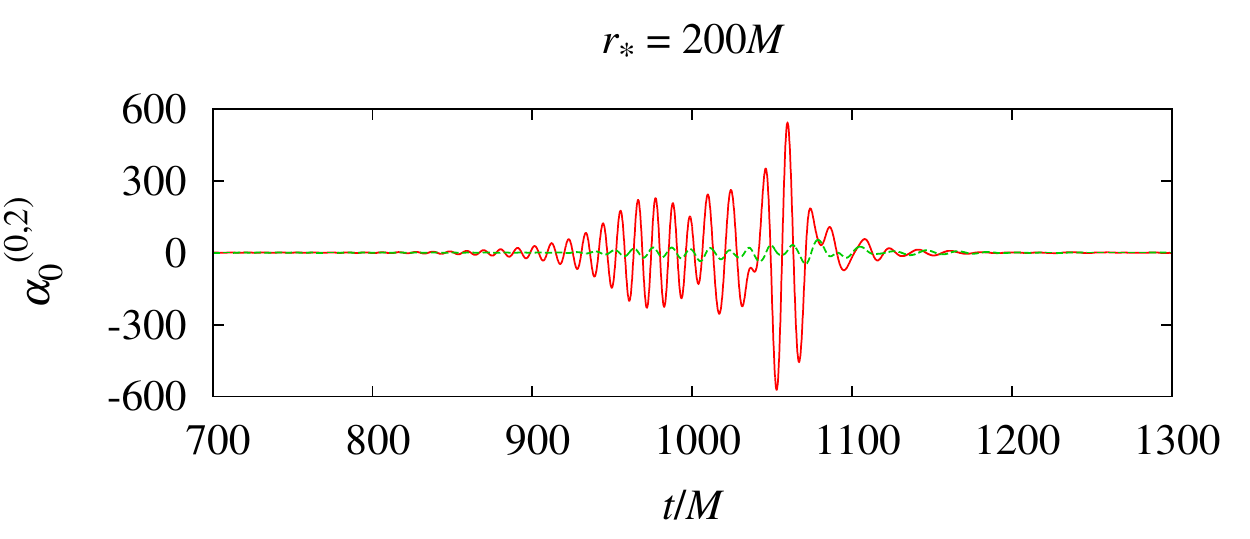}
\caption{Behaviour of real part of the spectral component $\alpha_2^{(0)}$ 
(solid line, red online) and $\alpha_0^{(0)}$ (dashed line, green online)
of the Teukolsky variable for a gravitational perturbation
as functions of time observed at the point $r_*=200M$ in the strongly nonlinear case.}
\label{Fig:Timeevolution-GW-amp05-n0m2-n0m0}
\end{figure}
%

\Fref{Fig:Timeevolution-GW-amp05-n0m2-n0m0} shows the value of
$\alpha_{\tilde{n}}^{\tilde{m}}$ for $(\tilde{n},\tilde{m})=(0,2)$ (solid line)
and $(\tilde{n},\tilde{m})=(0,0)$ (long dashed line) as functions of time
observed at the fixed position $r_*=200M$. We can confirm that
$\alpha_0^{(2)}$ shows the exponential growth and has quite
nontrivial wave form during the bosenova. After that, it decays
with frequencies few times smaller than $2\mu$.
The highly dynamical process of the bosenova generates the $\tilde{m}=0$ mode as well.
Generation of the $\tilde{m}=0$ mode reflects the rapid change
in the radial distribution of the axion cloud, and it shows frequencies different by a factor from
$2\mu$.

%
\begin{figure}[tbh]
\centering
\includegraphics[width=0.65\textwidth,bb= 0 0 360 252]{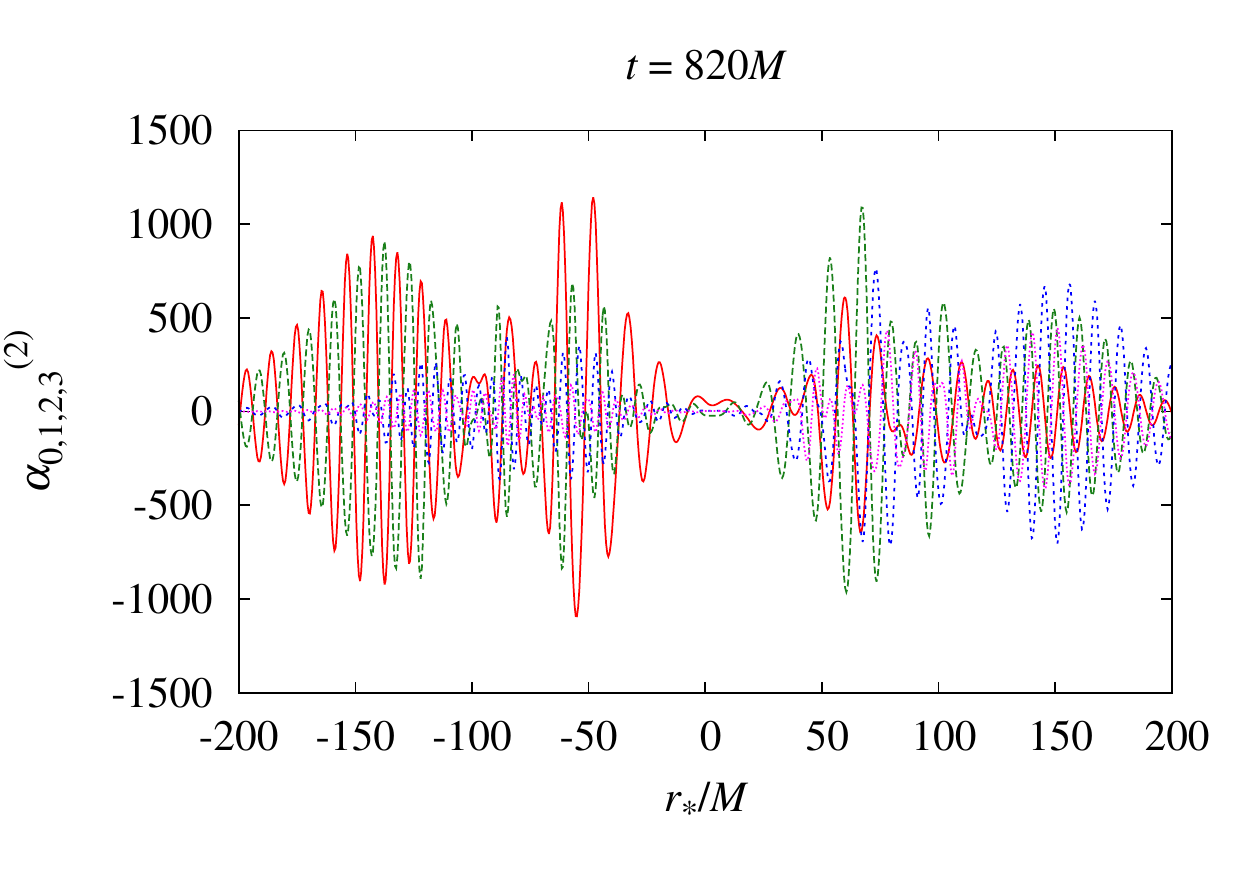}
\caption{A snapshot of the real part of the spectral components
$\alpha_{\tilde{n}}^{(\tilde{m})}$ of the Teukolsky variable for a gravitational perturbation
for fixed $\tilde{m}=2$ and $\tilde{n}=0$  (solid line, red online),
$1$ (long-dashed line, green online), $2$ (short-dashed line, blue online),
and $3$ (dotted line, purple online) at time $t=820M$ in the
strongly nonlinear case.}
\label{Fig:Snapshots-GW-amp05-n0n1n2n3m2}
\end{figure}
%

\Fref{Fig:Snapshots-GW-amp05-n0n1n2n3m2} shows a snapshot
for $t=820M$, but for $\alpha_{\tilde{n}}^{(\tilde{m})}$ with $\tilde{m}=2$ and
$\tilde{n}=0$ (solid line),
$1$ (long-dashed line), $2$ (short-dashed line),
and $3$ (dotted line).
This is the time when the bosenova phenomena is about to cease.
Similarly to the mildly nonlinear case,
the higher $\tilde{n}$ modes are also excited and show similar patterns
to the $\tilde{n}=0$ mode.

To summarize, in the highly nonlinear case,
the gravitational waves show quite interesting behaviour.
In particular, the gravitational wave amplitude grows exponentially
until it becomes $10^4$ times the initial amplitude.
Further, during bosenova, gravitational waves with a nontrivial wave forms are radiated
with frequencies smaller than $2\mu$. Also,
gravitational waves in higher $\tilde{\ell}$ modes are excited
with  patterns similar to the lower $\tilde \ell$ modes.

\subsection{Discussion on this section}

In this section, by numerical simulations, we have studied the emission of gravitational waves
sourced by the scalar field in the three cases: the case without self-interaction,
a mildly nonlinear case, and a strongly nonlinear case.
From the results of these simulations, the gravitational
wave emission from a realistic axion cloud around a rotating black hole would be expected
as follows.
When the scalar field amplitude is small, $\varphi\ll 1$,
continuous gravitational waves are emitted,
and as the field amplitude grows large, beating appears in them.
The bosenova happens at some critical amplitude of scalar field,
and associated with it, the amplitude of gravitational waves
grows exponentially, and then, burst-type gravitational waves with
nontrivial waveforms are emitted. Under the influence of
scalar field self-interaction, gravitational wave forms become complicated
and may not be modeled by a simple formula.
This would make it difficult
to search gravitational wave signals in the data analysis.
But the good news is that the self-interaction enhances the amplitude
of gravitational waves, even by a factor of $10^4$ during the bosenova,
which would be helpful to detect the signals. Since bosenovae happen
intermittently in a long time scale, burst-type gravitational waves would be
emitted like a geyser from the axion cloud around a black hole.

The growth in the amplitude by a factor of $10^4$ may seem too large.
Naively, the increase in the gravitational wave amplitude is
expected to be a factor of $\sim 10^2$,
because the field amplitude increases by a factor of $10$, and therefore,
the increase in the source term of the energy-momentum tensor in the
Teukolsky equation is $\sim 10^2$.
However, in the Klein-Gordon case, the source term is $O(10^4)$
while the emitted gravitational waves have amplitude of $O(10^{-2})$.
This means that the scalar field quasibound states has a configuration
to cause a significant phase cancellation. In the strongly nonlinear case,
the field configuration changes
due to the nonlinear self-interaction, and this makes the effect of the phase cancellation
weaker. This is the interpretation for the significant increase in the gravitational
wave amplitude during the bosenova.

In our approach, we calculated the Teukolsky variable ${}_s\Psi$ for $s=-2$.
For the choice $s=-2$, ${}_{-2}\Psi$ is related to $\psi_0$ of the Newman-Penrose variables.
While $\psi_0$ is very convenient to calculate the ingoing energy flux to the
black hole, it is not directly related the energy flux formula towards infinity.
In order to evaluate the outgoing flux, the conversion from $\psi_0$
to $\psi_4$ is necessary (see a similar discussion on conversion from $\psi_4$ to $\psi_0$
for incoming waves from infinity for \cite{Press:1973-2}).
Since the relation between the amplitudes
of $\psi_0$ and $\psi_4$ depends on the angular frequency $\tilde{\omega}$,
we have to calculate Fourier transform of emitted gravitational waves,
and reconstruct $\psi_4$. This procedure is necessary to predict the observed
gravitational wave forms and is under progress.

Although we do not know the exact waveform, a rough estimate on
the detectability of the gravitational wave burst emitted during the bosenova
can be given as follows.
Let us consider the case where the black hole candidate, Cygnus X-1, wears a cloud
of axion scalar field with mass $2.4\times 10^{-12}~\mathrm{eV}$, which
corresponds to $M\mu\approx 0.30$.
Up to the bosenova, the gravitational waves show
exponential growth with keeping $\tilde{\omega}\approx 2\mu$
that corresponds to $\approx 1200~\mathrm{Hz}$, and
there is a period $\Delta t \approx 100M$ during which the gravitational amplitude
is larger than the Klein-Gordon case by a factor of $10^{3}$ or more.
On the other hand, in our previous paper \cite{Yoshino:2014} we calculated
the amplitude of continuous gravitational waves from Cygnus X-1
in the case that it wears the quasibound state of a Klein-Gordon field,
using the black hole parameters determined by observations (see, e.g., \cite{Fragos:2014}).
Then, the gravitational wave amplitude from the bosenova is obtained just
by multiplying $10^3$ to our previous estimate for the Klein-Gordon case.
The result is
%
\begin{equation}
h_0\sim 10^{-19}\left(\frac{f_{\rm a}}{10^{16}~\mathrm{GeV}}\right)^2
\end{equation}
%
after substituting the mass $M$, the distance $d$, and axion mass $\mu$.
Then, multiplying $\sqrt{\Delta t}$,
the strain amplitude becomes $h_{\rm rss}\sim 10^{-20}/\sqrt{\mathrm{Hz}}$.
This is about $10$ times larger than the observation threshold of LIGO
if the GUT scale decay constant is adopted.
This shows the possible detectability of gravitational waves from the bosenova.

Here, we note that the reliability of the estimate
here based on the Schwarzschild case may be questioned,
because there is difference between the bosenovae
in the Schwarzschild case and the Kerr case.
In the Kerr case, the infall of positive energy happens
as a result of excitation of $m=-1$ mode, and it is a slower process
compared to the Schwarzschild case.
This may affect the amplitude of gravitational wave burst
during the bosenova. Therefore,
it is necessary
to extend our code to the Kerr background case in order to make our
calculations more realistic and to enable our simulation
results applicable to observational data analysis.

%
%
\section{Summary and discussion}
\label{Sec:VI}

In this paper, we have examined the three features of time evolution of an axion cloud
around a rotating black hole by numerical calculations/simulations.
In the superradiant phase, we calculated the growth rate by the superradiant
instability without self-interaction (i.e., a the massive Klein-Gordon field)
including the overtone modes as well as the fundamental modes (\sref{Sec:II}).
Our results indicate
that for $\ell = m = 3$, the overtone modes can have larger growth rates
compared to the fundamental mode when the black hole is rapidly rotating.
In \sref{Sec:III}, we studied the phase where the nonlinear self-interaction becomes important
by simulating the behaviour of the scalar field numerically. 
The bosenova collapse happens for $\ell = m = 1$ mode, and it becomes more violent
as $M\mu$ is decreased. For $\ell = m = 2$ mode, our simulation results
suggest that the bosenova collapse would not happen, but a steady outflow
will be formed instead. If we consider superposition of two modes, further rich phenomena
were observed. 
In \sref{Sec:IV}, we showed our preliminary simulations
of gravitational wave emission from an axion cloud
in the presence of axion nonlinear self-interaction
around a Schwarzschild black hole.
Although the extension to the Kerr case is necessary,
we have obtained a lot of indications for these phenomena.
In particular, gravitational wave burst is emitted when the
scalar field collapse to a black hole.
More detailed summary and discussion are given at the end of
each section.

Up to now, we took attention to the phenomena that happens in the
initial phase (superradiant instability) or in the time scale of $10^3M$
after that (bosenovae and gravitational wave emission).
In what follows, we consider the evolution of an axion cloud
and a black hole in a longer time scale. 
In a long time scale, the black hole parameters $M$ and $J$
change in time due to the backreaction of energy/angular momentum extraction
by the axion cloud. The evolution of the black hole parameters was discussed
in the original axiverse papers
\cite{Arvanitaki:2009,Arvanitaki:2010}, and the more detailed
adiabatic evolution was calculated in 
\cite{Brito:2014}, taking account of 
the supply of energy/angular momentum by the accretion disk,
the energy/angular momentum extraction by the scalar cloud,
and the loss of the energy/angular momentum 
of the axion cloud by the gravitational wave emission. 
When the superradiant instability becomes important
as the black hole is spun up by the accretion disk,
the scalar cloud rapidly grows to decrease the black hole spin
parameter $a_*$ to the marginal value satisfying $\omega=m\Omega_{\rm H}$.
After that, the spin parameter gradually grows due to the effect of the accretion disk.
If we switch on the self-interaction effect, the 
nonlinear effect becomes significant 
when the axion field becomes
order of the decay constant, $\Phi\sim f_{\rm a}$, and
the axion cloud energy at this time is 
$E/M \sim 10^{3}(f_{\rm a}/M_{\rm pl})^2$. For $f_{\rm a}$
with the GUT scale ($1/10$ of the GUT scale),
the corresponding energy is $\sim 0.1\%$ 
($\sim 0.001\%$) of the black hole mass. 
Therefore, the evolution is expected to 
highly depend on the value of $f_{\rm a}$,
and the scenario would become different from 
the ones in \cite{Arvanitaki:2009,Arvanitaki:2010,Brito:2014}.

One important additional quantity in the presence of the
nonlinear self-interaction is the energy/angular momentum loss rates of the 
axion cloud by 
 the bosenova ($\ell = m = 1$ case) or the axion outflow
($\ell  = m = 2$ case). Since the direct approach is
very difficult because the nonlinear self-interaction
makes the system highly dynamical, one 
has to consider an effective approximation. If we take an average over
the timescale much longer than the bosenova occurrence, 
the energy/angular momentum of the gravitationally bounded
axion field would be approximately constant, whose values depend only on $a_*$, $M\mu$, and $f_{\rm a}$. 
The decrease in the energy/angular momentum of the black hole due to the 
superradiant instability in short time scales will be erased out 
after the averaging because of the recycling processes by the bosenova.
Then, the system is regarded as a converter of the
black hole energy/angular momentum 
into scattered axion fields and gravitational waves. 
Due to this conversion, the black hole decreases
the spin parameter $a_*$ to the marginal value.
Because its time scale becomes longer as $f_{\rm a}$ is decreased,
we may be able to obtain constraints
on $f_{\rm a}$ from the fact that some solar-mass black holes
are rapidly rotating like Cygnus X-1. 
This would give a method for constraining
the axion model parameter different from the one of \cite{Yoshino:2014} 
that uses the direct observation of gravitational waves.

In order to carry out this program, we have to make 
our gravitational wave code applicable to the Kerr case and discuss
how to include the backreaction of gravitational wave
emission to the axion field dynamics.
Then, by long term simulations, we will be able to
determine the averaged conversion rate of the extracted
energy/angular momentum to the scattered axion fields and gravitational waves.
We are planning to challenge these problems as the next step.

\ack

We would like to thank Tetsutaro Higaki for useful comments. This work was
supported by the Grant-in-Aid for Scientific Research (A)
(No. 26247042)
from Japan Society for the Promotion of Science (JSPS).

\appendix

%
%
\section{Numerical method: Scalar field}
\label{Appendix:A}

In this Appendix, we give supplementary informations for the
numerical method for the scalar field simulations
with the pseudospectral approach.

\subsection{Equation}

The explicit form for the sine-Gordon equation for $\varphi$ in the
Kerr background spacetime is
%
\begin{eqnarray}
\fl
\left[(r^2+a^2)^2-\Delta a^2\sin^2\theta\right]\ddot{\varphi}=
(r^2+a^2)^2\varphi_{,r_*r_*}
+2r\Delta \varphi_{,r_*}
-4Mar\dot{\varphi}_{,\phi}
\nonumber\\
+\Delta\left(\varphi_{,\theta\theta}+\cot\theta\varphi_{,\theta}\right)
+\frac{\Delta - a^2\sin^2\theta}{\sin^2\theta}\varphi_{,\phi\phi}
-\mu^2\Sigma\Delta \sin\varphi.
\end{eqnarray}
%
Substituting
\eref{Eq:decompose-phi-scalar} and \eref{Eq:decompose-phi-sinvarphi},
we have
%
\begin{eqnarray}
\fl
\left[(r^2+a^2)^2-\Delta a^2(1-y^2)\right]\ddot{f}^{(m)}=
(r^2+a^2)^2f^{(m)}_{,r_*r_*}
+2r\Delta f^{(m)}_{,r_*}
\nonumber\\
-(4\rmi mMar)\dot{f}^{(m)}
+\Delta(1-y^2)f^{(m)}_{,yy}
-2(|m|+1)\Delta yf^{(m)}_{,y}
\nonumber\\
\qquad\qquad\qquad\qquad
+\left[m^2a^2-\Delta(m^2+|m|)\right]f^{(m)}
-\mu^2\Sigma\Delta s^{(m)}.
\end{eqnarray}
%
after equating the coefficients of $\rme^{\rmi m\phi}$.
We further substitute \eref{Eq:decompose-y-scalar} and \eref{Eq:decompose-y-sinvarphi}.
Equating the coefficients of $y^n$,
the equation for $a_{n}^{(m)}$ is found to be
%
\begin{eqnarray}
\fl
[(r^2+a^2)^2-a^2\Delta]\ddot{a}_{n}^{(m)} + (a^2\Delta)\ddot{a}_{n-2}^{(m)}=
(r^2+a^2)^2a_{n~,r_*r_*}^{(m)}
+2r\Delta a_{n~,r_*}^{(m)}
\nonumber\\
-\left(4\rmi mMar\right)\dot{a}_{n}^{(m)}
+(n+2)(n+1)\Delta a_{n+2}^{(m)}
\nonumber\\
\fl
+
\left[
m^2a^2
-\left(
n^2+n
+m^2+|m|+2|m|n\right)\Delta\right] a_{n}^{(m)}
-\mu^2\Delta\left(r^2 \sigma_n^{(m)}
+a^2 \sigma_{n-2}^{(m)}\right).
\end{eqnarray}
%
Note that $a_{n-2}^{(m)}$ on the left-hand side must be interpreted as
zero for $n=0$ and $1$.

\subsection{Nonlinear term}

%
\begin{figure}[tbh]
\centering
\includegraphics[width=0.65\textwidth,bb=0 0 360 252]{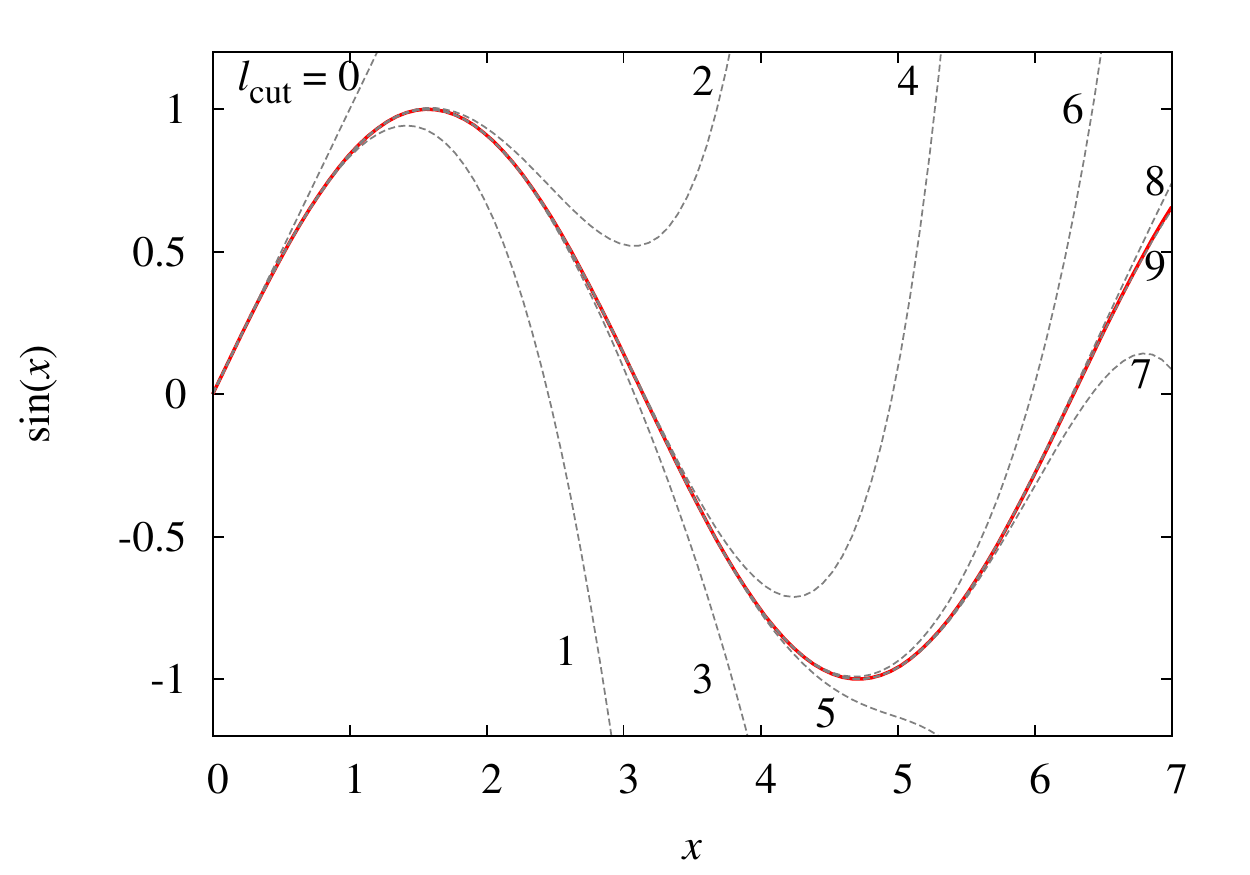}
\caption{Comparison between the value of $\sin x$ (solid line) and its Taylor series
up to the term $l=l_{\rm cut}$ for $l_{\rm cut}=0,...,9$ (dashed lines).}
\label{Fig:Taylor-sin}
\end{figure}
%

Here, we discuss how to compute the spectral components
$\sigma_{n}^{(m)}$ for the nonlinear term $\sin\varphi$.
This can be done with the Taylor expansion of $\sin\varphi$,
%
\begin{equation}
\sin\varphi = \sum_{l=0,1,2,...} \frac{(-1)^l}{(2l+1)!}\varphi^{2l+1}.
\end{equation}
%
In the numerical computation, we introduce a cut-off parameter $l_{\rm cut}$
for $l$.
The fortunate thing is that the convergence radius for this Taylor expansion is infinity.
This means that by increasing the value of $l_{\rm cut}$, we can approximate
$\sin\varphi$ in a larger range of $\varphi$. \Fref{Fig:Taylor-sin} compares
the values of $\sin x$ and its Taylor series up to $l=l_{\rm cut}$ for various $l_{\rm cut}$ values.
The Taylor series gives a good approximation from zero to a certain value of $x$, and
the approximation rapidly becomes bad beyond that point. The range
of good approximation becomes larger as $l_{\rm cut}$
is increased, and the difference is almost invisible for $l_{\rm cut}=9$ in the
range $0\le x\le 7$.

The spectral component of the Taylor series of $\sin\varphi$
can be calculated if we can compute each spectral
component of $\varphi^{2l+1}$. For this reason, let us derive the general relation
between the spectral components of two functions
and those of their product.
Let us consider $f$, $g$, and $h$, with
%
\numparts
\begin{eqnarray}
f =  \sum_{m=0,\pm1,\pm2,...} \sin^{|m|}\theta f^{(m)}(t,r,\theta) \rme^{\rmi m\phi},
\quad f^{(-m)}=f^{(m)*},
\\
g =  \sum_{m=0,\pm1,\pm2,...} \sin^{|m|}\theta g^{(m)}(t,r,\theta) \rme^{\rmi m\phi},
\quad g^{(-m)}=g^{(m)*},
\\
h =  \sum_{m=0,\pm1,\pm2,...} \sin^{|m|}\theta h^{(m)}(t,r,\theta) \rme^{\rmi m\phi},
\quad h^{(-m)}=h^{(m)*}.
\end{eqnarray}
\endnumparts
%
If $h=fg$ holds, we can derive the relation
%
\begin{equation}
h^{(m)} = \sum_{k=0}^{m}f^{(k)}g^{(m-k)} +
\sum_{k=1,2,...}(1-y^2)^k
\left(f^{(k+m)}g^{(k)*}+g^{(k+m)}f^{(k)*}\right).
\end{equation}
%
Next, let us consider the expansion with respect to $y$,
%
\begin{equation}
\fl
f^{(m)} = \sum_{n=0,1,...}a_n^{(m)}y^n, \quad
g^{(m)} = \sum_{n=0,1,...}b_n^{(m)}y^n, \quad \textrm{and} \quad
h^{(m)} = \sum_{n=0,1,...}c_n^{(m)}y^n.
\end{equation}
%
Then, we find the relation
%
\begin{equation}
\fl
c^{(m)}_n=
\sum_{k=0}^{m}\sum_{j=0}^{n}a_j^{(k)}b_{n-j}^{(m-k)}
+\sum_{k=1,2,...}\sum_{j=0}^{n}\sum_{i=0}^{j}
{}_kD_{n-j}\left(a_i^{(k+m)}b_{j-i}^{(k)*}+b_{i}^{(k+m)}a_{j-i}^{(k)*}\right),
\end{equation}
where we defined
\begin{equation}
{}_kD_{l}:=\left\{
\begin{array}{cl}
{}_{k}C_{l/2} & (l=0,2,4,...,2k),\\
0 & (l=1,3,5,...), \\
0 & (l=2k+2,2k+4,...).
\end{array}
\right.
\end{equation}
%
Using this relation, we can compute $\varphi^{2l+1}$ iteratively, and thus,
the spectral components $\sigma_{n}^{(m)}$ of $\sin\varphi$.

%
%
\section{Numerical method: Teukolsky variables}
\label{Appendix:B}

In this Appendix, we give supplementary informations for the
numerical method for the gravitational wave simulations.

\subsection{Equation}

We solve the Teukolsky equation for $s=-2$ by the pseudospectral decomposition
in the angular coordinates in a similar manner to the scalar case.
The equation for ${\psi}^{(\tilde{m})}$, the spectral components
of $\rme^{\rmi\tilde{m}\phi}$,  is presented in
\eref{Eq:Teukolsky-modify}. We further decompose the coordinate
$y$ with \eref{Eq:psi-y-decomposition}. By equating the coefficients
of $y^{\tilde{n}}$, we have
%
\begin{eqnarray}
\fl
[(r^2+a^2)^2-a^2\Delta]\ddot{\alpha}_{\tilde{n}}^{(\tilde{m})}
+ (a^2\Delta)\ddot{\alpha}_{\tilde{n}-2}^{(\tilde{m})}
=
(r^2+a^2)^2\alpha_{\tilde{n}~,r_*r_*}^{(\tilde{m})}
\nonumber\\
\fl\qquad
+{2(r^2+a^2)}
\left[\Delta_{,r}-\frac{\Delta}{r}+\frac{r\Delta}{r^2+a^2}\right] \alpha_{\tilde{n}~,r_*}^{(\tilde{m})}
-{4}\left[M(r^2-a^2)-\Delta r+\rmi\tilde{m}Mar\right]\dot{\alpha}_{\tilde{n}}^{(\tilde{m})}
\nonumber\\
+(4\rmi a)\Delta\dot{\alpha}_{\tilde{n}-1}^{(\tilde{m})}
+(\tilde{n}+2)(\tilde{n}+1)\Delta \alpha_{\tilde{n}+2}^{(\tilde{m})}
-(\tilde{n}+1)\left(|\tilde{m}-2|-|\tilde{m}+2|\right)\alpha_{\tilde{n}+1}^{(\tilde{m})}
\nonumber\\
\fl
+ \left\{
{\tilde{m}^2a^2}-{4\rmi\tilde{m}a(r-M)}
+{\Delta}\left(2-\frac{3}{r}\Delta_{,r}+\frac{2\Delta}{r^2}\right)
\right.
\nonumber\\
\fl
\left.\qquad
-\frac{\Delta}{2}\left(
2\tilde{n}^2-2\tilde{n}+\tilde{m}^2+|\tilde{m}^2-4|+3|\tilde{m}-2|+3|\tilde{m}+2|
\right)
\right\}\alpha_{\tilde{n}}^{(\tilde{m})} + t_{\tilde{n}}^{(\tilde{m})},
\end{eqnarray}
%
where we have assumed that the source term can be expressed in the
form of the series
%
\begin{equation}
4\pi\frac{2^{\frac{|\tilde{m}-2|+|\tilde{m}+2|}{2}}\Sigma r}{\Delta (1-y)^{|\tilde{m}-2|/2}(1+y)^{|\tilde{m}+2|/2}}
T^{(\tilde{m})}=\sum_{\tilde{n}=0,1,2,...}t_{\tilde{n}}^{(\tilde{m})}y^{\tilde{n}}.
\end{equation}
%
Note that $\alpha_{\tilde{n}-2}^{(\tilde{m})}$ on the left-hand side must be interpreted as
zero for $\tilde{n}=0$ and $1$.

\subsection{Regularity of the source term}

In the Schwarzschild case, the source term is reduced
to a relatively simple formula that guarantees the regularity at the two poles, $y=\pm 1$.
The explicit form of $T^{(m_1,m_2)}$, defined in \eref{Eq:T-decompoision-m1m2},
can be derived by substituting
$\varphi = \sum_{m_1} \sin^{|m_1|}\theta f^{(m_1)}\rme^{\rmi m_1\phi}=\sum_{m_2} \sin^{|m_2|}\theta f^{(m_2)}\rme^{\rmi m_2\phi}$ into \eref{Eq:T-components}, calculating
\eref{Eq:T-definition} with \eref{Eq:T4-definition}, and
extracting the coefficient $T^{(m_1,m_2)}$  of  $\rme^{\rmi (m_1+m_2)\phi}$.
In order to make the expression concise, it is useful to introduce
the two differential operators:
%
\numparts
\begin{eqnarray}
{\cal D}^{(1)} = \partial_t-\partial_{r_*},
\\
{\cal D}^{(2)} = \left({\cal D}^{(1)}\right)^2
-\frac{2(r-3M)}{r^2}{\cal D}^{(1)}.
\end{eqnarray}
\endnumparts
%
Using these operators, the following formula is derived:
%
\begin{eqnarray}
\fl
\frac{(4/r^2) T^{(m_1,m_2)}}{(1-y)^{|m_1+m_2-2|/2}(1+y)^{|m_1+m_2+2|/2}}  = (m_2^2-|m_2|)
(1+y)^{\frac{p_1^{(+)}}{2}}
(1-y)^{\frac{p_1^{(-)}}{2}}
f^{(m_2)}{\cal D}^{(2)}f^{(m_1)}
\nonumber\\
+
(m_1^2-|m_1|)
(1+y)^{\frac{p_2^{(+)}}{2}}
(1-y)^{\frac{p_2^{(-)}}{2}}
f^{(m_1)}{\cal D}^{(2)}f^{(m_2)}
\nonumber\\
\fl
+
2m_1m_2
(1+y)^{\frac{p_3^{(+)}}{2}}
(1-y)^{\frac{p_3^{(-)}}{2}}
\left(-{\cal D}^{(1)}f^{(m_1)}{\cal D}^{(1)}f^{(m_2)}+\frac{\Delta^2}{r^6}f^{(m_1)}f^{(m_2)}\right)
\nonumber\\
\fl
+
(1+y)^{\frac{p_4^{(+)}}{2}}
(1-y)^{\frac{p_4^{(-)}}{2}}
\left(f^{(m_2)}_{,yy}{\cal D}^{(2)}f^{(m_1)}+f^{(m_1)}_{,yy}{\cal D}^{(2)}f^{(m_2)}
-2{\cal D}^{(1)}f^{(m_1)}_{,y}{\cal D}^{(1)}f^{(m_2)}_{,y}+2\frac{\Delta^2}{r^6}f^{(m_1)}_{,y}f^{(m_2)}_{,y}\right)
\nonumber\\
\fl
+2m_2
(1+y)^{\frac{p_5^{(+)}}{2}}
(1-y)^{\frac{p_5^{(-)}}{2}}
\left(-f^{(m_2)}_{,y}{\cal D}^{(2)}f^{(m_1)}
+{\cal D}^{(1)}f^{(m_1)}_{,y}{\cal D}^{(1)}f^{(m_2)}-\frac{\Delta^2}{r^6}f^{(m_1)}_{,y}f^{(m_2)}\right)
\nonumber\\
\fl
+2m_1
(1+y)^{\frac{p_6^{(+)}}{2}}
(1-y)^{\frac{p_6^{(-)}}{2}}
\left(-f^{(m_1)}_{,y}{\cal D}^{(2)}f^{(m_2)}
+{\cal D}^{(1)}f^{(m_2)}_{,y}{\cal D}^{(1)}f^{(m_1)}-\frac{\Delta^2}{r^6}f^{(m_1)}f^{(m_2)}_{,y}\right),
\end{eqnarray}
%
where
%
\numparts
\begin{eqnarray}
p_1^{(\pm)} = |m_1|+|m_2|-2+4\theta(\pm m_2)-|m_1+m_2\pm 2|
\\
p_2^{(\pm)} = |m_1|+|m_2|-2+4\theta(\pm m_1)-|m_1+m_2\pm 2|
\\
p_3^{(\pm)} = |m_1|+|m_2|-2+2\theta(\pm m_1)+2\theta(\pm m_2)-|m_1+m_2\pm 2|
\\
p_4^{(\pm)} =|m_1|+|m_2|+2-|m_1+m_2\pm 2|
\\
p_5^{(\pm)} = |m_1|+|m_2|+2\theta(\pm m_2)-|m_1+m_2\pm 2|
\\
p_6^{(\pm)} = |m_1|+|m_2|+2\theta(\pm m_1)-|m_1+m_2\pm 2|
\end{eqnarray}
\endnumparts
%
Here, all $p_i^{(\pm)}$ are non-negative for arbitrary integers $m_1$ and $m_2$,
and hence, the regularity
of the source term at the two poles is guaranteed.
Substituting $f^{(m_1)} = \sum_{n}a_n^{(m_1)}y^n$ and $f^{(m_2)} = \sum_{n}a_n^{(m_2)}y^n$,
we can derive the expression for $t_{\tilde{n}}^{(m_1,m_2)}$
for each value of $(m_1,m_2)$.

\end{document}